\documentclass[12pt,preprint]{aastex}

\usepackage{graphicx}
\usepackage{enumerate}

\newcommand{\simgt}{\lower.5ex\hbox{$\;\buildrel>\over\sim\;$}}
\newcommand{\simlt}{\lower.5ex\hbox{$\;\buildrel<\over\sim\;$}}

\newcommand{\msun}{\ensuremath{M_\odot}}
\newcommand{\lsun}{\ensuremath{L_\odot}}
\newcommand{\hi}{\rm H{\sc i}}
\newcommand{\hii}{\rm H{\sc ii}}
\newcommand{\heii}{\rm He{\sc ii}}

\newcommand{\ntot}{N$_{\rm tot}$}

\newcommand{\nhtwo}{N$_{\rm H2}$}
\newcommand{\av}{A$_{\rm V}$}

\newcommand{\neii}{\rm [Ne\,{\sc ii}]}
\newcommand{\neiii}{\rm [Ne\,{\sc iii}]}
\newcommand{\arii}{\rm [Ar\,{\sc ii}]}
\newcommand{\ariii}{\rm [Ar\,{\sc iii}]}
\newcommand{\nev}{\rm [Ne\,{\sc v}]}
\newcommand{\siii}{\rm [S\,{\sc iii}]}
\newcommand{\sii}{\rm [S\,{\sc ii}]}
\newcommand{\SiII}{\rm [Si\,{\sc ii}]}
\newcommand{\siv}{\rm [S\,{\sc iv}]}
\newcommand{\oiv}{\rm [O\,{\sc iv}]}

\newcommand{\feiii}{\rm [Fe\,{\sc iii}]}
\newcommand{\feii}{\rm [Fe\,{\sc ii}]}
\newcommand{\htwo}{\rm H$_2$}
\newcommand{\water}{{\rm H}$_2${\rm O}}

\newcommand{\oh}{{\rm OH}}
\newcommand{\logoh}{12$+$log(O/H)}
\newcommand{\nfn}{$\langle\nu{\rm L}_\nu\rangle$}
\newcommand{\nfnfar}{$\langle\nu{\rm L}_\nu (71)\rangle + \langle\nu{\rm L}_\nu (160)\rangle$}
\newcommand{\nfncolor}{$\langle\nu{\rm L}_\nu (71)\rangle/\langle\nu{\rm L}_\nu (160)\rangle$}
\newcommand{\ltir}{$L_{\rm TIR}$}
\newcommand{\wmsq}{W\,m$^{-2}$}
\newcommand{\cmthree}{cm$^{-3}$}
\newcommand{\cmtwo}{cm$^{-2}$}
\newcommand{\pctwo}{pc$^{-2}$}

\newcommand{\sigpah}{$\Sigma$(PAH)}
\newcommand{\sightwo}{$\Sigma[{\rm H}_2(0-2)]$}
\newcommand{\cgcg}{CGCG\,$005-027$}

\newcommand{\hb}{H{$\beta$}}

\newcommand{\Ne}{$n_{\rm e}$}

\newcommand{\sbs}{SBS\,0335$-$052\,E}
\newcommand{\izw}{I\,Zw\,18}

\newcommand{\spitzer}{{\it Spitzer}}
\newcommand{\hers}{{\it Herschel}}

\newcommand{\iso}{{\it ISO}}
\newcommand{\iras}{{\it IRAS}}
\newcommand{\mopex}{{\it MOPEX}}
\newcommand{\spice}{{\it SPICE}}
\newcommand{\irsclean}{{\it IRSCLEAN}}

\shorttitle{The \spitzer\ View of Low-Metallicity Star Formation}
\shortauthors{Hunt et al.}

\begin{document}

\title{The \spitzer\ View of Low-Metallicity Star Formation: 
III. Fine Structure Lines, Aromatic Features, and Molecules}

\author{ Leslie~K.~Hunt$^1$, Trinh~X.~Thuan$^2$,
	Yuri~I.~Izotov$^3$ and 
	Marc~Sauvage$^4$ }


\altaffiltext{1}{INAF-Osservatorio Astrofisico di Arcetri,
Largo Fermi 5, I-50125 Firenze, Italy; hunt@arcetri.astro.it.}
\altaffiltext{2}{Department of Astronomy, University of Virginia,
PO Box 400325, Charlottesville, VA 22904-4325, USA;  txt@virginia.edu}
\altaffiltext{3}{Main Astronomical Observatory, National Academy of Sciences of Ukraine, 03680 Kiev, Ukraine; izotov@mao.kiev.ua}
\altaffiltext{4}{CEA/DSM/DAPNIA/Service d'Astrophysique, UMR AIM, CE Saclay, 
91191 Gif sur Yvette Cedex, France; msauvage@cea.fr}

\begin{abstract}
We present low- and high-resolution \spitzer/IRS spectra, 
supplemented by IRAC and MIPS measurements,
of 22 blue compact dwarf (BCD) galaxies. 
The BCD sample spans a wide range in oxygen abundance [\logoh\ between 7.4 and 8.3],
and hardness of the interstellar radiation field (ISRF).  
The IRS spectra provide us with a rich set of diagnostics 
to probe the physics of star and dust formation in very low-metallicity
environments. We find that metal-poor BCDs have harder 
ionizing radiation than metal-rich galaxies: \oiv\ emission is 
$\simgt$ 4 times as common as \feii\ emission. They also have 
a more intense ISRF, as indicated by the 71 to 160\,\micron\ luminosity ratio.
Two-thirds of the sample (15 BCDs) show PAH features, although the fraction 
of PAH emission normalized to the total infrared (IR) luminosity 
is considerably smaller in metal-poor BCDs ($\sim$0.5\% ) 
than in metal-rich star-forming galaxies ($\sim$10\%). 
We find several lines of evidence for a deficit of small PAH carriers
at low metallicity, and attribute this to destruction by a hard,
intense ISRF, only indirectly linked to metal abundance.
Our IRS spectra reveal a variety of \htwo\ rotational lines,
and more than a third of the objects in our sample (8 BCDs) have $\simgt 3\sigma$ 
detections in one or more of the four lowest-order transitions. 
The warm gas masses in the BCDs range from $10^3$ to $10^8$\,\msun, and
can be comparable to the neutral hydrogen gas mass; 
relative to their total IR luminosities, 
some BCDs contain more \htwo\ than SINGS galaxies.

\end{abstract}

{\it Facilities:} \facility{Spitzer ()}

\section{Introduction}

First \iras\ and \iso, then SCUBA and {\it COBE}, and most recently
\spitzer\ have convincingly shown that most of
the star formation in the universe is enshrouded in dust.
Large populations of infrared-luminous galaxies at $z\simlt$1.5
contribute to about 70-80\% of the far-infrared
and 30\% of the sub-millimeter backgrounds, 
and are probably responsible for most of the star-formation activity at
high redshift 
\citep[e.g.,][]{chary01,lefloch05,dole06}.
Indeed, half the energy and most of the photons in 
the universe come from the infrared spectral region
\citep[e.g.,][]{hauser01,franceschini08}.

That dust is so prominent in the high-redshift universe
may appear surprising, since it has been assumed that
dust would be absent in low-metallicity environments.
However, recent theoretical developments challenge the assumption
that dust is absent in metal-poor gas. 
Large amounts of dust can be created on short timescales
by supernovae (SNe)
and Asymptotic Giant Branch stars which evolve in a metal-free
ISM \citep[e.g.,][]{todini01,schneider04,nozawa07,bianchi07,valiante09}.
Observationally,
the copious dust emission observed at very high redshifts
($z\simgt6$) in quasars implies that indeed
dust can form rapidly at early epochs \citep{bertoldi03}.
Although luminous quasars can hardly be considered as 
typical examples of star-forming environments, observations 
of sub-millimeter galaxies \citep[SMGs,][]{chapman05} and 
dust-obscured galaxy populations \citep[DOGs,][]{dey08} 
also suggest that early star formation episodes can also be very intense
and relatively brief.
However, exactly how these massive starbursts occur and evolve
is not yet clear.
The short interval in which star formation and the ensuing chemical enrichment
and dust formation convert a dust-free metal-free environment to a dusty
metal-rich one by redshift $\sim$6 is as yet unobserved, and studies of 
such transitions remain a major observational challenge. 

The Local Universe is home to star-forming dwarf galaxies that are of much
lower metallicity than those observed thus far at high redshift. 
Over the last thirty years or so, only a handful
of blue compact dwarf (BCD) galaxies have been discovered
with \logoh$\sim$7.2. Despite intensive searches, only one 
BCD, SBS\,0335$-$052\,W, is known to host \hii\ regions with \logoh$\leq$7.1 \citep{izotov09}. 
Even at these extremely low metallicities,
BCD spectral energy distributions (SEDs) can be dominated by 
infrared dust reprocessing of ultraviolet radiation
from young massive stars
\citep[e.g., \sbs\ and \izw:][]{thuan99a,houck04,wu07}. 
Because such galaxies are chemically unevolved, they 
can provide a window on primordial galaxy formation and evolution.
This is, in some sense, a ``local'' approach to a cosmological problem.
If we can study the properties of a metal-poor ISM and its constituents
locally, we may be able to better understand the high-redshift transition from
metal-free Population\,III (Pop III) stars to the chemically evolved
massive galaxies typical of the current epoch.

We adopt here this ``local'' approach. 
With the aim of characterizing the properties of star and dust formation in 
a metal-poor ISM,
we have used the \spitzer\ Space Telescope to study 
a sample of nearby low-metallicity star-forming dwarf galaxies.
This is the third in a series of papers;
the first two papers have focused on two particular BCDs, Haro 3, the highest
metallicity object in our sample 
\citep{hunt06}, and Mrk 996, a BCD with an unusually dense and 
compact nuclear \hii\ region \citep{thuan08}. 
Here, we present for the first time the entire sample of 23 BCDs, and
report on its spectral properties, as derived from IRS observations\footnote{We
present IRS results for only 22 BCDs, since one, SBS\,0940$+$544, was observed
in a Guaranteed Time Program.}.
The sample has been carefully chosen to span a wide range in
oxygen abundance: \logoh\ varies between 7.4 and 8.3, with a median of
7.9. About a quarter 
of the objects in the sample are in the eXtremely Metal-Deficient (XMD) 
or Extremely Metal-Poor Galaxy (EMPG) range 
\citep[\logoh$\simlt$7.65][]{pustilnik07,brown08}.
Thus our sample comprises the largest number of XMDs/EMPGs to date with 
good signal-to-noise IRS spectra.

In $\S$\ref{sec:observations}, we describe the sample selection criteria,
together with the \spitzer\ observations and the data reduction methods.
In $\S$\ref{sec:analysis}, 
we derive fluxes of lines and features in the spectra by fitting them with 
PAHFIT \citep{smith07}. 
We present and discuss our results for 
the infrared (IR) fine-structure (FS) lines in $\S$\ref{sec:ionized}, 
the aromatic features in $\S$\ref{sec:aromatics}, 
and in $\S$\ref{sec:molecules}, the molecular lines. 
Our conclusions are summarized in $\S$\ref{sec:summary}.

\section{Sample Definition, Observations, and Data Reduction
\label{sec:observations}}


One of the most important factors in shaping the SED at low metallicity
is the hardness and intensity of the interstellar radiation field (ISRF).
The ISRF hardness and intensity govern
dust properties such as the grain size distribution and temperature. 
Hard and intense ISRFs may destroy small grains all together,
thus suppressing the aromatic or polycyclic aromatic hydrocarbon features
(PAHs) which are typical of metal-rich starburst galaxies
\citep[e.g.,][]{voit92,madden06,smith07}.
In the post-\iso\ and post-\spitzer\ eras, ratios of IR lines of neon 
have often been used
to quantify the hardness of the ISRF in star-forming
galaxies \citep[e.g.,][]{thornley00,dale06}.
However, since we have selected the objects in our sample on the basis
of their optical spectra,
we have quantified ISRF hardness optically, with the ratio 
of the nebular \heii\ $\lambda$\,4686 emission line relative to \hb. 
The \heii\ line, with an ionization potential of 54.4\,eV,
is a good indicator of a hard ultraviolet (UV) radiation field.
Nebular \heii\ lines are predicted to be very strong in metal-free 
(or nearly metal- free) primordial galaxies \citep{schaerer02,schaerer03},
although
no \heii-emitting high-redshift galaxy (z$\simgt$5) has yet been discovered 
\citep[e.g.,][]{nagao08}.

However, some low-redshift low-metallicity BCDs are known to show  
strong nebular \heii\ $\lambda$4686 emission with \heii/\hb$\simgt$3\%
\citep []{guseva00, izotovthuan04, thuanizotov05}.
We have selected 23 objects so that they span a wide range in \heii/\hb.  
Roughly 1/3 of the sample consists of objects having strong \heii\ emission,
with \heii/\hb$\simgt$2\%;
another third have intermediate \heii\ emission with
\heii/\hb$\sim$1-2\%, 
and the remaining third has weak or no \heii\ emission, with 
\heii/\hb$\simlt$1\%.
This wide range of \heii\ strength in our sample 
also provides a large range in metallicity as 
there is a loose correlation between the two quantities \citep{thuanizotov05}.
The general properties of the sample are given in Table \ref{tab:sample}.

The observations presented in this paper are part of our GO
\spitzer\ program (PID 3139) for 23 BCDs. In the course of this program, 
we have obtained IRS spectra
in the low- and high-resolution modules \citep[SL, SH, LH][]{houckirs},
IRAC images at 4.5 and 8\,\micron\ \citep{fazioirac},
and MIPS images at 24, 71, and 160\,\micron\ \citep{riekemips}.
For all instruments, the data reduction starts with
the Basic Calibrated Data ({\it bcd}). 
The corresponding masks (the DCE masks),
furnished by the \spitzer\ Science Center pipeline,
were then used to flag 
potential spurious features in the images, such as 
strong radiation hits, saturated pixels, or nonexistent/corrupted data.

\subsection{IRS spectroscopy}

Spectroscopy was performed in the staring mode, 
in both orders of the Short Low-resolution module (SL1, SL2)
and with the Short and Long High-resolution modules (SH, LH) \citep{houckirs}.
Thus, we have obtained low-resolution spectra from 5.2 to 14.5\,\micron\ (R$\simeq$64-128), and 
high-resolution spectra from 9.6 to 37.2\,\micron\ (R$\simeq$600).
We have used different acquisition schemes, depending on the 
brightness of the source as measured by its 
scaled \hb\ flux.
Sources were centered in the slits by peaking up on 2MASS stars.

The individual {\it bcd} frames were processed by the S13.2.0 version of the 
SSC pipeline, which performs ramp fitting, dark current 
subtraction, droop and linearity corrections,
flat-fielding, and wavelength and flux calibrations\footnote{See the IRS
Data Handbook, \url{http://ssc.spitzer.caltech.edu/irs/dh}.}.
Because the pipeline does not include background subtraction, we 
have constructed,
for the low-resolution long-slit spectra, a coadded 
background frame from the {\it bcd} observations 
with the source in the opposing nod and off-order positions
\citep[see also][]{weedman06}. 
Coadding was performed with the sigma-clipping option
of the {\it imcombine} task
in IRAF\footnote{IRAF is the Image Analysis and 
Reduction Facility made available to the astronomical community by the National Optical
Astronomy Observatory, which is operated by AURA, Inc., under
contract with the U.S. National Science Foundation.}. 
The inclusion of off-order and off-nod frames in the background image 
means that
the total integration time on the background is three times that on the source, 
thus improving the signal-to-noise of the two-dimensional (2D) subtraction.

We did not acquire a separate background spectrum
for the high-resolution SH and LH spectra, and the slit size precludes
using the same procedure as for the SL module. 
Therefore we subtracted the background from the SH and LH observations
using the one-dimensional (1D) spectra as described below.
The SH and LH 2D {\it bcd} images were coadded in the same way 
as for the SL modules,
and corrections for sporadic bad pixels and cosmic ray
hits were carried out manually by inspection of the images at
the separate nod positions.

We extracted the source spectra with \spice, the post-pipeline IRS package
furnished by the SSC.
Before extraction, all spectra at separated nod positions were cleaned with the \irsclean\
algorithm (http://ssc.spitzer.caltech.edu/postbcd/irsclean.html).
The automatic point-source extraction window was used for all modules.
Automatic extraction uses a variable-width extraction window which
scales with wavelength in order to recover a constant fraction of 
the diffraction-limited instrument response.
For the SL spectra, this gives a 4-pixel (7\farcs2) length at 6\,\micron, and
an 8-pixel length (14\farcs4) at 12\,\micron; 
the slit width is 3\farcs6 for both SL modules.
At high resolution, the \spice\ extraction was performed over the entire slit
(4\farcs7$\times$11\farcs3 SH; 11\farcs1$\times$22\farcs3 LH).
Orders were spliced together by averaging, ignoring the noisy
regions at the red end of each order \citep[e.g.,][]{armus04}.
Then the individual spectra were box-car smoothed to a resolution element, 
and clipped in order to
eliminate any remaining spikes in the high-resolution data.
Finally, the two spectra for each module (one for each nod position) 
were averaged.

Background was subtracted from the 1D SH spectra by
defining a level from the overlap region with the SL spectra,
from which we had subtracted a 2D background.
Specifically, the
difference between the 2D background-subtracted SL and SH spectra 
was minimized over
the substantial region where they overlap in wavelength ($\sim$5\,\micron). 
The spectral shape of the background was given by 
the model of Reach and coworkers\footnote{See
\url{http://ssc.spitzer.caltech.edu/documents/background}.}, and 
the multiplicative constant by the minimization process. 
Subsequently,
the LH background was subtracted in an analogous
way, by minimizing the difference
between the SH and LH spectra over their overlap region ($\la$1\,\micron).

This ``bootstrapping'' procedure for background subtraction
in the high-resolution modules should in principle 
eliminate the need for scaling adjustments.  
However, its accuracy depends on the unresolved nature of the sources
in the spectral apertures.
We measured the source sizes by fitting Gaussians to the 
IRAC 7.9\,\micron\ and MIPS 24\,\micron\ images. 
The mean (and median) image FWHM is 3\farcs7 at 7.9\,\micron, and 6\farcs2 at 24\,\micron.
The SL1 slit width is 3\farcs6, so most of the sources are sufficiently
compact to not significantly exceed the SL1 slit size.
The resolution of MIPS24 is $\sim$6\arcsec, making the BCDs essentially unresolved
at these longer wavelengths;
in fact, the characteristic diffraction pattern is seen in all the MIPS24 images.
The FWHM of the largest source at 24\,\micron\ (Tol\,1924$-$416) is 10\arcsec,
so even in the most extreme case,
the LH slit width of 11\farcs1 completely encompasses the BCD light
at 24\,\micron.
Our assumption of point-like morphologies at \spitzer\ wavelengths thus
appeared to be reasonably justified.
Nevertheless, for a few (of the largest) sources,
the level of the background needed to be adjusted
slightly to provide a smooth transition among the modules.
In any case, the 24\,\micron\ flux as measured from MIPS is generally
consistent with the continuum level derived for the IRS spectra (see below).

We have obtained IRAC and MIPS photometry for all 23 BCDs in our
sample.
However, we have IRS observations for only 22 BCDs. 
One BCD, SBS\,0940$+$544, has an IRS observation in the Guaranteed Time 
program (PID\,85). Thus, we present here IRS results for only 22 objects, and
will incorporate SBS\,0940$+$544 into our sample in a future paper.
 

\subsection{IRAC and MIPS Imaging}

Although the focus of this paper is on the IRS spectra of our sample
objects, we briefly
describe their \spitzer\ photometric data because we need photometric
quantities such as the 
total IR luminosity (\ltir) to properly interpret their spectral properties.
A more detailed description of the photometry and reduction 
procedures can be found in \citet[][Haro\,3]{hunt06} and
\citet[][Mrk\,996]{thuan08}. The discussion of the photometric data of
the whole sample is deferred to a future paper 
on SEDs.

Briefly, we acquired IRAC images in two filters, 4.5 and 7.9\,\micron,
each with two sets of four frames in the high dynamic range mode.
The individual {\it bcd} frames were processed with  
the S14.0.0 version of the SSC pipeline
(see the IRAC Data Handbook\footnote{Available from the SSC website
\url{http://ssc.spitzer.caltech.edu/irac/dh/}}).
The {\it bcd} frames (corrected for ``banding'' effects for bright sources)
were coadded using \mopex, the image mosaicing and source-extraction package
provided by the SSC \citep{mopex}.
Pixels flagged by the masks were subsequently ignored.
Additional inconsistent pixel values were removed by means of the \mopex\ outlier
rejection algorithms, in particular 
the dual-outlier technique, together with the multiframe algorithm.
The frames were corrected for geometrical distortion and
projected onto a ``fiducial'' (refined) coordinate system
with pixel sizes of 1\farcs20, roughly equivalent to the original pixels.
Standard linear interpolation was used for the mosaics.
The noise levels in our post-pipeline \mopex\ IRAC mosaics are comparable to
or lower than those in the SSC products.

Our MIPS images were acquired in the Fixed Cluster-Offset mode in all
three channels, with offsets of 12\arcsec\ in two additional
pointings.
The individual {\it bcd} frames were processed by the 
the S14.4.0 version of the SSC pipeline,
which converts the integration ramps inherent to the MIPS detectors into 
slopes, and corrects for temporal variations of the slope images
\citep{gordon05}.
As for the IRAC images, we processed the dithered {\it bcd} frames in 
the spatial domain with \mopex.
The DCE masks and the static masks were used to flag pixels which were 
subsequently ignored.
The \mopex\ outlier rejection was used to remove any additional spurious pixel
values, with the dual-outlier and multiframe algorithm as for the IRAC frames. 
Geometrical distortion was corrected before projecting the frames onto 
a fiducial coordinate system with pixel sizes of 1\farcs20 for MIPS24,
roughly half of the original pixel size of 2\farcs5.
Pixel sizes of the final mosaics at 71 and 160\,\micron\ are also 
approximately half
the originals, i.e. 4\farcs95 at 71\micron\ and 8\farcs0 at 160\,\micron.
Unlike the IRAC coadds,
we incorporated the sigma-weighting algorithm because 
it gave less noisy MIPS mosaics than without.
Standard linear interpolation was used in all cases.
In all three channels, our \mopex\ mosaics are superior to 
those provided by the automated post-pipeline reduction.

\subsection{\label{sec:phot} IRAC and MIPS Photometry}

We have performed aperture photometry on the IRAC and MIPS images
with the IRAF photometry package {\it apphot}, taking care to convert
the MJy/sr surface brightness units of the images to integrated flux units.
The background level was determined by averaging several 
adjacent empty sky regions. 
Fluxes were determined at radii where the growth curve has
become asymptotically flat.
In many cases, there are multiple sources within the IRS apertures,
and we used subjective judgment to match photometry with the spectroscopic
slit.
Following \citet{draine07}, these fluxes were used to derive 
the ``color temperature'' between 71 and 160\,\micron, 
\nfncolor,
and the total infrared flux (TIR):
$$L_{\rm TIR}\,=\,0.95\,\langle\nu L_\nu\rangle_{7.9}
+ 1.15\,\langle\nu L_\nu\rangle_{24}
+ \langle\nu L_\nu\rangle_{71}
+ \langle\nu L_\nu\rangle_{160} $$
\noindent
Both quantities will be pertinent for the discussion of the spectra.

In order to check our 
reduction and calibration procedures for both the IRS and MIPS,
we have compared the MIPS 24\,\micron\ total fluxes with the 
continuum levels in the IRS spectra. 
Such a comparison (see below) 
shows good general agreement, implying that both the
IRS flux calibration (in the case of a point-source) and background
subtraction and the derivation of the total MIPS flux have been done correctly.
The only exception is UM\,311, which is in a crowded field (there are several
point sources in close proximity to a spiral galaxy), and thus 
its photometry is probably contaminated by other sources.

All targets were detected at 24 and 71\,\micron.
However, at 160\,\micron, we obtained only upper limits (ULs) for 6 of them.
As a consequence of a \spitzer\ queue problem,
SBS\,1030$+$583 has no MIPS24 observations.

\section{Spectral Analysis \label{sec:analysis}}

The flux emitted by aromatic (or PAH) features constitutes  a large fraction
of the IR energy budget in a typical star-forming galaxy spectrum.
However, these features
are known to decrease in intensity or even disappear at low
metallicities \citep{engelbracht05,madden06,wu06,engelbracht08}.
Our sample contains many metal-deficient objects and 
is thus uniquely able to address the behavior of
PAH emission at low metallicities. In order to directly compare the results for 
our low-metallicity sample with those for 
an existing well-defined set of more metal-rich galaxies,   
we wanted to analyze our IRS spectra in the same manner as this latter set.  
This consideration dictated the use of PAHFIT \citep{smith07},
an IDL procedure which was developed for and
applied to the Spitzer Nearby Galaxy Survey \citep[SINGS,][]{kennicutt03}.

PAHFIT is particularly suited for separating emission lines
and PAH features (e.g., the PAH blend at 12.6--12.7\,\micron\ and the \neii\
line at 12.8\,\micron), as well as measuring faint PAH features superimposed
on a strong continuum.
Broad ``plateaus'' underlying the 8\,\micron\ 
and 16-17\,\micron\ emission \citep{peeters02,peeters04b}
and strong silicate absorption at 9.7\,\micron\
can make the measurement of broad PAH bands exceedingly difficult in
those wavelength regions.
PAHFIT overcomes this problem by fitting simultaneously the spectral features, 
the underlying continuum, and the extinction, and gives
quantitative estimates and uncertainties for all parameters.
Drude profiles are fitted to the PAH features, and Gaussian profiles   
to the molecular hydrogen and fine-structure lines.
In addition to PAHFITting the spectra, 
we also measured emission line fluxes over the entire IRS range
by fitting Gaussian profiles. 
Our strategy for deriving quantitative measurements from the IRS spectra is described in the following.

\subsection{PAHFIT \label{sec:pahfit}}

Our spectra have a higher resolution than those normally fitted with PAHFIT,
specifically for wavelengths $\geq$14.5\,\micron\ where we have no
low-resolution spectra.
Hence, after binning the spectra to a 0.03\,\micron\ resolution,
and fitting the entire spectrum, we use only  
the spectral lines measured with PAHFIT having $\lambda$ $\leq$ 
14.5\,\micron.
In contrast, we adopt all the dust features fitted by PAHFIT 
over the entire IRS range.
We modified the PAHFIT IDL code to allow the central wavelengths 
and widths of the Drude profiles to vary, and 
experimented with different weighting schemes.
We also tried several different combinations of dust continuum 
temperatures, but these did not influence the fit results.
PAHFIT was run with three different options, including
mixed extinction, screen extinction, and fixing extinction to zero.

The nominally best fits (those with lowest $\chi_\nu^2$) for 11 BCDs
had non-zero extinction.
However, in every case but Haro\,3 and Mrk\,996,
the statistical significance of the fit was comparable to similar fits,
but with zero extinction.
Hence, when possible, we opted for the fit with no extinction
[$\tau(9.7)$\,=\,0].

\begin{figure}
\centerline{
\hbox{ 
\includegraphics[angle=0,width=0.33\linewidth,bb=40 251 587 713]{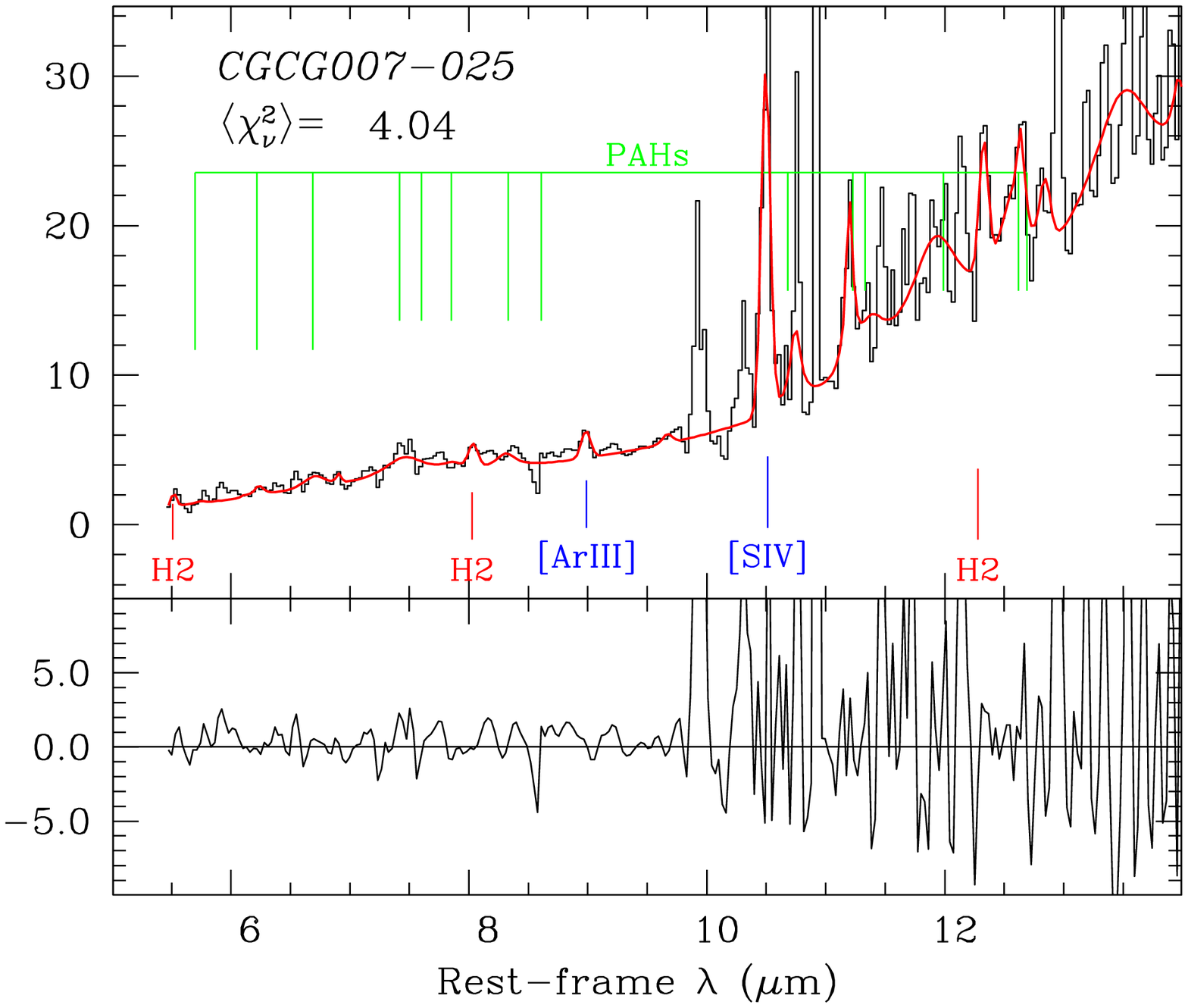} 
\hspace{-0.1cm}
\includegraphics[angle=0,width=0.33\linewidth,bb=40 251 587 713]{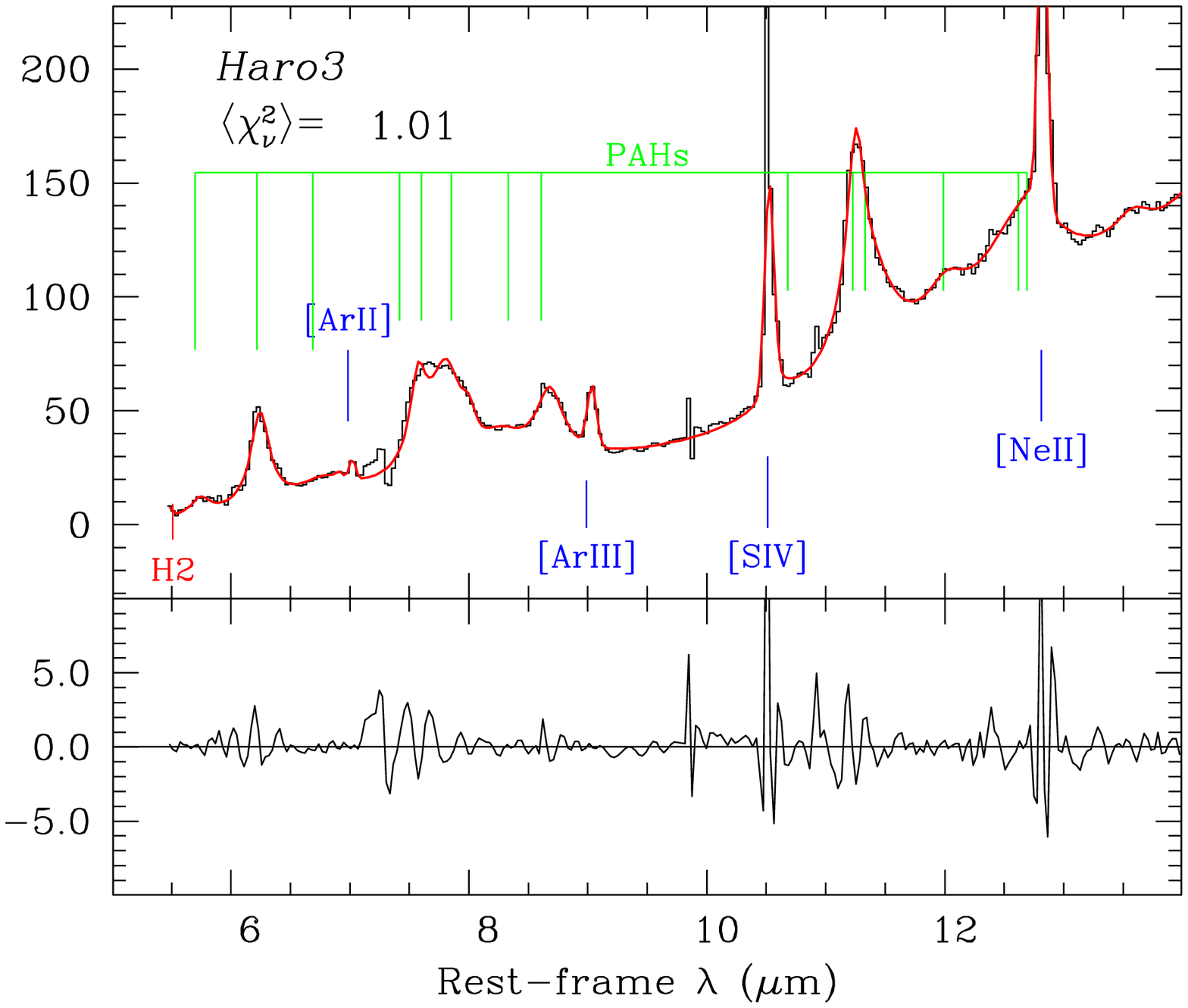}
\hspace{-0.1cm}
\includegraphics[angle=0,width=0.33\linewidth,bb=40 251 587 713]{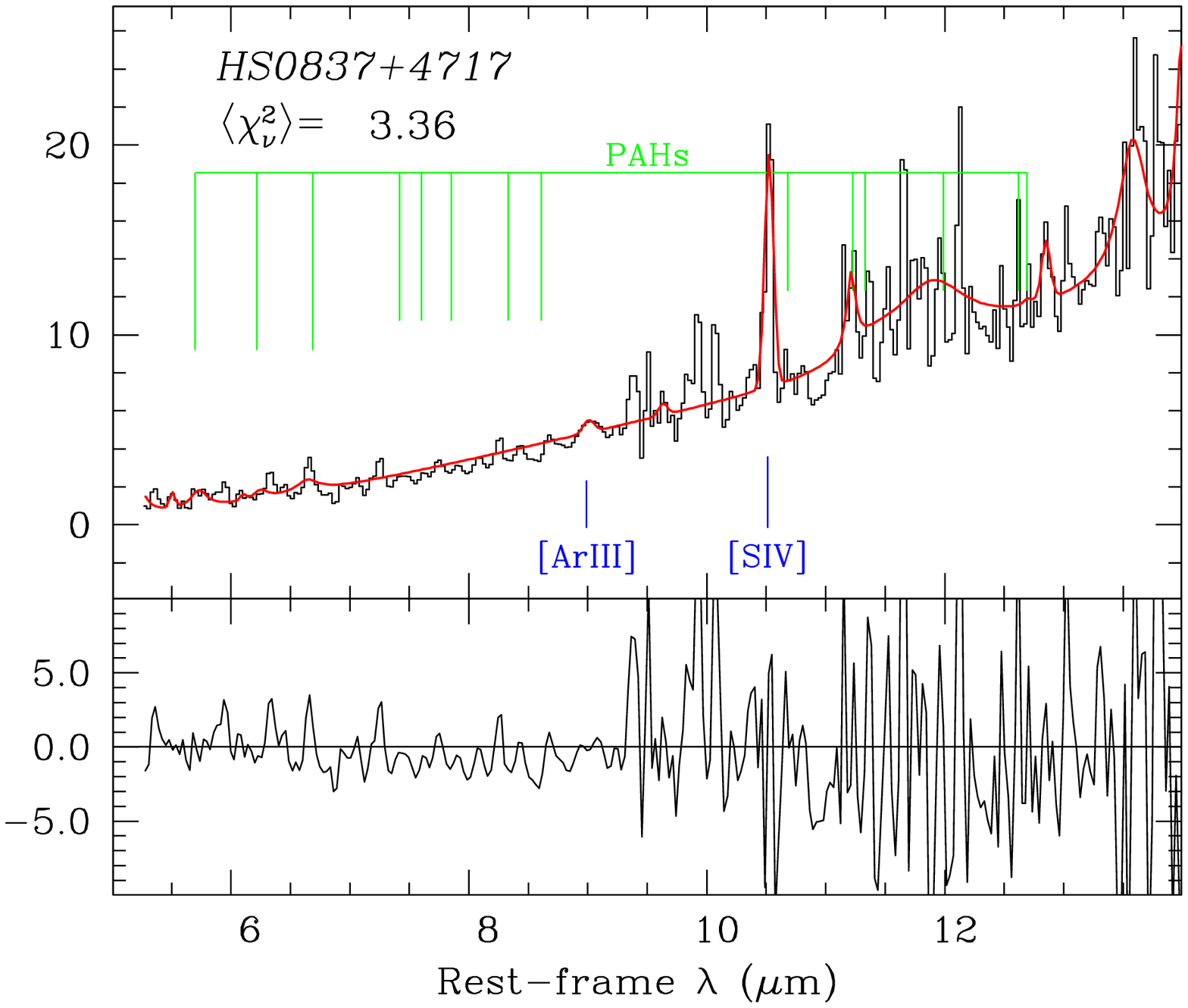} 
} 
}
\vspace{0.3cm}
\centerline{
\hbox{ 
\includegraphics[angle=0,width=0.33\linewidth,bb=40 251 587 713]{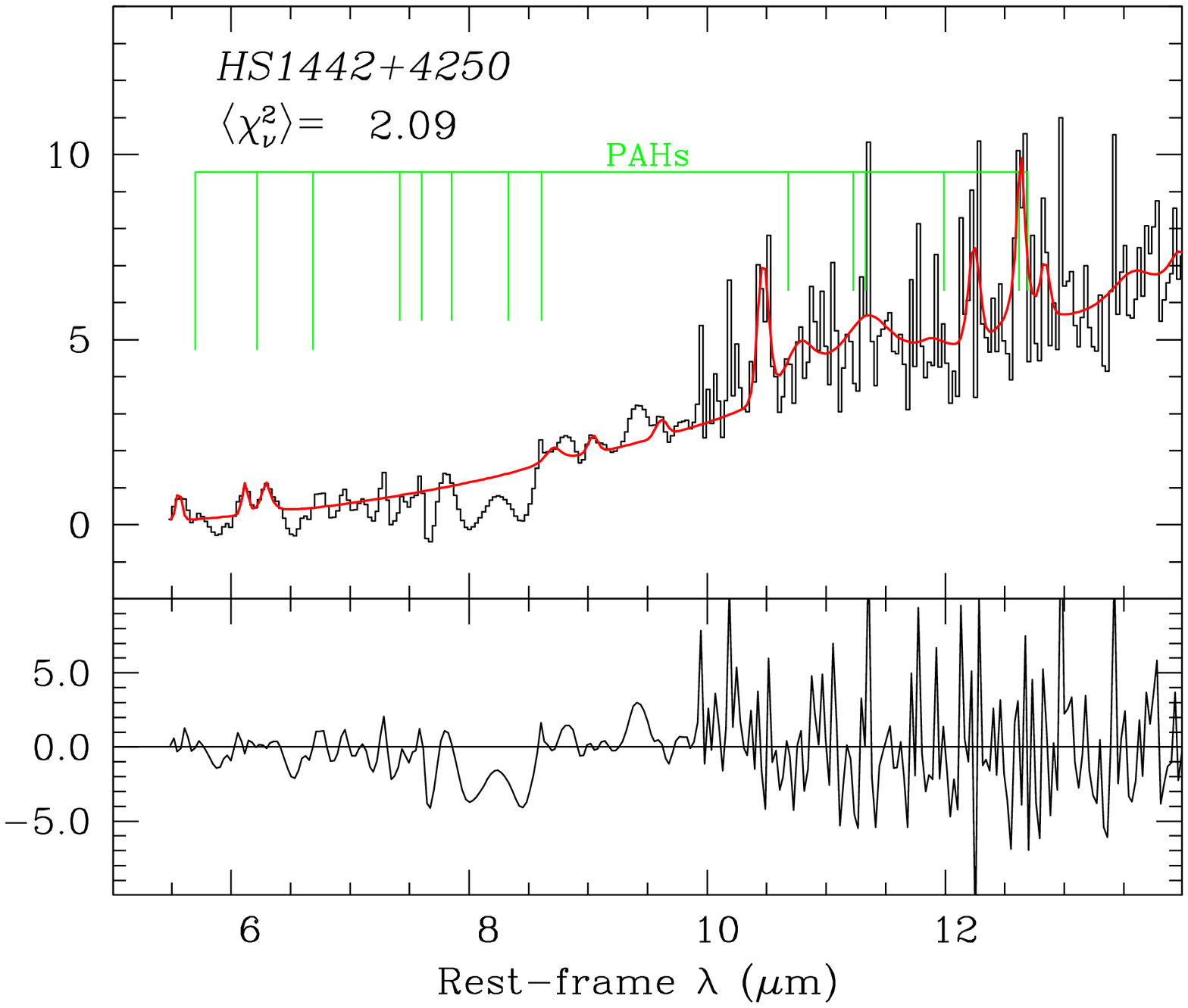}  
\hspace{-0.1cm}
\includegraphics[angle=0,width=0.33\linewidth,bb=40 251 587 713]{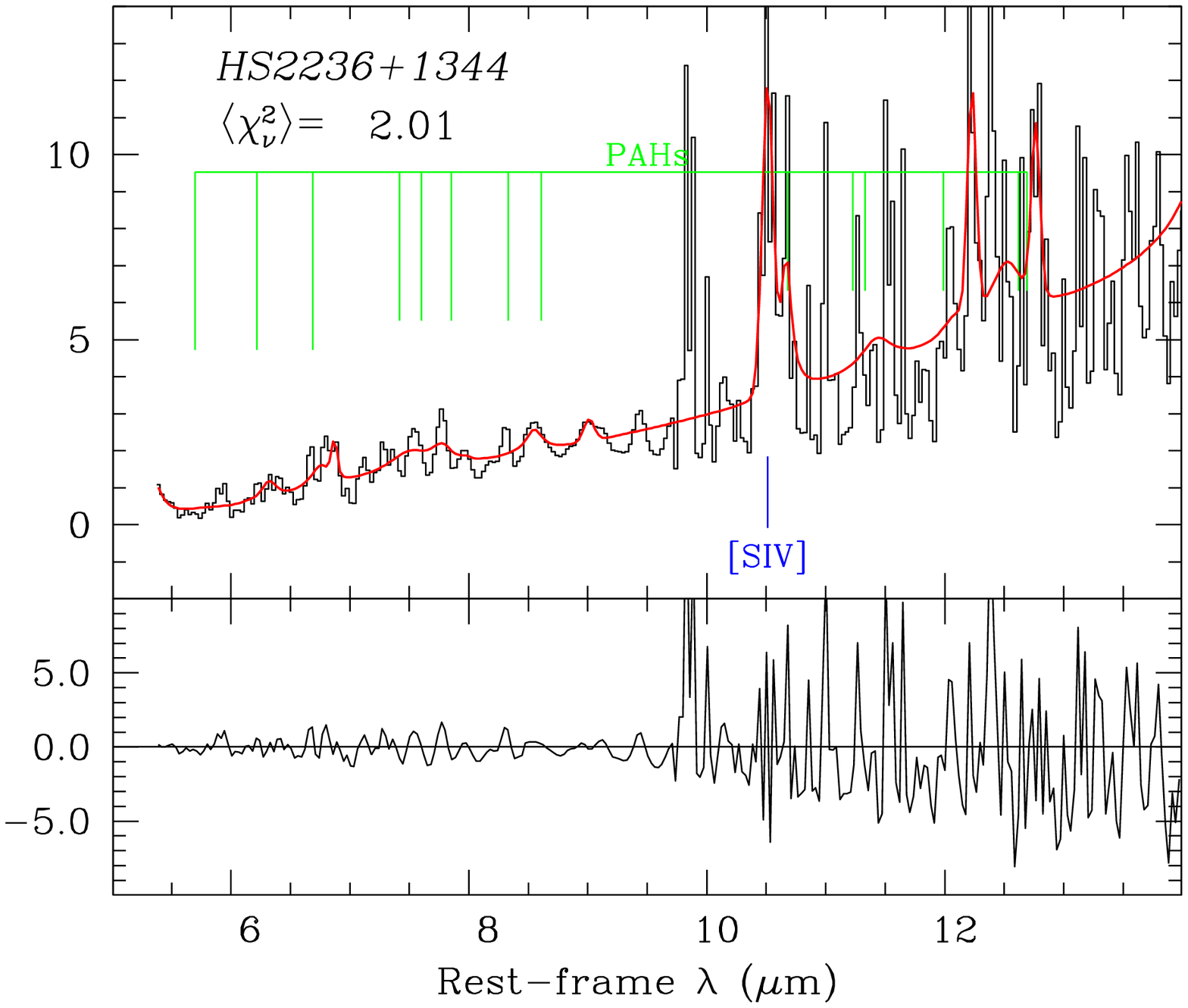} 
\hspace{-0.1cm}
\includegraphics[angle=0,width=0.33\linewidth,bb=40 251 587 713]{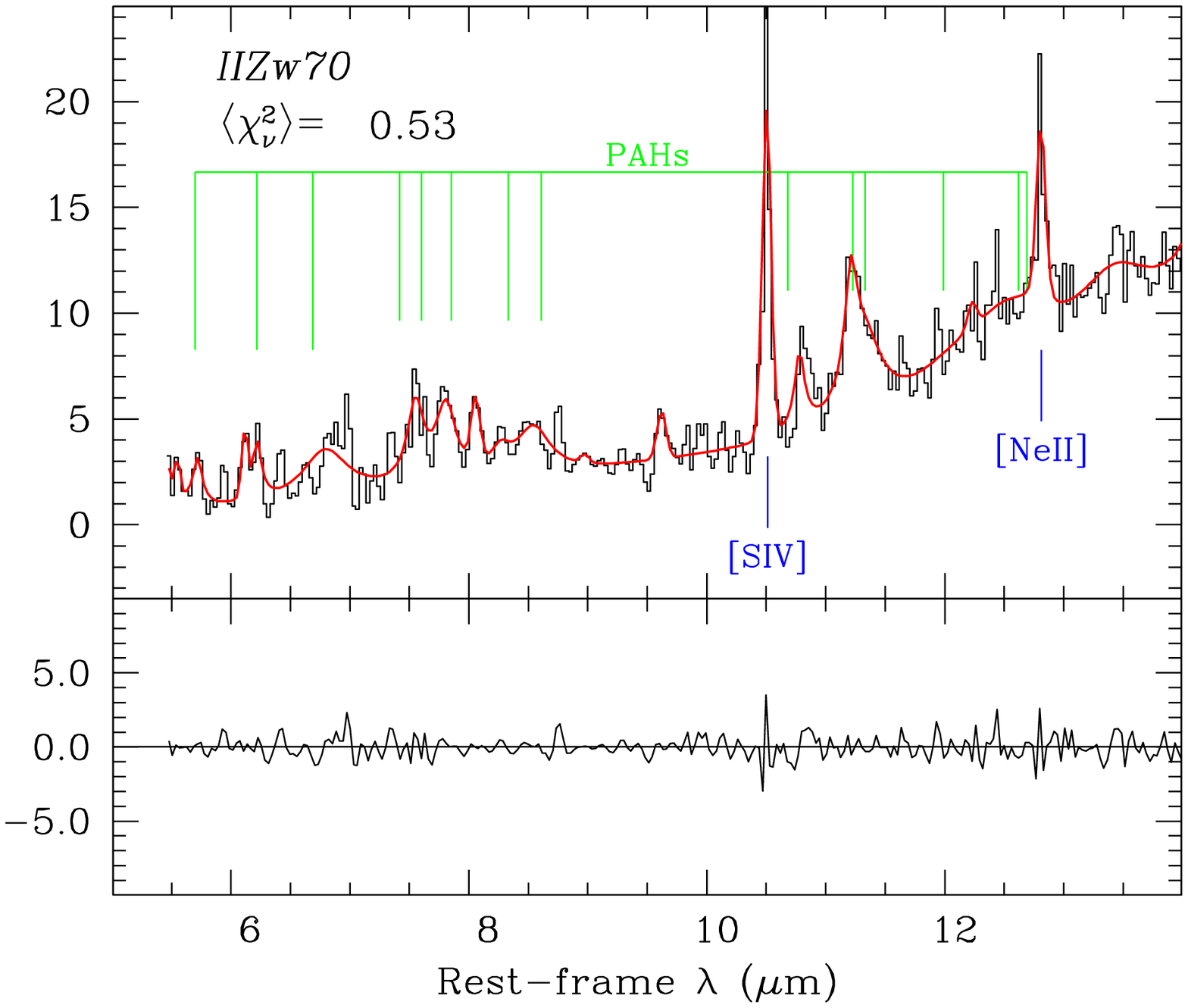} 
} 
}
\vspace{0.3cm}
\centerline{
\hbox{ 
\includegraphics[angle=0,width=0.33\linewidth,bb=40 251 587 713]{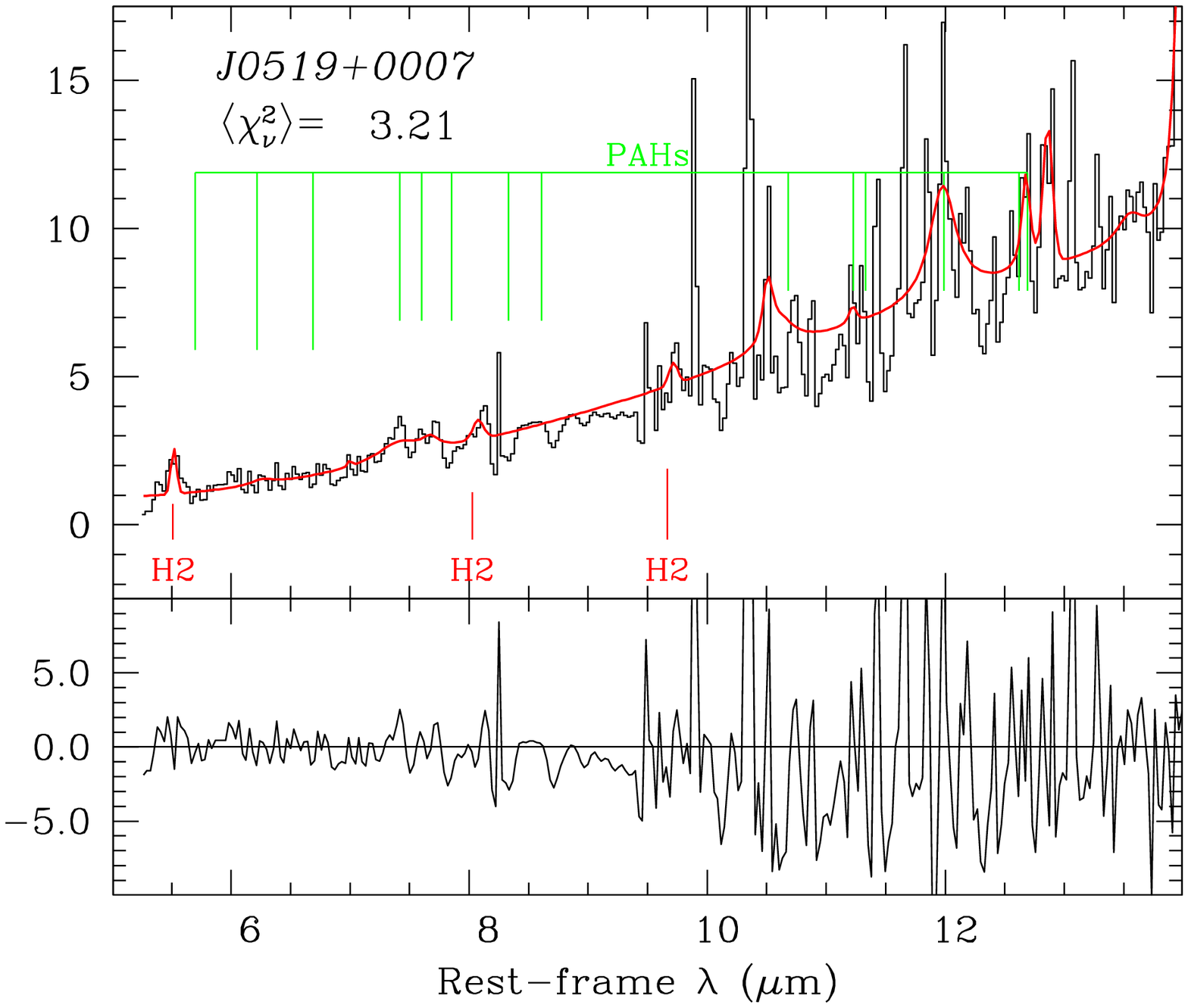}  
\hspace{-0.1cm}
\includegraphics[angle=0,width=0.33\linewidth,bb=40 251 587 713]{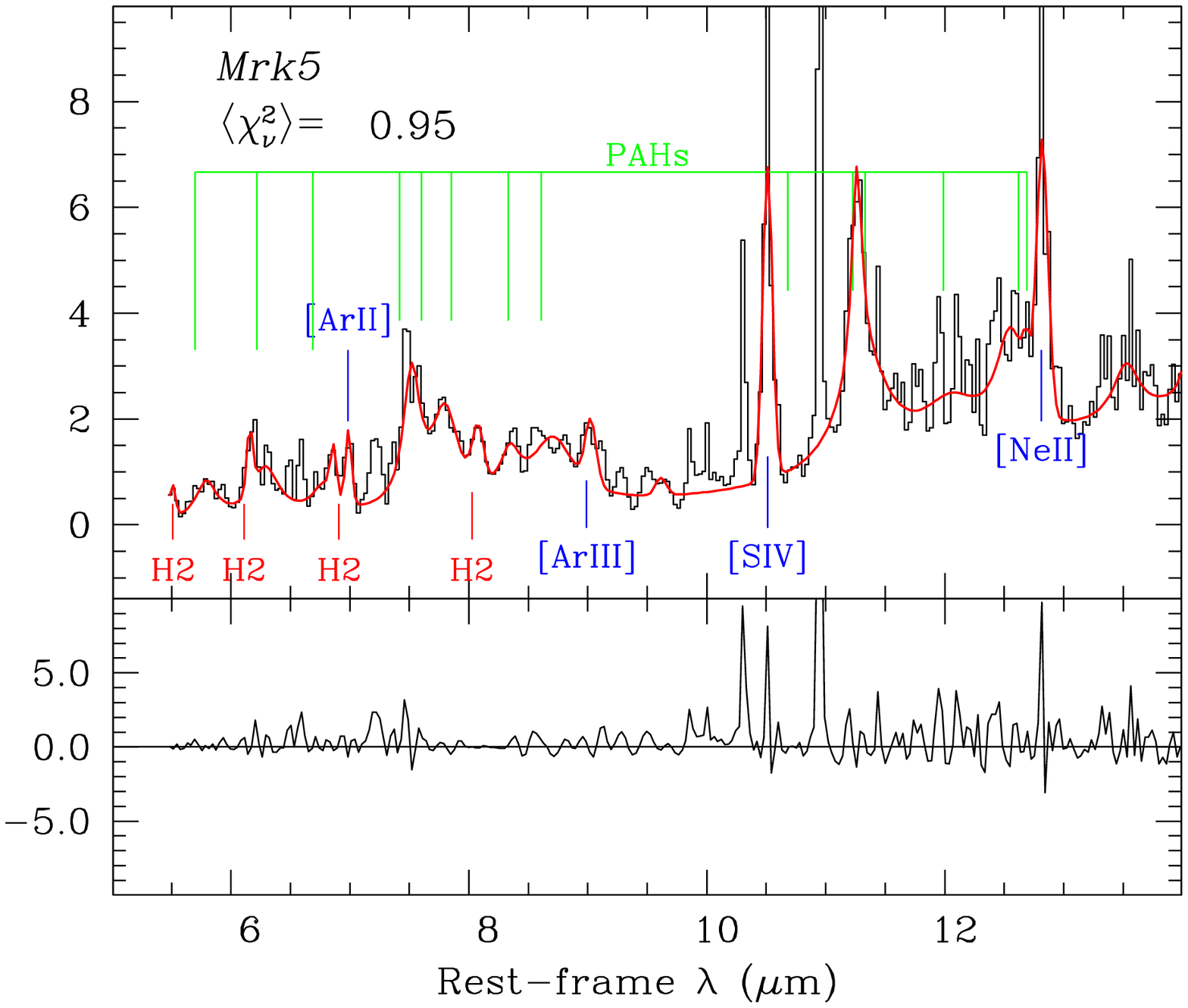}  
\hspace{-0.1cm}
\includegraphics[angle=0,width=0.33\linewidth,bb=40 251 587 713]{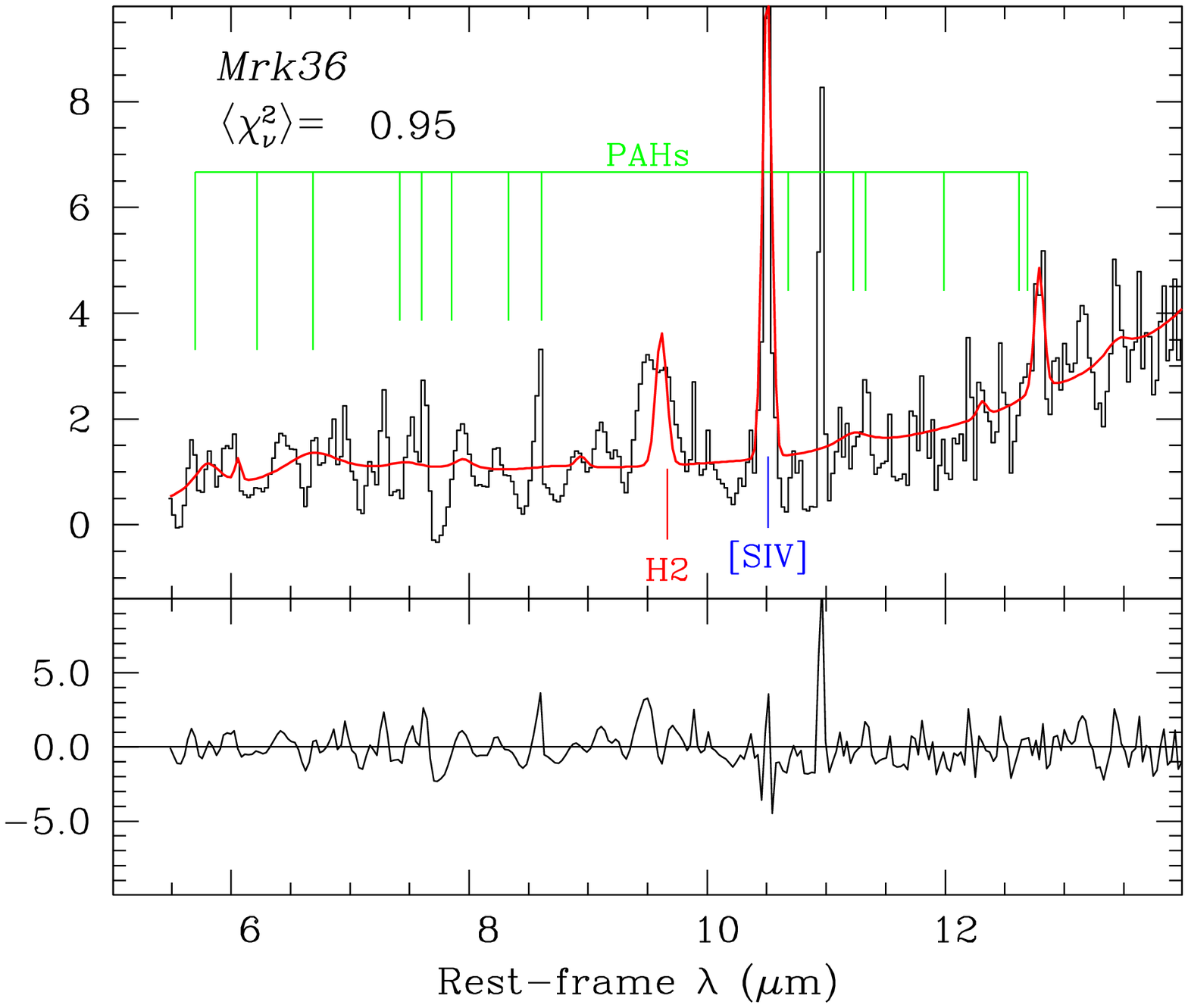} 
}}
\caption{IRS SL$+$SH spectra from 1 to 15\micron, with the best-fitting PAHFIT
model superimposed shown as a red curve.
The vertical axis is in units of mJy.
Significant emission-line identifications are shown as labeled vertical lines.
The top panels report the best-fit reduced $\chi^2_{\nu}$, obtained over the wavelength
region shown. 
The bottom panels show the residuals from the PAHFIT models.
\label{fig:pahfit1}}
\end{figure}

The results are shown in Figure \ref{fig:pahfit1}.
The top panels show the best-fit PAHFIT model as a red curve,
and the bottom panels give the residuals.
The reduced $\chi_\nu^2$ given in each panel in the figures is
calculated from 
the residuals of the fit up to 14.5\,\micron, normalized by the
mean spectral uncertainty over the same spectral range.
The figures show that the residuals are generally
quite well-behaved, and that $\chi_\nu^2$ is generally sufficiently
small to imply a good fit.
Uncertainties on the integrated line fluxes were estimated in two ways.
The standard deviation $\sigma$ per pixel
of the continuum was measured individually in a relatively clean region of each SL,
SH, and LH spectrum. 
A first estimate of the uncertainty in the line flux,
assumed to be spread over 3 pixels,
is taken to be $3\times3\sigma$. This is compared with a 
second uncertainty 
estimate, the one given by PAHFIT for each spectral feature. The largest
of the two estimates was then chosen to be the final uncertainty of  
the measurement.
Table \ref{tab:fspahfit} 
reports all FS lines with $\lambda\leq$14.5\,\micron\ 
detected at the $\simgt 3\sigma$ level with PAHFIT, and  
Table \ref{tab:dfpahfit} lists the dust features.

\subsection{Gaussian Fitting of Emission Lines \label{sec:gauss}}

We fitted all possible molecular and FS emission lines with Gaussian profiles,
taking the integrated fitted flux as the measure of the total flux in the line. 
When necessary, we used multiple Gaussian profiles to deblend
adjacent lines (e.g., \oiv\ and \feii\ at $\sim$26\,\micron).
Linear continua were fitted simultaneously, taking care to define 
the continuum regions well beyond the Gaussian wings.

In order for a spectral feature to be identified with a known emission
line, its rest-frame wavelength shift from the laboratory wavelength had to  
to be $\leq$0.065\,\micron\ 
(for each object, we correct the observed spectrum to its rest frame by 
using the optical redshift listed in Table \ref{tab:sample}).
Each emission line was measured several times, varying the
continuum positions and blending parameters, to obtain an additional 
estimate of the uncertainty.
The uncertainty on the integrated flux was then taken to be the larger
of the two following estimates:
the standard deviation of the ensemble of different measurements, or 
9 times the standard deviation $\sigma$ of the continuum fluctuations, 
as described above (assuming the line is spread over 3 pixels).
To guard against false detections from bad pixels or spectral glitches,
we also required that the average of the two nod positions have a signal-to-noise
$\geq$3.
Table \ref{tab:fsgauss}
reports the long-wavelength FS fluxes we obtain from these measurements.

For the objects where \oiv\ or \feii\ was detected, 
blown-up parts of the spectra showing the 26\micron\ region 
are given in Figure \ref{fig:oiv}. 
Although we detect \oiv\ at 3$\sigma$ (2$\sigma$) in 7 (9) galaxies, 
\feii\ is detected only in 2 (5 at 2$\sigma$).
This result will be discussed in more detail in $\S$\ref{sec:oxygen_iron}.
The extremely high-excitation lines [Ne{\sc v}] (ionization potential
of 97.1\,eV) at 14 and 24\,\micron\ are not detected in our data for any BCD.
The MIPS 24\,\micron\ continuum level 
is also shown in Fig. \ref{fig:oiv} by an open circle.
As stated in $\S$\ref{sec:phot}, the IRS spectra and MIPS 24\,\micron\
total fluxes are consistent, except for UM\,311 which is in a crowded
field. 

In addition to the definite presence of \htwo, 
a few spectra shown in Fig. \ref{fig:h2o_oh}
also tentatively suggest the presence of two other 
molecules, \oh\ ($\sim$28.9, 30.3, 30.7\,\micron)
and \water\  ($\sim$29.8 and 29.9\,\micron). 
Table \ref{tab:h2} gives the \htwo\ line fluxes (including those measured
by PAHFIT for $\lambda\la$ 14.5\,\micron, see above),
and Table \ref{tab:molecules} reports the other molecules.
These detections will be discussed further in $\S$\ref{sec:molecules}.

\subsection{Comparison of Fitting Techniques \label{ref:comparison}}

To judge the reliability of our line flux measurements, we have compared the 
fluxes of those emission lines that 
fall in the short-wavelength region of the spectra  
($\lambda\leq$\,14.3\,\micron), and that have been 
measured in two different ways: 
first by PAHFIT and second by individual Gaussian fits. 
For strong lines with fluxes $\sim 2\times10^{-17}$\,\wmsq, 
the difference between these two methods is $\simlt$4\%;
for faint lines with fluxes $\sim 5\times10^{-18}$\,\wmsq,
the difference can be as large as 25\%.
The larger discrepancy for the fainter lines 
can be attributed to the choice of different continuum levels in the
two methods.
This is the main reason why we preferred to use PAHFIT 
for the short-wavelength
region. However, the overall agreement between the two methods is 
sufficiently good to conclude that we can reliably 
compare short- and long-wavelength
line fluxes, and that we do not introduce systematic differences 
in  measuring them with slightly different methods.

Although PAHFIT corrects for dust extinction, we have not 
applied an analagous correction to the line
fluxes for $\lambda \geq$14.5\,\micron.
Such corrections would be applicable only to two sources,
Haro\,3 and Mrk\,996.
Haro\,3, with $\tau(9.7)$\,=\,0.39, would have a correction 
$\simlt$25\% for wavelengths longer than 15\,\micron;
Mrk\,996 would have a 2\% correction.
Hence, because the correction for both objects are comparable to 
or smaller than uncertainties in the line fluxes, we did not
apply them to the long-wavelength measurements.

\section{The Ionized Component of the ISM \label{sec:ionized}}

The spectral region covered by \spitzer/IRS provides a
wealth of diagnostics to probe the physics of the ISM at low metallicity.
The FS lines allow us to assess the physical conditions of the ISM,
including the spectral shape or hardness of the ISRF and the density of the ionized gas.
These quantities affect the dust properties by causing changes
in the temperature of the ``classical'' grains, and by altering
the stochastic emission from the aromatic features.
They constrain the molecular content through dissociation
and excitation processes. 

\subsection{The Hardness of the Interstellar Radiation Field \label{sec:hardness}}

The IR FS lines span a wide range of ionization potentials
and are thus sensitive diagnostics of the ionized ISM in BCD \hii\ regions. 
We focus first on diagnostics of the
ISRF hardness. In their discussion, we will attempt 
to distinguish among the interdependent effects of
ISRF {\it hardness}, ISRF {\it intensity}, and nebular {\it metallicity}. 

Traditionally, flux ratios of high-ionization species have been 
used to measure the ISRF hardness, or equivalently the UV slope of its 
spectrum, which is sensitive 
to the effective temperatures of the ionizing stars. 
In the IR range, the \neiii/\neii\ and \siv/\siii\ flux ratios are
usually adopted \citep[e.g.,][]{thornley00,giveon02}. 
Such flux ratios also depend on the ionization parameter $U$, defined
as $U\,=\,Q/4\pi R_* n_{\rm e} c$, where $Q$ is the number of ionizing
photons, \Ne\ is the electron density, and $R_*$ is the radius of the
star-forming region \citep[e.g.,][]{thornley00}.
We first examine whether these ratios depend on metallicity as shown
in Fig. \ref{fig:nes_oh}. 
The top panel shows that within our sample there may be a very slight dependence 
of the \siv/\siii\ ratio on metallicity, although with a very large
scatter (there are no SINGS data for the variation of this flux ratio
with abundance).
The bottom panel shows 
that there is very little correlation of the \neiii/\neii\ flux 
ratio with oxygen abundance within our sample alone. However, when the 
SINGS data are included, it is clear that 
the low-metallicity BCDs show considerably higher neon ratios
than the more metal-rich SINGS galaxies \citep{dale09}. 
The SINGS galaxies span an abundance range of
roughly a factor of 10, from \logoh\,$\sim$8.4 to 9.4, 
and our sample fills in the decade below, down to \logoh\,$\sim$7.5.
Considering the two samples together, 
the bottom panel shows an increase of the \neiii/\neii\ flux ratio by a
factor of $\sim$10 from
\logoh\,$\sim$9.4 down to $\sim$8.2. Then the curve 
 flattens out 
into a ``plateau'' for lower abundances, with \neiii/\neii\ $\sim$5. 
Again, the scatter of  the \neiii/\neii\ flux 
ratio at a given oxygen abundance is large, indicating that 
metallicity cannot be the main parameter controlling the hardness of the
ISRF.
Similar behavior is shown also in previous metal-poor samples
\citep{wu06,hao09}. 

Our sample was defined on the basis of the optical \heii/\hb\
ratio which gives an independent estimate of the ISRF hardness.
However, within our sample, there is no correlation
between the \neiii/\neii\ and \siv/\siii\ ratios and 
the \heii/\hb\ ratio as shown in  Fig. \ref{fig:nes_heii}. 
A possible reason for this lack of correlation
is that the IR and optical ratios do not probe the same excitation energy.
The ionization potentials for \neiii\ and \neii\ are respectively
41.0\,eV and 21.6\,eV, while they are respectively 34.8\,eV and 23.3\,eV 
for \siv\ and \siii.
These ionization potentials are well below that of  He$^+$ which is 54.4\,eV.
Thus, the ISRF UV energies probed 
by the neon and sulfur lines are too soft compared to those  
probed by the \heii\ ratio. 

We have therefore examined the relation between
the \oiv/\SiII\ and \heii/\hb\ ratios; 
the ionization potential of \oiv\ is 
54.9\,eV, similar to that of He$^+$ (54.4\,eV), while that of 
\SiII\ is only 8.2\,eV.
Since both silicon and oxygen are products of 
core-collapse supernovae (SNe),  
the \oiv/\SiII\ ratio should not be subject to 
abundance anomalies. 
The left panel of Fig. \ref{fig:oiv_heii_oh} shows a weak
correlation between the two ratios for the BCDs in our sample: 
a high \oiv/\SiII\ ratio is associated with a higher \heii/\hb,
at a confidence level of $\sim$93\% (one-tailed).
Higher \oiv/\SiII\ is associated with lower nebular O/H
abundance (right panel of Fig. \ref{fig:oiv_heii_oh}), at a 
$\sim$98\% confidence level (one-tailed). 
This last correlation is similar to the one 
between \heii/\hb\ and O/H \citep{thuanizotov05}. 
Evidently, lower metallicity galaxies tend
to have generally harder ionizing radiation.


Although the BCDs by themselves show little correlation between
the \oiv/\SiII\ ratio and the \neiii/\neii\ and \siv/\siii\ ratios, 
when SINGS galaxies are included, there is a definite correlation.
Fig. \ref{fig:nes_oiv} illustrates this, and suggests
that in star-forming galaxies the presence of hard radiation
as probed by the \oiv/\SiII\ ratio implies also the presence of
less hard radiation as probed by the \neiii/\neii\ and \siv/\siii\ ratios.

\subsection{The Intensity of the Interstellar Radiation Field \label{sec:intensity}}

We now explore diagnostics for the ISRF intensity, as
distinct from the ISRF hardness, since it does not depend on the 
spectral shape of the ionizing radiation.
The ISRF intensity is related to the ionization parameter
which measures the number of ionizing photons per hydrogen atom.
MIPS observations are relevant here, because the ratio of
71 to 160\,\micron\ fluxes is sensitive to the temperature of
the large (``classical'') grains, and can thus act as an indicator of the 
ISRF intensity \citep{draine07}.
Essentially, we will use the MIPS luminosity ratio (``color temperature'')
\nfncolor\ as a proxy for the ionization parameter.
The upper panel of Fig. \ref{fig:p24oh_r71} shows the relation between this
``color temperature'' \nfncolor\
and the quantity \nfn $_{24}$/(\nfn $_{71}+$ \nfn $_{160}$); 
the latter represents the relative contribution
of the 24\,\micron\ emission to the long-wavelength component of \ltir,
the total IR luminosity as defined by \citet[][see $\S$\ref{sec:phot}]{draine07}.
The lower panel of Fig. \ref{fig:p24oh_r71} shows the variation
of the same 71 to 160\,\micron\ luminosity ratio  
with oxygen abundance.
SINGS galaxies are also shown in Fig. \ref{fig:p24oh_r71}
using data taken from \citet{dale07}.
The correlations in both panels are highly significant even within the BCD sample
alone: 
the correlation in the top panel is at a $\simgt$99\% confidence level
(one-tailed), while the bottom one is at a $\simgt$99.9\% confidence
level (one-tailed).
This indicates that the 24\micron\ fraction of \nfnfar\ increases 
with increasing large-grain
temperature (and decreasing oxygen abundance).  
Hence, the relative strength of the 24\,\micron\ luminosity is a 
good tracer of  the intensity of the ISRF.
As dust gets warmer, more large-grain
emission is shifted toward shorter wavelengths, resulting in 
increased 24\micron\ flux \citep[e.g.,][]{dale05}.

Apparently, at low metallicities, \logoh $\simlt$7.8, 
large grains can be heated to extreme 
\nfncolor\ ratios, reaching values of 10--30.
The effect could be even more extreme since the lowest
metallicity BCDs in our sample are not plotted; the MIPS 160\,\micron\
emission for the four galaxies with \logoh$\simlt$7.6 is not detected. 
Compared to the SINGS sample \citep{dale05,dale07}, the ISRF intensity
as measured by \nfncolor\ is $\simgt$ 5 times greater on average in the BCDs than
in even the most intense SINGS galaxy\footnote{To do the comparison, 
we have converted the flux ratios used by 
\citet{dale05} to the \nfn\ ratios used here.}. 

Using the 71/160\micron\ luminosity ratio (\nfncolor) as a tracer of the ISRF
intensity, we next investigate whether any of the FS line ratios,
indicators of the ISRF hardness as shown above, are correlated
with it.
Figure \ref{fig:neoiv_r71} shows the \neiii/\neii\ and \oiv/\SiII\ line
ratios as a function of the 71/160\micron\ ratio;
SINGS galaxies are also plotted combining data from \citet{dale07} and \citet{dale09}.
The \neiii/\neii\ flux ratio shows no correlation, although there is a 
weak one in the BCD sample alone (at the $\simgt$97\% confidence level) for \oiv/\SiII;
the \oiv/\SiII\ ratio is larger for warmer large grains. 
In the SINGS sample, most of the highest \neiii/\neii\ and \oiv/\SiII\ 
flux ratios are associated with AGN, so any correlation could be masked by
AGN high-excitation emission-line ratios in galaxies with a relatively low-intensity ISRF. 
In any case, 
we conclude that the hardness of the ISRF, as measured by \oiv/\SiII\
is weakly related to its intensity, as traced by the 
71/160\micron\ luminosity ratio. In lower metallicity BCDs, the ISRF
tends to be both harder and more intense. 

To summarize, in our low-metallicity BCD sample,
the \oiv/\SiII\ ratio is correlated with \heii/\hb, \logoh,
and the large-grain dust temperature (71/160\micron\ ratio). This implies 
that star-forming galaxies with a lower nebular metallicity 
as compared to more metal-rich galaxies, 
generally possess an ISRF which is both harder, as indicated 
by higher \oiv/\SiII\ or \heii/\hb\ flux ratios, and 
more intense as indicated by a larger 71/160\micron\ luminosity ratio. 
The relatively large scatter in some of these relations,
and the ``plateau-like'' behavior of others, implies 
that these quantities do not depend on only one parameter, but that 
many parameters are involved. 
The ISRFs in the SINGS galaxies are of considerably lower intensity
than those generally found in BCDs.
This is probably because the star-forming regions are more extended,
less extreme, and have a lower ionization parameter $Q$. 
Such factors are probably as important as ISRF hardness and metallicity.



\subsection{Electron densities\label{sec:densities}}

We have derived electron densities \Ne\ in the ionized gas from the ratio
of the temperature-insensitive \siii\ IR lines \citep{draine09}.
Fig. \ref{fig:densities} shows these densities as compared with those
derived using the optical 
\sii\ $\lambda$6717/$\lambda$6731 ratio.
If the IR emission were to originate in regions suffering from higher
extinction than the optical regions,
we might expect the IR \Ne\ to be higher than those derived from 
the optical lines.
Fig. \ref{fig:densities} shows that this is true for some BCDs,
but not all. 
On the contrary, a few galaxies 
show higher optical than IR densities. 
The reason for this behavior is unclear;
beam dilution could play a role 
since the IRS aperture is several times larger than the optical one and
thus includes lower density outer regions. 

In the optical, high-ionization radiation responsible for ions such as \nev\ is found
to be associated with extremely massive compact super star clusters
and high optical densities \citep{thuanizotov05}.
The idea that compact star-forming regions were associated with
higher electron densities \Ne\ was first proposed on theoretical
grounds by \citet{hirashita04}.
The size-density correlation in \hii\ regions 
in which smaller regions are denser
is observed over six orders of magnitude
from Galactic objects to distant BCDs \citep{hunt09}.
If fast radiative shocks were responsible for the ionizing radiation
as proposed by \citet{thuanizotov05}, then we would expect such
processes to be more efficient at high densities.
However, in the current BCD sample,
only two (three) of 7 (9) galaxies with a 3$\sigma$ (2$\sigma$) \oiv\ detection
has a \Ne\ above the low-density limit of $\sim$30\,\cmthree.
The electron densities as derived from the \siii\ IR lines 
are apparently not correlated with the hardness
of the ISRF, as measured by \oiv/\SiII.
One reason for this non-correlation 
may be that \oiv\, with an ionization potential 
of 54.9 eV is not probing sufficiently high ionization
energies (the \nev\ line has an ionization potential of 97.1\,eV).
Another reason may be that the \siii\ IR lines arise from a region that
is less dense than the \oiv\ region.
More work on a larger sample is needed to explore the connection between
density, geometry, and the ISRF.

\subsection{\oiv\ and \feii\ in BCDs \label{sec:oxygen_iron}}

The two FS lines \oiv\ and \feii\ lie at adjacent wavelengths
($\Delta\lambda\,=\,$0.098\micron) near 26.0\,\micron. They are, however,   
sufficiently separated to be resolved by LH IRS spectra.
While \oiv\ traces high-excitation gas, \feii\ has an ionization
potential of 7.9\,eV, more similar to \SiII\ (see $\S$\ref{sec:SiII}).
Both lines may be excited by photoionization and shocks
\citep{lutz98,ohalloran06,allen08}.
By fitting the two lines with Gaussian profiles, we are able to deblend 
them in our spectra.
We detect \oiv\ at 3$\sigma$ in 7/22 observed galaxies, but
\feii\ at 3$\sigma$ in only Mrk\,5 and Mrk\,996 (see Fig. \ref{fig:oiv}). 
At low metallicity, \oiv\ emission is almost 4 times as common as 
\feii\ (32\% vs. 9\%).

This conclusion contrasts with the findings of \citet{ohalloran06,ohalloran08}.
Using \spitzer\ archival IRS data, 
they have published 26.0\,\micron\ \feii\ fluxes for a sample of starbursts,
including the 22 BCDs reported here.
While the uncertainties in the spectra estimated by us (see $\S$\ref{sec:analysis})
are generally more conservative
than those reported by \citet{ohalloran08}, these
authors find \feii\ emission in all objects, but report no \oiv\ emission.
The \feii\ emission fluxes they find are also larger than ours, suggesting 
contamination from the \oiv\ emission.
Figure \ref{fig:oiv} shows the identifications of the
two lines; it is evident that the \feii\ flux can be securely measured 
only when \oiv\ is taken into account.
No spectra are shown by \citet{ohalloran08}, so we cannot directly
compare their line identifications with ours to resolve the source of 
discrepancy.

\subsection{\SiII\ Emission \label{sec:SiII}}

The \SiII\ line is detected in more than half the BCDs in our sample.
At an ionization potential of only 8.2\,eV, \SiII\ traces
the neutral gas, rather than the more highly excited ionized gas traced
by the neon, sulfur, and oxygen lines. 
Figure \ref{fig:tir} shows the \SiII\ emission normalized to the TIR luminosity.
The mean \SiII/TIR ratio
is more than a factor of two lower than that of the SINGS galaxies,
$8\times10^{-4}$ versus $2\times10^{-3}$ (compare the dotted and dashed
horizontal lines in Fig. \ref{fig:tir}).
This slight difference is probably due to the lower metal abundance of the BCDs,
and the consequently less efficient cooling of the gas through FS lines.
It could also be due to a more intense ISRF, and thus a lower Si$^+$ abundance.
In any case,
the ratio of \SiII\ emission to \ltir\ is relatively constant with luminosity
(excluding UM\,311 as discussed previously).
These considerations suggest that the \SiII\ line is an efficient coolant,
comparable to the [C{\sc ii}] line \citep[e.g.,][]{malhotra97}, 
and could be used as a proxy for \ltir\ to within a factor of two, even at low
metallicities.

\section{Aromatic Features \label{sec:aromatics}}

The wealth of spectral data obtained by \iso\ 
combined with much theoretical work and laboratory measurements
has shown convincingly that aromatic features 
are associated with stochastic or 
single-photon emission from large molecules called 
polycyclic aromatic hydrocarbons or PAHs
\citep{joblin95,verstraete01,hony01,peeters02,vandiedenhoven04,peeters04a,bauschlicher08,bauschlicher09}.
PAH emission dominates the mid-infrared spectra of star-forming
galaxies \citep{brandl06,smith07}, and
contributes importantly to their IR energy budget. 
Through the photoelectric effect, PAHs heat
the gas in the ISM \citep[e.g.,][]{hollenbach97}, thereby 
reinforcing the coupling of the gas and dust components.
Because they are so dominant in the MIR spectral range,
PAHs have been used to identify galaxies with intense star
formation at redshifts of $\sim2-3$, and to estimate their
bolometric luminosities \citep[e.g.,][]{houck07,weedman08,dey08}.

In most star-forming galaxies,
PAH emission shows a surprisingly narrow range of properties,
well characterized by a ``standard set'' of 
features with fixed wavelengths and FWHMs \citep{smith07}.
However, first \iso\ and later \spitzer\ have shown that metal-poor
star-forming galaxies are deficient in PAH emission
\citep{thuan99a,engelbracht05,madden06,wu06,engelbracht08}.
Although it has generally been concluded that the reason for
this deficiency is low metallicity, it is not yet clear whether
it is the metallicity directly, or rather the 
indirect effects of a low metallicity environment. 
In the first case, there would be a consequent
lack of raw material (carbon and nitrogen) with which to form PAHs; 
in the second,
the increased ISRF hardness or intensity would lead to their destruction. 
Here, we examine in detail the PAH emission in our sample to 
examine this question, and
show for the first time that some PAHs can survive even in 
extremely metal-poor environments.

\subsection{PAH Properties at Low Metallicity\label{sec:pahtrends}}

The 7.7\,\micron\ blend is the most common PAH feature in our sample,
being detected in 15 of 22 objects.
The 7.7\,\micron\ feature is also the strongest one, comprising
on average 49\% of the total PAH power. 
The remaining PAHs are significantly weaker, and less frequently detected:
13 BCDs show the 11.3\,\micron\ blend, 9 the 8.6\,\micron\ feature,
and 7 have a detection of the 6.2\,\micron\ PAH.
No 17\,\micron\ PAH features were detected, and only
Mrk\,1329 showed a significant detection at 16.4\,\micron.
The 6.2, 7.7, 8.6, 11.3, and 12.6\,\micron\ features 
by themselves constitute $\sim$72\% of the total PAH power in BCDs
(see Table \ref{tab:dfpahfit}); 
the SINGS sample has about $\sim$85\% of the total PAH power
in these features \citep{smith07}.
Similarly to the SINGS galaxies,
other weaker features (not listed in Table \ref{tab:dfpahfit})
which contribute to the PAH luminosity 
include the 
5.7, 6.6, 6.7, 8.3, 10.6, and 10.8\,\micron\ features.
In our sample, these features comprise roughly 28\% of the PAH luminosity,
but in SINGS galaxies, only 15\%.
Part of the reason for this difference 
may be the difficulty in detecting long-wavelength
PAHs in BCDs; only 1 BCD has the 16.4\,\micron\ feature detected,
while in SINGS galaxies the 16-17\,\micron\ blend makes up
$\sim$6\% of the total PAH power.

The fractional power of the four strongest features
relative to the total PAH luminosity is illustrated in
Figure \ref{fig:pahfrac}.
The horizontal lines in each panel correspond to the SINGS medians
(dashed) and the BCD means (dotted).
The BCD means are calculated taking into account all galaxies
with 7.7\,\micron\ detections; thus they are a sort of weighted
average which considers frequency of detection together with intensity.
This is the reason that the mean for the 6.2\,\micron\ PAH lies
below most of the BCD data points: that feature was detected only in
7 galaxies, while the other features in Fig. \ref{fig:pahfrac} were
detected with a frequency more similar to the 7.7\,\micron\ PAH.
The relative PAH fractions for the BCD and SINGS 6.2 and 7.7\,\micron\ features 
are virtually indistinguishable: 0.10 vs. 0.11 for 6.2\,\micron, 
and 0.49 vs. 0.42 for 7.7\,\micron\ \citep{smith07}.
However, for the longer wavelength PAHs at 8.6 and 11.3\,\micron,
the BCD fractions are $\simgt$ 40\% larger: 
0.10 vs. 0.07 (8.6) and 0.17 vs. 0.12 (11.3).
Although there is considerable scatter, the BCD mean PAH relative fractions for these
features tend to exceed even the 10 to 90th percentile spreads of the 
SINGS galaxies.

Figure \ref{fig:pahfwhm} shows the mean Drude profile widths 
and central wavelengths of six aromatic features detected
at $\simgt 3\sigma$ in our sample; the numbers of BCDs with each
feature are given in parentheses. 
The analogous quantities for the SINGS galaxies are also plotted. 
In no case, were the PAHFIT profiles fixed to the SINGS galaxy
parameters able to fit the BCD data; every BCD was better fitted by allowing
the Drude profile widths (FWHMs) and central wavelengths to vary.
In some cases, the differences between the $\chi_\nu^2$
with the SINGS ``standard'' profile parameters
and the best-fit parameters were small, $\simlt$15\%;
but in others, the improvement of $\chi_\nu^2$ was almost a factor of 2.
The only {\it systematic} difference between the two samples
is the narrower profile width for
the 8.6\,\micron\ feature; only the very broadest BCD profiles are
as broad as those in more metal-rich systems.

The larger relative intensities of the 8.6 and 11.3\,\micron\
feature in BCDs and the narrow profile width
of the 8.6\,\micron\ PAH could be due to a different size distribution 
of PAH populations at low metal abundances.
\citet{bauschlicher08} find that the 8.6\,\micron\ band arises from
large PAHs, with $N_{\rm min} \simgt$ 100 carbon atoms.
Hence, they suggest that the relative intensity of
the 8.6\,\micron\ band can be taken as an indicator of
the relative amounts of large and small PAHs in a given population.
The large 8.6\,\micron\ intensity relative to total PAH power in BCDs
could be a signature of fewer small PAHs (or more larger ones) at low metallicity.
The relative lack of small PAHs could also explain the
low detection rate of the 6.2\,\micron\ feature,
since most of its intensity comes from PAHs with less than 100 C atoms
\citep{schutte93,hudgins05}.
The size of a PAH molecule grows with the number of carbon
atoms in it \citep[e.g.,][]{draine07}.
For a given maximum number of carbon atoms $N_{\rm max}$, a {\it larger
minimum number} $N_{\rm min}$ theoretically gives narrower profiles \citep{joblin95,verstraete01}.
Although fitting intrinsically asymmetric PAH profiles by symmetric Drude
profiles in PAHFIT is a simplification, 
the narrower width of the 8.6\,\micron\ feature 
could be a signature of relatively larger PAHs at low metallicity.
The difference disappears at 7.6\,\micron\ and is less pronounced
at longer wavelengths than 8.6\,\micron\ because
both small and large PAH sizes contribute to
these bands\footnote{The 7.8\,\micron\ band may be dominated by larger PAHs 
only \citep{bauschlicher09}; but
this band is present in only two of the BCDs in our sample, Haro\,3 and
II\,Zw\,70.}
\citep{schutte93,peeters02,vandiedenhoven04,bauschlicher09}.

Another indication that low-metallicity BCDs may be lacking the
smallest size PAHs comes from a correlation analysis.
In a detailed study of Galactic \hii\ regions, young stellar objects,
reflection and planetary nebulae (RNe, PNe), and evolved stars, \citet{hony01} 
found only a few correlations among PAH features.
One of these is between 6.2 and the 12.7\,\micron\
emission, which arises from the CC stretch mode in ionized PAHs, 
in contrast to the 11.3\,\micron\ band which is attributed to
the neutral CH out-of-plane bending mode \citep{hony01,peeters02}.
The comparison between the 6.2/11.3 and 12.7/11.3 flux ratios
is shown for our sample of BCDs in Fig. \ref{fig:pah62127};
we show galaxies with $\simgt3\sigma$ detections in at least two features.
With the exception of Mrk\,1315, galaxies with 6.2\,\micron\ PAH detections
and \logoh$\geq$8.1 (filled circles)
are similar to the \hii\ regions studied by \citet{hony01}.
On the other hand, BCDs with lower O/H 
occupy a region of the plot with small 6.2/11.3 ratios.
Although the statistics are poor,
small 6.2/11.3 are found for most of the lowest-metallicity
BCDs. They lie in the same region as the RNe and PNe studied by \citet{hony01}.
Because the 11.3\,\micron\ feature arises from non-adjacent or ``solo''
CH groups, dominant 11.3\,\micron\ emission implies large
PAH complexes, with $\simgt$100-200 carbon atoms and long straight
edges \citep{hony01,bauschlicher08,bauschlicher09}. 
Moreover, fewer small PAHs reduce the intensity
of the 6.2\,\micron\ emission, 
as the bulk of this feature comes from
PAHs containing less than 100 C atoms \citep{schutte93,hudgins05}.
Therefore, small 6.2/11.3\,\micron\ band ratios at low metallicites
could imply an overall deficit of small PAHs.

A third indication that there is a deficit of small
PAHs in a metal-poor ISM is provided by the intensity of the
8.6\,\micron\ feature relative to the 7.7\,\micron\ blend.
As mentioned above,
\citet{bauschlicher08} find that 8.6\,\micron\ band arises from
large PAHs, with $N_{\rm min} \simgt$ 100 carbon atoms.
On the other hand, the 7.7\,\micron\ feature 
contains emission from both small and large PAH sizes 
\citep{schutte93,peeters02,vandiedenhoven04}.
For the 9 BCDs with the 7.7 and 8.6\,\micron\ 
PAH features detected at $\simgt 3\sigma$,
the mean 8.6/7.7 flux ratio is 0.48, and the median is 0.33.
This ratio is at least double that in the more metal-rich SINGS galaxies,
with 8.6/7.7 $\sim$0.18 in the mean, and a range of 0.11$-$0.21
\citep{smith07}; 7 of 9 BCDs are outside the upper 90th percentile
limit of 0.21.
Again, although the number statistics are small,
the large 8.6/7.7\,\micron\ flux ratios of
the PAHs in these low-metallicity BCDs
appear to be dominated by the largest complexes.

In summary, there are three different lines of evidence 
which tentatively suggest that the PAH populations in low-metallicity BCDs 
are deficient in the smallest sizes ($N_{\rm min} \simlt$50 C atoms).
The IRS spectra show flux ratios typical of
the largest PAHs modeled so far, with $N_{\rm min} \simgt$100 C atoms.
Apparently, the ISM environment in the BCD \hii\ regions 
is not propitious to the existence of the smallest PAHs and   
has led to their destruction.
Perhaps only the PAH population with the largest complexes can
survive the extreme physical conditions in the low-metallicity \hii\
regions in the BCDs.

\subsection{PAHs and the Energy Budget\label{sec:budget}}

PAHs emit about 10\% of the total infrared power in typical
solar metallicity star-forming
galaxies (\logoh\ $\sim$ 8.7-9.0), with a maximum of $\sim$20\% \citep{smith07}.
In our low-metallicity BCD sample, with oxygen abundances \logoh\ ranging from
7.4 to 8.3, the fraction of PAH emission ranges from $\sim$0.1\% to
$\sim$1.6\%, with a median (and mean) of $\sim$0.5\%\footnote{Because of
crowding and source structure, the TIR luminosity in UM\,311 may be overestimated;
nevertheless even without UM\,311, the mean is only slightly larger, 0.57\% vs. 0.54\%.}.
Figure \ref{fig:pahtot_tir} shows how the sum of all $\simgt 3\sigma$
PAH emission features normalized to the total IR luminosity \ltir, 
\sigpah/TIR, varies as a function of 
\ltir\ for our BCD (filled circles) and SINGS galaxies 
(non-AGN open squares, AGN open triangles) \citep{smith07}.
For the SINGS galaxies,  \ltir\ has been calculated {\it within the spectral
extraction aperture}, while for the BCDs, \ltir\ is taken to be the
total luminosity as described in $\S$\ref{sec:phot}.
Because most of the 
BCDs are point-like at 24\,\micron\ and are generally unresolved with MIPS, 
this is usually comparable to \ltir\ within the spectral aperture (see Fig.
\ref{fig:oiv}).
Were the source sizes larger at 160\,\micron\ \citep[e.g., Haro3: ][]{hunt06},
\ltir\ could be overestimated, and the PAH fractions somewhat higher.
The infrared luminosities \ltir\ of the SINGS sample\footnote{The normalizing 
factor, \ltir, is calculated by different groups with slightly different 
algorithms; however, the general
consistency of the data suggests that our conclusions do not depend
on the exact method used to derive \ltir.}
vary by a range of more than three
orders of magnitude, while the BCD luminosities occupy the middle portion of
this range. 
Figure \ref{fig:pahtot_tir} shows clearly the 
segregation of the SINGS points, with high \sigpah/TIR, and the BCD points
with low \sigpah/TIR. The AGN, with intermediate \sigpah/TIR, lie between 
the metal-rich star-forming SINGS galaxies and BCDs.
Figure \ref{fig:pahtot_tir} also shows that there is no trend of the PAH 
fraction \sigpah/TIR with \ltir\ in either sample. 

We might expect the fraction of IR luminosity in PAHs to vary with metallicity.
We explore this possibility in the left panel of Fig. \ref{fig:pahtot_ohne},
where \sigpah/TIR is plotted against oxygen abundance. 
Although 
there is no correlation of PAH fraction with metallicity within the BCD sample, 
the inclusion of SINGS galaxies does show a trend of 
increasing PAH fraction with increasing metallicity.
We next investigate other physical factors which may control the PAH
fraction, such as the ISRF hardness. 
The right panel of Fig. \ref{fig:pahtot_ohne} shows \sigpah/TIR as a function
of the IR neon line ratio, \neiii/\neii\ (only those BCDs with at least 
a 3$\sigma$
detection in the neon lines are shown).
As discussed in $\S$\ref{sec:ionized}, the \neiii/\neii\ ratio is a 
good tracer of ISRF hardness, up to an intermediate energy range.
There is a weak correlation between \sigpah/TIR and \neiii/\neii, 
significant at the 95\% (one-tailed) confidence level. 
Thus, 
over the range in UV energies $\simlt$41\,eV, the fraction
of \ltir\ emerging as PAH emission may depend weakly on the hardness of the ISRF.
This suggests that the PAH emission fraction in BCDs is not directly 
controlled by metallicity, but 
by the hardness of the radiation field. The latter could also be responsible
for the destruction of the smallest PAH particles, as discussed
in the previous section.
We have checked for a correlation of \sigpah/TIR with ISRF intensity
as measured by \nfncolor, but found none.


To conclude our discussion of PAHs and the IR energy budget, we examine
the diagnostic proposed by \citet{houck07}: the quantity 
$\nu L_\nu$(7.7\micron),
where $L_\nu$(7.7\,\micron) is determined from the flux density at the
peak of the 7.7$+$7.8\,\micron\ blend, with no correction for the underlying
continuum. 
Because it arises from a range of sizes, 
the 7.7\,\micron\ band is arguably the most reliable PAH diagnostic, 
as it avoids preferential selection of either small or large PAH populations 
\citep{schutte93,peeters02}.
Figure \ref{fig:pah77} compares $\nu L_\nu$(7.7\micron)/\ltir\
for our BCDs with that for a sample of local starbursts \citep{brandl06,houck07}.
Considering only those BCDs with a 160\,\micron\ detection,
the BCDs have a mean $\nu L_\nu$(7.7\micron)/\ltir\
roughly four times lower than that of the starbursts. 
The means of the two samples are shown by horizontal dotted lines.
Part of this difference may be
due to a luminosity effect since the BCDs are more than 100 times less
luminous in the mean than the starburst galaxies 
(see for example the SINGS galaxies in Fig. \ref{fig:pahtot_tir}).
It may also result from variations of physical conditions
in the ISM, since the BCDs with the highest $\nu
L_\nu$(7.7\micron)/\ltir\ clearly overlap with   
the range observed in local starbursts that are however much
more luminous in \ltir.

\subsection{PAH Diagnostics for AGN/Starbursts Revisited\label{sec:pahdiagnostics}}

Several diagnostics based on IR spectra have been proposed to separate AGN from
star-forming regions in galactic nuclei.
Previous work has also focused on the comparison of the flux ratios of PAH
features and emission line ratios from FS transitions.
In this section, we examine our BCD sample in the context of such diagnostics
and assess their usefulness in the low metallicity regime. 

\subsubsection{7.7/11.3\,\micron\ Band Ratios}

We saw in the previous section that the total PAH fraction \sigpah/\ltir\
is correlated with the IR neon ratio \neiii/\neii\ (Fig. \ref{fig:pahtot_ohne}).
Here, following \citet{smith07},
we examine the variations of the 7.7/11.3\,\micron\ band ratios with
\neiii/\neii\ for both the BCD and SINGS samples as shown in
Fig. \ref{fig:pah_ne}.
Among the SINGS galaxies, there is a weak correlation, mostly for AGN, 
in the sense that a smaller 7.7/11.3 ratio is associated with 
a larger \neiii/\neii ratio \citep[e.g.,][]{smith07}.
The BCDs extend the correlation to considerably 
higher neon ratios (by one order of magnitude), that is to harder
ISRFs. 
The plot shows that the radiation fields in the low metallicity 
BCDs can be as extreme as those in the SINGS AGN.  
Such hard, intense ISRFs could suppress the ionized 7.7\,\micron\ feature, relative to the
neutral 11.3\,\micron\ feature.  
This may be due to the dominant contribution of 
large PAHs to the 11.3\,\micron\ band,
as compared to the mixed size distribution 
thought to be responsible for the 7.7\,\micron\ band;
the small PAH molecules  
contributing to the 7.7\,\micron\ emission could be destroyed while 
the large PAH molecules responsible for the  11.3\,\micron\ emission are not 
(see $\S$\ref{sec:pahtrends}). 

\subsubsection{AGN/Starburst Diagnostics}

Because a hard, intense radiation 
field changes the PAH properties, a deficit of PAH emission,
combined with FS line indicators of excitation, has often been 
used as a AGN diagnostic in solar metallicity star-forming galaxies. 
However, we have seen above that the radiation fields in low metallicity BCDs
can be just as intense as
those in AGN (see Fig. \ref{fig:pah_ne}), possibly destroying the 
smallest PAHs. Thus a PAH emission deficit 
may not necessarily be a reliable diagnostic for AGN.

We compare our BCD sample with AGN taken from \citet{genzel98}, 
using the diagnostics proposed by the latter authors and \citet{laurent00}.
Figure \ref{fig:genzel_diag} shows the
\oiv/\neii\ ratio \citep[or 1.7$\times$33.5\micron\ \siii, see][]{genzel98}
as a function of the ``strength'' of the 7.7\,\micron\ PAH band.
\oiv/\neii\ is another measure of ISRF hardness (see $\S$\ref{sec:ionized}
where we used \oiv/\SiII).
The ``strength'' of the 7.7\,\micron\ PAH feature was
determined by estimating the continuum level between 5 and 14\,\micron\,
and dividing the PAH flux by this level;
this method is the same as that used by \citet{genzel98}, so we preferred it
to the simpler equivalent width measurement given by PAHFIT. 
For metal-rich systems, plotting these two ratios against one another
results in a clear distinction between star-forming galaxies 
(stars and squares)
and AGN (triangles), the latter having a higher \oiv/\neii\ ratio and 
a weaker 7.7\,\micron\ PAH feature than the former.  
However, the situation for  
the BCDs in our sample is different. They do not fall 
in the star-forming but rather closer to the AGN region. 
The starburst with the highest \oiv\ fraction in the \citet{genzel98} sample
is NGC\,5253 (open star symbol), a BCD-like galaxy
with an oxygen abundance of \logoh$\sim$8.2 \citep{kobulnicky99}.
Two BCDs in our sample would 
have been classified as pure AGN on the basis of this diagnostic diagram.
Thus, the strength of \oiv\ combined with the 7.7\,\micron\ PAH deficit
in metal-poor starbursts makes them difficult to distinguish from AGN.

The second AGN diagnostic diagram often used is illustrated in Fig. \ref{fig:laurent_diag}.
This diagnostic, proposed by \citet{laurent00},
exploits the steeply rising MIR continuum in starbursts,
compared to the flatter continua in AGN and quiescent star-forming regions
(or PDRs).
Most star-forming galaxies have both PDR and \hii\ region components in their
spectra; the PDR component tends to have a flat MIR continuum and be
dominated by PAH bands, while the \hii\ regions have 
a deficiency of PAHs and a steeply rising MIR continuum from warm dust. 
To analyze our BCD sample, we have not used the original 
diagnostic diagram as proposed by \citet{laurent00}, but a 
slightly modified version of it as formulated by \citet{peeters04a}. 
It differs from the original diagram in the continuum wavelength range that 
enters in the denominator of the labels on both axes.
Fig. \ref{fig:laurent_diag} plots the 14$-$15\,\micron\ continuum against
the 6.2\,\micron\ PAH feature, both normalized by the 5.3$-$5.8\,\micron\
continuum.
In this diagram, because of their steep MIR continua,
the BCDs (filled circles) are clearly distinguished 
from the AGN (triangles), ULIRGs (squares), and metal-rich
star-forming galaxies (stars), lying above them. 
A few BCDs are consistent with metal-rich Galactic \hii\ regions 
($\times$),
but others are more extreme, both in their steeper continuum slope and in their
enhanced PAH strength (or faint 5\,\micron\ continuum).
The four BCDs with the steepest continua and most enhanced PAH strength
are also the most metal-rich galaxies in our sample
(Haro\,3, Mrk\,450, Mrk\,1315, and UM\,311).
We conclude that the AGN/starburst diagnostic diagram proposed 
by \citet{laurent00} based on the continuum MIR slopes of these
types of objects effectively separates AGN from starbursts, even
at low metallicity. 
 
\section{Molecules \label{sec:molecules}}

A long-standing puzzle of metal-poor star formation has been
the apparent lack of cool molecular gas from which to form stars.
Numerous attempts
to detect carbon monoxide in metal-poor BCDs, 
with nebular oxygen abundances 
\logoh $\simlt$8.2, have failed. 
Despite vigorous on-going star formation, 
there seems to be very little CO in BCDs
\citep{sage92,taylor98,gondhalekar98,barone00,leroy05}.
Intriguingly, \logoh $\simlt$8.2 appears to be 
the same metallicity ``threshold'' below
which PAH emission is thought to be suppressed  
\citep{engelbracht05,wu06,madden06,engelbracht08}.
Here we show that warm molecular gas in the form of \htwo\ 
does exist at very low metallicity. 
Our new IRS spectra reveal a variety of \htwo\ rotational lines,
and more than a third of the objects in our sample (8 BCDs) have $\simgt 3\sigma$ 
detections in one or more of the four lowest-order transitions of \htwo.

\subsection{\htwo\ Emission \label{sec:h2}}

Rotational transitions of \htwo\ are important diagnostics of the
warm neutral phase of the ISM, at temperatures between $\sim$100 and 1000\,K.
These transitions, observable from space between 5 and 28\,\micron,
are a main coolant of the warm molecular component.
Massive stars are the main source of excitation \citep{hollenbach97}, 
as \htwo\ molecules can be pumped by non-ionizing far-ultraviolet (FUV) photons 
into excited states.
They then return to the ground state through fluorescence or collisional 
deexcitation.
The pure fluorescence mechanism is likely to be responsible for 
the near-infrared roto-vibrational transitions
\citep[e.g.,][]{puxley88},
because of the higher critical densities needed for collisional deexcitation.
Critical densities of the lower-order rotational transitions are sufficiently low 
($\simlt\,10^3$\,\cmthree\ for S(0) to S(3)) 
that the lower rotational levels should be thermalized.
For the pure rotational \htwo\ transitions studied here, $\Delta v$ is 0;
we will refer to the $\Delta J$ series of rotational transitions as S(0), S(1), ..., S(7)
\citep[for a complete tabulation, see][]{roussel07}.
 
We first examine the \htwo\ emission in terms of its contribution to the
energy budget in the neutral ISM.
Following \citet{roussel07}, we use the sum of the S(0) to S(2) transitions to 
assess \htwo\ cooling.
The bulk of \htwo\ cooling occurs through these lines. 
Figure \ref{fig:h2tir} shows the sum of the S(0) to S(2) 
\htwo\ emission, normalized to the total IR luminosity, \ltir.
As before, both the SINGS and BCD samples are plotted;
the SINGS galaxies are confined to the hatched region. 
The \sightwo/TIR ratios range from $1\times10^{-4}$ to $2\times10^{-3}$,
with a mean of $\sim 5\times10^{-4}$ for the BCD sample
\footnote{We have excluded UM\,311 because of the probable
overestimate of its IR flux.}.
The \sightwo/TIR for the SINGS galaxies range from 
$2.5\times10^{-4}$ to $7.5\times10^{-4}$ \citep{roussel07}.
Relative to \ltir,
the warm molecular content of the BCDs is comparable to that of
the SINGS galaxies. 

Because warm \htwo\ emission originates in PDRs, the UV-illuminated
parts of molecular clouds, \htwo\ and PAH emission should
be related, since both species are excited by UV photons.
Although our sample contains too few points to explore such
a correlation, 
the mean ratio of \sightwo\ to  \sigpah\ of $\sim$6\% is roughly ten times
larger than that for the SINGS galaxies \citep{roussel07}.
Because the \htwo\ emission of BCDs, relative to TIR, is similar to the SINGS sample,
this difference arises mainly from the deficit in PAH emission at low metallicites.
Models show that the intensity of rotational transitions of \htwo\ peaks inside 
molecular clouds at
extinctions \av $\sim$2 \citep{hollenbach97}, while PAHs are expected to
be excited predominantly on the cloud surface layers.
Deeper into the cloud, densities are higher and temperatures lower,
and PAHs would be coagulated onto grain mantles \citep{boulanger90}.
The similarity of \htwo/\ltir\ over a wide range of metallicity
implies that \htwo\ might be somewhat self-shielded in BCDs. 
Although PAHs can survive the impact
of FUV photons with energies between $\sim$11 and 13.6\,eV,
\htwo\ would be dissociated without self-shielding.
It is also difficult to understand how, without self-shielding, the \htwo\
could survive the intense UV radiation field in low metallicity BCDs 
that is destroying the smallest PAHs (see $\S$\ref{sec:pahtrends}).

Although more than a third of our sample 
show definite low-order \htwo\ emission in their IRS spectra, all
attempts to detect \htwo\ in BCDs in UV {\it FUSE} absorption
spectra\footnote{Half a dozen have been observed, none of which are in the present sample.} 
have failed \citep{thuan05}. 
The absence of \htwo\ absorption lines in {\it FUSE} spectra 
implies that the warm \htwo\ detected through IR emission must be quite
clumpy. The {\it FUSE} observations are not sensitive to such a clumpy 
\htwo\ distribution because they can only probe 
the transparent UV sight lines, not being able to penetrate dense clouds
with $N(H_2)\simgt10^{20}$\,cm$^{-2}$ \citep{hoopes04}.

\subsubsection{\htwo\ Excitation Temperatures \label{sec:excitation}}

Excitation diagrams for the \htwo\ lines help infer the
excitation temperatures for molecular hydrogen.  
For four galaxies in our sample, 
there were sufficient data to construct excitation diagrams. 
These diagrams show, as a function of the upper level energy $kT_i$,
the natural logarithm of the column density of the species $N_i$ in
the upper level of each transition, divided by the statistical
weight, $g_i$. 
Assuming local thermodynamic equilibrium (LTE),
the total column density \ntot\ then follows from the Boltzmann equation:
\begin{equation}
\frac{N_i}{N_{\rm tot}}\,=\,\frac{g_i}{Z(T_{ex})} \exp \left( - \frac{T_i}{T_{ex}} \right) 
\end{equation}
where
$T_{ex}$ is the excitation temperature for the $i^{th}$ level,
and $Z(T_{ex})$ is the partition function.
The statistical weight is given by $g_i\,=\,(2I+1)/(2J+1)$;
the spin number takes the values $I=0$ for
even $J$ (para) transitions, and $I=1$ for odd $J$ (ortho) transitions.
\citet{herbst96} gives a convenient expression for the partition function:
$Z(T_{ex})\,=\,0.0247\,T/[1-\exp(-6000\,{\rm K}/T)]$, valid for temperatures $\simgt$40\,K.
The column density for a given transition $N_i$ is calculated from the
observed \htwo\ flux $F_i$ in that transition: 
\begin{equation}
N_i\,=\, \frac{F_i 4\pi}{ h\nu_i A_i \Omega}
\end{equation}
where $A_i$ is the transition's spontaneous emission coefficient,
$\nu_i$ is the frequency of the transition, 
and $\Omega$ is the beam solid angle. 

Figure \ref{fig:h2excitation} shows the resulting excitation diagrams for the
four BCDs which have at least one detection at the 3$\sigma$ level or
greater in a low-order transition
(except for SBS\,1152$+$579 for which we have used 2.5$\sigma$
detections), and at least two other significant ($\simgt3\sigma$)
detections in higher-order lines.
Since the inferred column density for a given transition depends
on the beam size, we have experimented  
using different beam sizes of the IRS modules for the various transitions.
The main result of this exercise is that the most appropriate beam size
is the largest LH beam, equal to 11\farcs1$\times$22\farcs3.
Because the scale of our spectra was ultimately set 
by the LH background subtraction
(see $\S$\ref{sec:observations}), and it was never necessary to multiply
by a scaling factor larger than the ratios of the apertures,
this is a reasonable conclusion.
However, the sources observed at 24\,\micron\ with MIPS
are either  
unresolved or only slightly extended, which would imply a slightly 
different beam dilution for each galaxy at longer wavelengths;
applying such a correction would have been highly problematic. 
Nevertheless, since we have used 
the IRS point-source flux calibration, this may not be a serious problem,
and, because of the way our spectra have been scaled, 
the excitation diagrams should not be affected
significantly by this approximation.
With a beam size that is too small,
column densities are overestimated and total masses are underestimated.
Hence, in Fig. \ref{fig:h2excitation}, we have taken the most conservative
approach for column densities, using the LH beam as discussed above.
At the median distance of 21\,Mpc for our sources, 
this corresponds to a region of $\sim$1.6\,kpc in diameter.

The excitation temperature of the line-emitting gas is the reciprocal of the
slope of the excitation curve;
in the absence of fluorescence,
this temperature would be the gas kinetic temperature in LTE.
Assuming thermal emission\footnote{This assumption is almost
certainly invalid for the higher-order transitions, because that 
would require densities $\simgt 10^5$\,\cmthree.
In any case, the excitation diagrams remain a useful tool for
inferring physical conditions of the line-emitting gas.},
Fig. \ref{fig:h2excitation} implies 
a range of temperatures for the molecular gas.
As a crude approximation to the probable continuous temperature
distribution of the gas (see the parabolic fit for \cgcg),
we have fitted the data points in each diagram by two linear segments 
\citep[see also][]{rigopoulou02,higdon06,roussel07}.
The temperature inferred is a strong function of the set of transitions
from which it is derived:  
lower-order transitions probe lower-temperature gas than
the higher-order ones.
The implied temperature for the warm gas (from the lower-order transitions) is 
$\sim$245\,K (not considering Mrk\,996 at 98\,K, 
the only galaxy with a significant S(0) detection).
The temperatures of the hotter molecular component (from the high-order transitions)
range from $\sim$820\,K to $\sim$1600\,K.
Considering the transitions from which they are derived,
these temperatures are consistent with those found for starbursts 
\citep{rigopoulou02}, SINGS galaxies \citep{roussel07}, and 
ULIRGs \citep{higdon06}.

We have not considered any departure from the standard
ortho-para ratio of 3 for LTE \cite[e.g.,][]{burton92,roussel07}.
Departure from thermalization of ortho and para levels would be
suggested by a non-monotonic increase of inferred excitation
temperatures with increasing rotational transition.
The temperatures derived from mixed-parity \htwo\ transitions for the 
BCDs in our sample are 
generally consistent, indicating no departure from 
thermalization for most objects, 
although Fig. \ref{fig:h2excitation} does show
some sign of a discrepancy for SBS\,1152$+$579.
A detailed discussion of this topic at low metallicity is
precluded by the paucity of the BCD data presented here. 

\subsubsection{Column Densities and Gas Masses \label{sec:h2mass}}

Once we have an estimate of the gas temperature for each galaxy,
we can derive the
\htwo\ column density and total \htwo\ mass within the IRS beam. 
The derived masses and column densities 
depend on the size of the IRS beam used to relate
the observed flux to the column densities.
As stated above,
we adopted a conservative approach for the column densities, and used the 
large LH beam, which may give masses that are slightly overestimated.

In our discussion of \htwo\ column densities and masses, 
we will consider only those objects 
in which ``secure'' temperatures have been derived, i.e. those that have 
two relatively reliable detections (at the $2.5\sigma$ level or greater) 
of lower-order transitions\footnote{Now including S(3) to consider
correctly parity in the derivation of temperature.}.
Four objects meet these criteria:  \cgcg, HS\,0837$+$4717, Mrk\,996, 
and SBS\,1152$+$579.
Mrk\,996 and \cgcg\ have
a total warm (T$\sim$100-120\,K) \htwo\ column density \nhtwo\ $\sim
3\times10^{21}$\cmtwo\ or 53\,\msun\,\pctwo.
At a higher temperature of $\sim$340\,K,
SBS\,1152$+$579 has a much lower molecular column density
with \nhtwo\ $\sim 3\times 10^{18}$ (0.04\,\msun\,\pctwo). 
With a warm/total \htwo\ fraction of $\sim$5\% \citep{rigopoulou02,roussel07},
the total molecular column density 
in SBS\,1152$+$579 would be $\simlt$1\,\msun\,\pctwo.
The column density for HS\,0837$+$4717 is more uncertain,
because it has only 2.5$\sigma$ detections; the inferred 
\nhtwo\ is high, $\sim 5\times 10^{21}$ (78\,\msun\,\pctwo), 
at a temperature of $\sim$95\,K.
With the exception of SBS\,1152$+$579,
the inferred \htwo\ column densities derived for the BCDs 
are rather high, compared to the values normally thought to hold  
for metal-poor galaxies.

Converting to masses, the two most extreme cases
are Mrk\,996 with a total warm (98\,K) \htwo\ mass  
in the IRS beam of 1.5$\times10^8$\,\msun\
and \cgcg\ (at 120\,K) of 1.3$\times10^8$\,\msun. 
Compared to the SINGS galaxies, these are rather high warm \htwo\ masses. 
There are only three SINGS galaxies with warm ($\simgt$100\,K)
\htwo\ masses which exceed $10^8$\,\msun, and only one (NGC\,7552)
with a warm \htwo\ mass greater than that of Mrk\,996.
The other BCDs in our sample, with at least one
low-order detection have warm (250-470\,K) \htwo\ masses ranging from
$4\times10^3$\,\msun\ (Mrk\,209) to $5\times10^7$\,\msun\ (HS\,0837+4717).
The median SINGS warm \htwo\ mass is $\sim3.4\times10^6$\,\msun,
and ranges from $\sim10^3$ to $3\times10^8$\,\msun\ \citep{roussel07}.
Most of the BCDs have $\simgt 10^3$ times more warm \htwo\ mass than the SINGS
low-metallicity galaxies with CO detections, NGC\,6822 and NGC\,2915
\citep{roussel07}. 

As stated above, these mass estimates may be overestimates. 
The true beam size is
probably smaller than that of LH. Moreover, molecular hydrogen is not 
distributed uniformly within that beam but in dense clumps. 
However, even if we reduced the mass estimates by a factor of five 
(roughly the difference in beam area between LH and SH),
the BCDs' molecular content would still be quite high.
This is contrary to the conventional view that \htwo\ molecules 
do not exist in a low-metallicity environment, because of the lack of
grains on which to form them. 
Clearly, metal abundance is not the only factor guiding \htwo\ 
formation in low-metallicity BCDs. 


Finally, we can compare the warm \htwo\ mass to that of the neutral atomic gas. 
According to the star formation models by \citet{krumholz09}, 
the \htwo\ column density of Mrk\,966
would correspond to an ISM molecular fraction $\sim$50\%, given its metal abundance.
Mrk\,996 has an \hi\ mass of 1.5$\times10^8$\,\msun \citep{thuan99b};
thus its \htwo\ warm mass is comparable to that of 
its neutral atomic gas mass, in agreement with the theoretical
predictions of \citet{krumholz09}.

Surprisingly, our data suggest that
metallicity and \htwo\ content are not directly linked. 
Of the four BCDs discussed above with significant low-order \htwo\ detections, 
three are below the sample median \logoh\ of 7.9. 
Other factors must play a role in determining the \htwo\ fraction.
In star-forming regions where substantial amounts of dust are concentrated
in high-density regions, self-shielding can promote \htwo\ formation;
thus the region can cool more effectively and star-formation rate 
is enhanced \citep{hirashita04}. 
Thus \htwo\ content may be more correlated with compactness (size and density)
than with metal abundance. 

\subsection{Water and the Hydroxyl Radical \label{sec:water}}

Our IRS spectra tentatively suggest the presence of \water\ and OH
in six objects (see Table \ref{tab:molecules}).
Figure \ref{fig:h2o_oh} shows the $\simgt 3\sigma$ detections of
\water\ in three BCDs and OH in four. 
The \water\ lines at 29.836\,\micron\ and 29.885\,\micron\ result
from rotational transitions of ${\rm H_2\,^{16}O}\ 7_{25}-6_{16}$ and
$5_{42}-4_{13}$, respectively.
The OH emission lines correspond to rotationally excited levels 
$^{2}\Pi_{1/2}\rightarrow ^{2}\Pi_{3/2} 7/2-5/2$ (28.940\,\micron),
$^{2}\Pi_{3/2} 19/2-17/2$ (30.346\,\micron), and
$^{2}\Pi_{1/2} 17/2-15/2$ (30.657\,\micron).
Because some of the spectra also show spurious spectral features
near the tentative detections, not associated with 
known emission lines, these molecule detections are uncertain.
Nevertheless, if real, the transitions of ``hot'' \water\ and OH
would be, to the best of our knowledge, 
the first such detections in extragalactic objects.

Molecular emission from \water\ and OH is perhaps not unexpected,
given the physical conditions in some BCDs.
\water\ emission from vapor is thought to arise from
slow nondissociative, or C-type shocks \citep[e.g.,][]{draine80}.
In such a situation, at a shock velocity of $\simgt$10\,km\,s$^{-1}$,
temperatures would exceed 300\,K, and subsequent reactions could rapidly
convert all gas-phase oxygen not bound in CO into \water\ \citep{elitzur78}.
If there is a connection between 
the dissociative shocks responsible for \water\ vapor emission
and the fast radiative shocks possibly responsible for \oiv\ emission 
(see $\S$ \ref{sec:oxygen_iron}), we would expect both types of emission.
Indeed, all three galaxies with \water\ emission in our sample 
also show $\simgt 3\sigma$ \oiv\ detections.
Nevertheless, to be consistent with the shock scenario for \water\
production, the \htwo\ density would need to be rather high,
$\simgt 10^5$\cmthree\ \citep{elitzur78}.
The electron densities in the ionized gas 
inferred from the \siii\ line ratio ($\S$ \ref{sec:densities}) 
are considerably smaller than this, but the \oiv\ and \water\ emission
could arise from different regions that are much denser. 
We have no way of verifying this condition with the present data.
With \hers, it may be possible to further pursue observations of
\water\ vapor at low metallicity, and better understand the necessary
physical conditions and formation scenarios for this molecule.

OH emission could be associated
with the photo-dissociation of \water\ at photon energies $\simgt$9\,eV, 
similar to the situation in Galactic outflows \citep{tappe08}.
Interestingly, the four galaxies with OH detections 
also have high-order \htwo\ detections, 
implying the presence of hot dense gas, thought to be necessary
for the neutral reactions leading to OH and \water\ formation
\citep[e.g.,][]{hollenbach79}.

\section{Summary and Conclusions\label{sec:summary}}

We have presented low- and high-resolution IRS spectra, 
supplemented by IRAC and MIPS measurements,
of 22 BCD galaxies, obtained during
our Cycle 1 GO \spitzer\ program (PID 3139). The BCD sample was chosen
to span a wide range in oxygen abundance [\logoh\ between 7.4 and 8.3],
and in ISRF hardness as measured
by the intensity of the nebular \heii\ $\lambda$\,4686 emission line
relative to \hb. 
The IRS spectra provide a variety of fine-structure 
lines, aromatic features, and molecular lines which
probe the physical conditions in a metal-poor ISM,  
and enable a study of
the dust properties as function of metallicity, hardness and
intensity of the ISRF. 
We have used the PAHFIT routine to fit simultaneously the 
spectral features, the underlying continuum and
the extinction in the IRS spectra. Fluxes have also been derived by
fitting Gaussian profiles to all FS and molecular lines. To place the
results for our low-metallicity BCD sample in
perspective, we have compared its properties to those of the SINGS  
galaxy sample \citep{kennicutt03}.     

We have obtained the following results: 

\begin{enumerate}[(1)]
\item
The \neiii/\neii\ and \siv/\siii\ flux ratios are
good diagnostics for the softer UV radiation ($\leq$ 40 eV). 
The \neiii/\neii\ ratio depends on metallicity, 
being 1--2 orders of magnitude greater in our metal-poor 
BCD sample than in the more metal-rich SINGS galaxies, but
the ratio flattens out at \logoh$\simlt$8.3.
The \oiv/\SiII\ ratio is an effective measure of harder radiation,
and
there is a strong correlation of the \oiv/\SiII\ ratio with 
oxygen abundance, implying that lower metallicity galaxies have harder 
ionizing radiation than more metal-rich ones.
\item
We detect at 3$\sigma$ \oiv\ (ionization potential of 54.9\,eV)
in 7/22 observed galaxies, but
\feii\ (ionization potential of 7.9\,eV) in only 2. 
At low metallicity, \oiv\ emission is almost 4 times as common as 
\feii\ (32\% vs. 9\%). This is another indication that the ISRF of 
low-metallicity galaxies is very hard.  
\item
Electron densities derived from the IR \siii\ line ratio do not
correlate with those inferred from the optical \sii\ ratio, 
or with the presence of \oiv\ emission.
We would expect a correlation of \oiv\ intensity with electron density, 
if fast radiative shocks were responsible for the ionizing radiation 
that produces \oiv.
Perhaps the \siii\ IR lines arise in and probe a region that
is less dense than the \oiv\ region.
\item
The ratio of
71 to 160\micron\ fluxes \nfncolor\ is sensitive to the temperature of
the large (``classical'') grains, and is thus a good indicator of the 
ISRF intensity. The latter is $\simgt$5 times greater in the BCDs than
in the SINGS galaxy with the most intense ISRF, implying 
that metal-poor star-forming galaxies have not only a harder ISRF,
but also a more intense one than metal-rich galaxies.  
\item
Two-thirds of the BCDs show PAH features. 
The flux ratios of the different bands
are typical of the largest PAHs modeled so far, 
with $N_{\rm min} \simgt$100 C atoms.
We interpret these trends as an
indication that low-metallicity BCDs contain   
relatively larger PAHs than more metal-rich environments. 
Apparently, only these large PAHs are able to survive the hard and 
intense ISRF in a low-metallicity ISM. 
\item
The fraction \sigpah/TIR of PAH emission \sigpah\ normalized to the total 
IR luminosity TIR is considerably smaller in 
low-metallicity BCDs ($\sim$0.5\%) than in the more metal-rich SINGS 
galaxies ($\sim$10\%). 
At low metallicity, just as in metal-rich galaxies,
the PAH fraction of \ltir\ is constant.
\item
There is a good correlation between \sigpah/TIR and the \neiii/\neii 
flux ratio, but not with metallicity or with ISRF intensity.  
Evidently, over the range in UV energies $\simlt$41\,eV, the fraction
of TIR emerging as PAH emission depends on the hardness of the ISRF.
This suggests that the PAH fraction in BCDs is not directly 
controlled by metallicity, but rather
by the hardness of the radiation field, also responsible
for the destruction of the smallest PAH particles.
\item
The hardness of the ISRF in BCDs can be comparable to 
that in AGN. Because of this, the \oiv/\neii\
flux ratio, often used as a diagnostic to distinguish
star-forming galaxies from AGN at solar abundances, cannot play that
role in the low-metallicity regime: in the \oiv/\neii\ vs. 7.7\,\micron\ 
PAH band diagnostic diagram, low-metallicity BCDs occupy the same region
as metal-rich AGN. 
On the contrary, the AGN/starburst diagnostic diagram proposed 
by \citep{laurent00}, based on the continuum MIR slopes of these
types of objects, does a good job at separating AGN from starbursts, even
at low metallicity. 
\item
Our IRS spectra reveal a variety of \htwo\ rotational lines,
and more than one third of the objects in our sample (8 BCDs) have $\simgt 3\sigma$ 
detections in one or more of the four lowest-order transitions of \htwo. 
The BCDs contain more \htwo\ than most SINGS galaxies, relative to \ltir.
This is quite contrary to the usual assertion that \htwo\ molecules 
do not exist in a low-metallicity environment, because of the lack of
grains on which to form them. 
Clearly, metal abundance cannot be the only factor guiding \htwo\ 
formation in low-metallicity BCDs.  
\item
The mean ratio of \htwo\ to PAH emission is $\sim$6\%, or 10 times
larger than that for the SINGS galaxies (but comparable to SINGS AGN).
This difference arises mainly from the BCD deficit in PAH emission.
While PAHs can survive the impact
of FUV photons with energies between $\sim$11 and 13.6\,eV,
\htwo\ would be dissociated without self-shielding,
implying that \htwo\ could be somewhat self-shielded in BCDs.
\item
The mean excitation temperature derived for the warm gas 
from the lower-order \htwo\ transitions is 
$\sim$245\,K.
The temperatures of the hotter molecular component from the high-order
\htwo\ transitions
range from $\sim$820\,K (in Mrk\,996 and SBS\,1152$+$579) to $\sim$1600\,K. 
These temperatures are consistent with those found for starburst 
SINGS galaxies and ULIRGs.
The warm molecular gas masses in our BCDs range from $10^3$ to $10^8$\,\msun,
and can be comparable to the neutral hydrogen gas mass.
\item
Some of our IRS spectra suggest tentatively the presence of molecules 
other than \htwo, such as \water\ and OH. 
All three galaxies with \water\ emission 
also show a $\simgt 3\sigma$ \oiv\ line, 
consistent with a shock scenario for \water\ production.
All four galaxies with OH emission also have high-order \htwo\ detections, 
implying
the presence of hot, dense gas which would favor OH and \water\ formation.
\end{enumerate}     
 


Our IRS data have shown that the ISRF in metal-poor BCDs can
be as extreme in hardness as the radiation in AGN.
In fact, the deficit of PAH emission at low metallicity 
and their hard radiation as indicated by \oiv\ luminosity 
would have placed some BCDs 
as AGN in diagnostic diagrams such as that proposed by \citet{genzel98}.
Our analysis suggests that in the absence of an AGN,
low metal abundance is a necessary condition for such extreme fields. 
However, we have also shown that 
low metallicity alone does not guarantee an extreme ISRF
in terms of hardness and intensity. 
Other factors such as compactness of the star-forming region
must play a role.

We have inferred a deficit of small PAHs, and find that
PAH destruction mechanisms are more closely related 
to the hardness of the radiation field, rather than to metallicity.
This is a rather difficult exercise, since most of the \oiv\
detections are in galaxies where we have little information on the
PAH population, and the converse is also true.
Nevertheless, there appears to be a close connection between
hardness of the ISRF and the properties of the PAHs;
this connection is tighter than that with metallicity.
In fact, we find correlations of PAH properties with ISRF
hardness, but almost none with metallicity, in spite of the
good correlations between hardness tracers and metal abundance.

It is not clear whether the origin of such hard radiation
can be stellar.
In their detailed modeling of the \oiv\ line in Mrk 996,
\citet{thuan08} ran Costar models by \citet{schaerer97}
using stars with the highest effective temperature
(T$_{eff}$\,=53,000 K) and the hardest radiation possible
(corresponding to the hottest O3 stars). 
They failed to reproduce the observed intensity of the \oiv\ line 
by a factor of several. They then considered Wolf-Rayet stars of type
WNE-w. According to calculations by \citet{crowther99}, models for
WNE-w stars (early nitrogen Wolf-Rayet stars with weak lines) show a
strong ionizing flux above 54.4 eV, in contrast to the WCE and WNL
stars, which show negligible fluxes above that energy. The line 
intensity of \oiv\ in Mrk 996 can be reproduced with such a population
of WNE-w stars. However, 
\citet{thuan08} do not favor such an hypothesis because WNE-w stars are
very rare. Rather, those authors favor fast radiative shocks 
\citep{thuanizotov05}   
as the origin of the high-ionization radiation.
Shocks are a more realistic expectation in dense compact regions
with intense star
formation because of SN explosions and massive stellar
winds.
At low metallicity, the less efficient cooling may intensify
the effect of the shock.
We postpone an in-depth discussion of the origin of 
the hard ionizing radiation for the whole of our sample to a future paper.
There, we will present detailed photoionization and shock models to account
for both optical$+$ IR emission line intensities.

\acknowledgments

This work is based on observations made with the {\it Spitzer Space Telescope},
which is operated by JPL/Caltech under NASA contract 1407.
Support for this work was provided by NASA \spitzer\ GO grant
JPL-1263707.
We acknowledge financial contribution from contract ASI-INAF
I/016/07/0.


\begin{deluxetable}{lrrcrr}
\tablecaption{Sample Galaxies \label{tab:sample}}
\tablewidth{0pt}
\tablehead{
\colhead{Name} &
\colhead{Distance\tablenotemark{a}} &
\colhead{Redshift} &
\colhead{12$+$log(O/H)} &
\colhead{$c_{H\beta}$\tablenotemark{b}} &
\colhead{He{\sc ii}/H$\beta$} \\
}
\startdata
CGCG007-025  &  20.8 &  0.00483 &   7.76 &   0.29 &  0.0136\\
HS0837+4717  & 174.3 &  0.04195 &   7.60 &   0.29 &  0.0228\\
HS1442+4250  &  12.6 &  0.00211 &   7.63 &   0.11 &  0.0297\\
HS2236+1344  &  86.4 &  0.02062 &   7.47 &   0.16 &  0.0106\\
Haro3        &  17.4 &  0.00323 &   8.32 &   0.24 &  0.0000\\
IIZw70       &  21.8 &  0.00394 &   8.04 &   0.21 &  0.0056\\
J0519+0007   & 181.8 &  0.04476 &   7.43 &   0.30 &  0.0249\\
Mrk5         &  15.4 &  0.00264 &   8.04 &   0.42 &  0.0387\\
Mrk36        &   7.6 &  0.00215 &   7.81 &   0.02 &  0.0244\\
Mrk209       &   4.8 &  0.00094 &   7.81 &   0.00 &  0.0117\\
Mrk450       &  15.1 &  0.00288 &   8.15 &   0.14 &  0.0031\\
Mrk724       &  20.0 &  0.00402 &   8.03 &   0.12 &  0.0000\\
Mrk996       &  21.8 &  0.00541 &   8.10 &   0.53 &  0.0000\\
Mrk1315      &  13.1 &  0.00282 &   8.25 &   0.16 &  0.0000\\
Mrk1329      &  13.1 &  0.00544 &   8.25 &   0.16 &  0.0000\\
SBS0917+527  &  35.4 &  0.00776 &   7.90 &   0.09 &  0.0221\\
SBS0946+558  &  25.5 &  0.00517 &   8.04 &   0.18 &  0.0149\\
SBS1030+583  &  35.3 &  0.00757 &   7.83 &   0.00 &  0.0234\\
SBS1152+579  &  74.5 &  0.01720 &   7.85 &   0.26 &  0.0152\\
SBS1415+437  &  11.7 &  0.00203 &   7.61 &   0.00 &  0.0227\\
Tol1924-416  &  38.4 &  0.00945 &   7.94 &   0.26 &  0.0190\\
UM311        &  23.0 &  0.00559 &   8.31 &   0.15 &  0.0000\\
\enddata
\tablenotetext{a}{Distance in Mpc, taken from NED.} 
\tablenotetext{b}{Extinction coefficient: A$_{\rm H\beta}$=2.5,$c_{H\beta}$.} 
\end{deluxetable}

\begin{deluxetable}{lrrrr}
\tablecaption{Short-Wavelength Fine-Structure Lines\tablenotemark{a} \label{tab:fspahfit}}
\tabletypesize{\footnotesize}
\tablewidth{0pt}
\tablecolumns{ 5}
\tablehead{
\colhead{Name}  
  & \colhead{\arii} 
  & \colhead{\ariii} 
  & \colhead{\siv} 
  & \colhead{\neii} 
\\ 
  & \multicolumn{1}{c}{  6.985\micron} 
  & \multicolumn{1}{c}{  8.991\micron} 
  & \multicolumn{1}{c}{ 10.511\micron} 
  & \multicolumn{1}{c}{ 12.814\micron} 
\\ 
  & \multicolumn{1}{c}{ 15.759\,eV} 
  & \multicolumn{1}{c}{ 27.629\,eV} 
  & \multicolumn{1}{c}{ 34.830\,eV} 
  & \multicolumn{1}{c}{ 21.564\,eV} 
}
\startdata 
CGCG\,007$-$025  & \nodata (0.33) &   0.66 (0.18) &   5.98 (0.43) & \nodata (1.29) \\ 
Haro\,3          &   3.44 (0.97) &  10.60 (1.67) &  26.80 (3.70) &  27.50 (3.59) \\ 
HS\,0837$+$4717  & \nodata (0.13) &   0.28 (0.09) &   3.07 (0.41) & \nodata (1.23) \\ 
HS\,1442$+$4250  & \nodata (0.21) & \nodata (0.21) & \nodata (1.60) & \nodata (1.60) \\ 
HS\,2236$+$1344  & \nodata (0.12) & \nodata (0.12) &   2.50 (0.56) & \nodata (1.68) \\ 
II\,Zw\,70       & \nodata (0.72) & \nodata (0.72) &   4.03 (0.26) &   1.34 (0.30) \\ 
J0519$+$0007     & \nodata (0.22) & \nodata (0.22) & \nodata (1.17) & \nodata (1.17) \\ 
Mrk\,5           &   0.54 (0.08) &   0.52 (0.08) &   1.64 (0.13) &   1.08 (0.13) \\ 
Mrk\,36          & \nodata (0.27) & \nodata (0.27) &   2.36 (0.14) & \nodata (0.43) \\ 
Mrk\,209         &   1.26 (0.33) &   1.58 (0.26) &   8.63 (0.18) & \nodata (0.45) \\ 
Mrk\,450         & \nodata (0.21) &   1.46 (0.13) &   4.45 (0.35) &   0.91 (0.29) \\ 
Mrk\,724         & \nodata (0.28) & \nodata (0.28) &   3.92 (0.44) & \nodata (1.31) \\ 
Mrk\,996         & \nodata (1.23) &   1.87 (0.40) &   3.90 (0.32) &   4.55 (0.32) \\ 
Mrk\,1315        &   0.73 (0.05) &   1.36 (0.12) &  10.20 (0.48) & \nodata (1.43) \\ 
Mrk\,1329        & \nodata (0.46) &   1.80 (0.15) &  12.30 (0.27) &   1.27 (0.27) \\ 
SBS\,0917$+$527  & \nodata (0.12) & \nodata (0.12) &   1.46 (0.12) & \nodata (0.36) \\ 
SBS\,0946$+$558  & \nodata (0.15) & \nodata (0.15) & \nodata (1.63) & \nodata (1.63) \\ 
SBS\,1030$+$583  & \nodata (0.10) & \nodata (0.10) &   1.93 (0.59) & \nodata (1.77) \\ 
SBS\,1152$+$579  & \nodata (0.40) &   0.92 (0.13) &   6.32 (0.12) & \nodata (0.36) \\ 
SBS\,1415$+$437  & \nodata (0.18) & \nodata (0.18) &   1.02 (0.22) & \nodata (0.49) \\ 
Tol\,1924$-$416  & \nodata (0.80) &   2.56 (0.27) &  11.40 (0.34) &   1.88 (0.34) \\ 
UM\,311          & \nodata (0.29) &   1.71 (0.23) &   5.22 (0.69) & \nodata (2.08) \\ 
\enddata
\tablenotetext{a}{Fluxes are measured by integrating Gaussian profiles
as fit by PAHFIT, and
are in units of $10^{-17}$\,W\,m$^{-2}$.
$1\sigma$ uncertainties are given in parentheses; when there is no
flux available, the value reported in parentheses is the $3\sigma$ upper limit.} 
\end{deluxetable}

\begin{deluxetable}{lrrrrrrrc}
\rotate
\tablecaption{Dust Features\tablenotemark{a} \label{tab:dfpahfit}}
\tabletypesize{\footnotesize}
\setlength{\tabcolsep}{0.04in}
\tablewidth{0pt}
\tablecolumns{9}
\tablehead{
\colhead{Name}  
  & \colhead{DF\_5}  
  & \colhead{DF\_6}  
  & \colhead{DF\_7\tablenotemark{b}}  
  & \colhead{DF\_8}  
  & \colhead{DF\_11}  
  & \colhead{DF\_12}  
  & \colhead{DF\_17}  
  & \colhead{$\Sigma$PAH} \\
  & \multicolumn{1}{c}{  5.7\micron} 
  & \multicolumn{1}{c}{  6.2\micron} 
  & \multicolumn{1}{c}{  7.7\micron} 
  & \multicolumn{1}{c}{  8.6\micron} 
  & \multicolumn{1}{c}{ 11.3\micron} 
  & \multicolumn{1}{c}{ 12.6\micron} 
  & \multicolumn{1}{c}{ 17\micron} 
  & Total\tablenotemark{c}\\
  & & & & & & & Complex 
}
\startdata 
CGCG\,007$-$025  & \nodata (0.33) & \nodata (0.33) &   6.34 (0.82) & \nodata (0.33) &   2.81 (0.51) & \nodata (1.28) & \nodata (0.33) &  12.79 (1.12) \\ 
&  \nodata &  \nodata &  7.459 &  \nodata & 11.199 &  \nodata \\ 
&  \nodata &  \nodata &  0.470 &  \nodata &  0.067 &  \nodata \\ 
Haro\,3          &  20.30 (3.53) & 101.00 ( 10.60) & 188.50 ( 31.85) &  56.30 ( 14.40) & 148.90 ( 41.95) & 111.00 ( 19.30) & \nodata (1.39) & 680.60 ( 60.62) \\ 
&  5.744 &  6.245 &  7.808 &  8.682 & 11.335 & 12.673 \\ 
&  0.188 &  0.194 &  0.362 &  0.279 &  0.544 &  0.798 \\ 
HS\,0837$+$4717  & \nodata (2.91) & \nodata (0.13) & \nodata (0.13) & \nodata (0.13) &   1.37 (0.41) & \nodata (1.23) & \nodata (0.13) &   2.95 (0.57) \\ 
&  \nodata &  \nodata &  \nodata &  \nodata & 11.214 &  \nodata \\ 
&  \nodata &  \nodata &  \nodata &  \nodata &  0.090 &  \nodata \\ 
HS\,1442$+$4250  & \nodata (0.21) &   0.94 (0.24) & \nodata (0.21) & \nodata (0.57) &   3.46 (0.53) & \nodata (1.60) & \nodata (0.21) &   6.15 (0.79) \\ 
&  \nodata &  6.294 &  \nodata &  \nodata & 11.347 &  \nodata \\ 
&  \nodata &  0.094 &  \nodata &  \nodata &  0.545 &  \nodata \\ 
HS\,2236$+$1344  & \nodata (0.12) & \nodata (0.12) & \nodata (0.12) & \nodata (0.12) & \nodata (1.68) & \nodata (1.68) & \nodata (0.12) & \nodata (0.12) \\ 
&  \nodata &  \nodata &  \nodata &  \nodata &  \nodata &  \nodata \\ 
&  \nodata &  \nodata &  \nodata &  \nodata &  \nodata &  \nodata \\ 
II\,Zw\,70       & \nodata (0.72) & \nodata (3.78) &  10.81 (1.98) & \nodata (0.72) & \nodata (0.66) & \nodata (0.66) & \nodata (0.72) &  12.39 (2.03) \\ 
&  \nodata &  \nodata &  7.808 &  \nodata &  \nodata &  \nodata \\ 
&  \nodata &  \nodata &  0.207 &  \nodata &  \nodata &  \nodata \\ 
J0519$+$0007     & \nodata (0.22) & \nodata (0.22) &   2.45 (0.55) & \nodata (0.22) & \nodata (1.17) & \nodata (1.17) & \nodata (0.22) &   4.24 (0.67) \\ 
&  \nodata &  \nodata &  7.405 &  \nodata &  \nodata &  \nodata \\ 
&  \nodata &  \nodata &  0.467 &  \nodata &  \nodata &  \nodata \\ 
Mrk\,1315        & \nodata (0.15) &   3.97 (0.12) &   1.91 (0.10) &   3.47 (0.48) &   1.45 (0.48) & \nodata (1.43) & \nodata (0.15) &  15.24 (0.75) \\ 
&  \nodata &  6.201 &  7.500 &  8.648 & 11.256 &  \nodata \\ 
&  \nodata &  0.279 &  0.165 &  0.431 &  0.088 &  \nodata \\ 
Mrk\,1329        & \nodata (5.70) & \nodata (3.03) &   9.31 (0.99) &   4.70 (0.87) &   6.94 (0.54) &   1.89 (0.27) &   1.85 (0.49) &  24.69 (1.53) \\ 
&  \nodata &  \nodata &  7.520 &  8.537 & 11.272 & 12.627 \\ 
&  \nodata &  \nodata &  0.474 &  0.166 &  0.541 &  0.265 \\ 
Mrk\,209         & \nodata (2.82) & \nodata (0.32) & \nodata (2.37) & \nodata (0.32) & \nodata (0.45) & \nodata (0.45) & \nodata (0.32) &   7.41 (1.28) \\ 
&  \nodata &  \nodata &  \nodata &  \nodata &  \nodata &  \nodata \\ 
&  \nodata &  \nodata &  \nodata &  \nodata &  \nodata &  \nodata \\ 
Mrk\,36          & \nodata (3.00) & \nodata (0.27) & \nodata (0.27) & \nodata (0.27) & \nodata (0.43) & \nodata (0.43) & \nodata (0.27) &   4.73 (1.32) \\ 
&  \nodata &  \nodata &  \nodata &  \nodata &  \nodata &  \nodata \\ 
&  \nodata &  \nodata &  \nodata &  \nodata &  \nodata &  \nodata \\ 
Mrk\,450         &   2.06 (0.25) &   6.16 (0.23) &  16.40 (1.42) &   5.95 (0.85) &   5.46 (1.61) & \nodata (0.76) & \nodata (0.21) &  36.03 (2.33) \\ 
&  5.800 &  6.291 &  7.520 &  8.520 & 11.331 &  \nodata \\ 
&  0.169 &  0.283 &  0.659 &  0.255 &  0.388 &  \nodata \\ 
Mrk\,5           &   2.17 (0.20) &   3.27 (0.14) &   7.47 (0.71) &   3.45 (0.39) &   5.42 (0.64) & \nodata (0.39) & \nodata (0.24) &  24.33 (1.08) \\ 
&  5.791 &  6.287 &  7.801 &  8.708 & 11.326 &  \nodata \\ 
&  0.214 &  0.283 &  0.281 &  0.440 &  0.544 &  \nodata \\ 
Mrk\,724         & \nodata (0.28) & \nodata (0.28) & \nodata (0.28) & \nodata (2.16) & \nodata (1.31) & \nodata (1.31) & \nodata (0.28) & \nodata (0.28) \\ 
&  \nodata &  \nodata &  \nodata &  \nodata &  \nodata &  \nodata \\ 
&  \nodata &  \nodata &  \nodata &  \nodata &  \nodata &  \nodata \\ 
Mrk\,996         &   7.03 (1.45) &  15.50 (4.32) &  17.90 (4.35) & \nodata (9.63) &  11.80 (2.02) &   3.53 (0.71) & \nodata (0.66) &  60.66 (6.71) \\ 
&  5.757 &  6.236 &  7.493 &  \nodata & 11.276 & 12.684 \\ 
&  0.101 &  0.179 &  0.472 &  \nodata &  0.250 &  0.082 \\ 
SBS\,0917$+$527  & \nodata (0.12) & \nodata (0.12) & \nodata (0.12) &   1.42 (0.23) &   1.00 (0.29) &   0.37 (0.12) & \nodata (0.12) &   2.79 (0.39) \\ 
&  \nodata &  \nodata &  \nodata &  8.559 & 11.230 & 12.601 \\ 
&  \nodata &  \nodata &  \nodata &  0.167 &  0.539 &  0.082 \\ 
SBS\,0946$+$558  & \nodata (0.15) & \nodata (0.15) &   0.89 (0.28) & \nodata (0.15) &   2.74 (0.54) & \nodata (1.63) & \nodata (0.15) &  12.11 (1.03) \\ 
&  \nodata &  \nodata &  7.520 &  \nodata & 11.230 &  \nodata \\ 
&  \nodata &  \nodata &  0.474 &  \nodata &  0.539 &  \nodata \\ 
SBS\,1030$+$583  & \nodata (0.10) & \nodata (0.10) &   4.75 (0.40) &   0.67 (0.14) & \nodata (1.77) & \nodata (1.77) & \nodata (0.10) &   5.42 (0.42) \\ 
&  \nodata &  \nodata &  7.520 &  8.534 &  \nodata &  \nodata \\ 
&  \nodata &  \nodata &  0.474 &  0.166 &  \nodata &  \nodata \\ 
SBS\,1152$+$579  &   3.30 (0.70) & \nodata (0.40) &  12.20 (0.67) &   3.41 (0.46) & \nodata (0.36) & \nodata (0.36) & \nodata (0.40) &  24.15 (1.11) \\ 
&  5.800 &  \nodata &  7.520 &  8.710 &  \nodata &  \nodata \\ 
&  0.305 &  \nodata &  0.474 &  0.510 &  \nodata &  \nodata \\ 
SBS\,1415$+$437  & \nodata (0.18) & \nodata (0.18) &  12.84 (0.25) & \nodata (2.04) & \nodata (0.49) & \nodata (0.49) & \nodata (0.18) &  12.84 (0.25) \\ 
&  \nodata &  \nodata &  7.782 &  \nodata &  \nodata &  \nodata \\ 
&  \nodata &  \nodata &  0.258 &  \nodata &  \nodata &  \nodata \\ 
Tol\,1924$-$416  & \nodata (0.80) & \nodata (0.80) &   6.39 (1.40) &   2.08 (0.62) &   2.32 (0.58) &   3.09 (0.41) & \nodata (0.80) &  13.88 (1.69) \\ 
&  \nodata &  \nodata &  7.507 &  8.510 & 11.230 & 12.643 \\ 
&  \nodata &  \nodata &  0.473 &  0.166 &  0.180 &  0.266 \\ 
UM\,311          & \nodata (2.37) &  12.70 (2.14) &  38.76 (5.37) &   4.27 (1.29) &   5.53 (0.69) & \nodata (2.08) & \nodata (0.29) &  64.98 (6.00) \\ 
&  \nodata &  6.216 &  7.822 &  8.634 & 11.237 &  \nodata \\ 
&  \nodata &  0.163 &  0.207 &  0.196 &  0.067 &  \nodata \\ 
\enddata
\tablenotetext{a}{Dust features are integrated over the entire
best-fit Drude profile as performed by PAHFIT. 
In the first line, fluxes are in units of $10^{-17}$\,W\,m$^{-2}$.
$1\sigma$ uncertainties are given in parentheses; when there is no
flux available, the value reported in parentheses is the $3\sigma$ upper limit.
The PAHFIT best-fit central wavelengths in \micron\ are given in the second line,
and the best-fit FWHM in \micron\ in the third.  } 
\tablenotetext{b}{This complex includes the sum of features 
from 7.4 to 7.8\,\micron; the FWHM corresponds to the longest wavelength feature.}
\tablenotetext{c}{This total includes the sum of features
in the table, together with features identified as PAHs by PAHFIT
with wavelengths from 
6.6 to 6.8\,\micron,
8.3\,\micron, and
10.6 to 10.8\,\micron.
}
 
\end{deluxetable}

\begin{deluxetable}{lrrrrrrrrrr}
\rotate
\tablecaption{Long-Wavelength Fine-Structure Lines\tablenotemark{a} \label{tab:fsgauss}}
\tabletypesize{\footnotesize}
\setlength{\tabcolsep}{0.04in}
\tablewidth{0pt}
\tablecolumns{ 9}
\tablehead{
\colhead{Name}  
  & \colhead{\neiii} 
  & \colhead{\siii} 
  & \colhead{\oiv} 
  & \colhead{\feii} 
  & \colhead{\feiii} 
  & \colhead{\siii} 
  & \colhead{\SiII} 
  & \colhead{\neiii} 
\\
  & \multicolumn{1}{c}{ 15.555\micron} 
  & \multicolumn{1}{c}{ 18.713\micron} 
  & \multicolumn{1}{c}{ 25.890\micron} 
  & \multicolumn{1}{c}{ 25.988\micron} 
  & \multicolumn{1}{c}{ 33.038\micron} 
  & \multicolumn{1}{c}{ 33.481\micron} 
  & \multicolumn{1}{c}{ 34.815\micron} 
  & \multicolumn{1}{c}{ 36.014\micron} 
\\ 
  & \multicolumn{1}{c}{ 40.962\,eV} 
  & \multicolumn{1}{c}{ 23.330\,eV} 
  & \multicolumn{1}{c}{ 54.934\,eV} 
  & \multicolumn{1}{c}{  7.870\,eV} 
  & \multicolumn{1}{c}{ 16.180\,eV} 
  & \multicolumn{1}{c}{ 23.330\,eV} 
  & \multicolumn{1}{c}{  8.151\,eV} 
  & \multicolumn{1}{c}{ 40.962\,eV} 
}
\startdata 
CGCG\,007$-$025  &   5.55 (0.49) & \nodata (1.28) & \nodata (1.49) & \nodata (1.49) & \nodata (1.77) &   3.48 (0.67) &   2.89 (0.50) & \nodata (1.49) \\ 
Haro\,3          &  79.58 (0.27) &  40.99 (0.27) &   5.38 (0.93) & \nodata (2.80) &  27.13 (0.93) & 116.41 (0.93) &  53.47 (0.93) &  12.11 (0.93) \\ 
HS\,0837$+$4717  &   1.54 (0.41) & \nodata (1.23) & \nodata (0.92) & \nodata (0.92) & \nodata (0.92) & \nodata (0.92) & \nodata (0.92) & \nodata (0.92) \\ 
HS\,1442$+$4250  & \nodata (1.60) & \nodata (1.60) & \nodata (1.00) & \nodata (1.00) & \nodata (1.00) & \nodata (1.00) & \nodata (1.00) & \nodata (1.00) \\ 
HS\,2236$+$1344  & \nodata (1.86) & \nodata (1.68) & \nodata (0.54) & \nodata (0.54) & \nodata (0.54) & \nodata (0.54) & \nodata (0.54) & \nodata (0.54) \\ 
II\,Zw\,70       &   6.78 (0.35) &   3.95 (0.22) &   0.77 (0.17) & \nodata (0.60) & \nodata (0.51) &  10.09 (0.17) &   6.65 (0.53) & \nodata (0.51) \\ 
J0519$+$0007     & \nodata (1.17) & \nodata (1.17) & \nodata (0.67) & \nodata (0.67) & \nodata (0.67) & \nodata (0.67) & \nodata (0.99) & \nodata (0.67) \\ 
Mrk\,5           &   4.31 (0.13) &   1.90 (0.13) & \nodata (0.21) &   0.24 (0.07) & \nodata (0.21) &   5.02 (0.07) &   2.45 (0.07) &   1.01 (0.07) \\ 
Mrk\,36          &   2.70 (0.14) &   1.53 (0.14) &   1.11 (0.30) & \nodata (0.91) & \nodata (0.91) &   2.63 (0.30) &   2.97 (0.30) & \nodata (0.91) \\ 
Mrk\,209         &   5.57 (0.15) &   2.69 (0.15) &   1.46 (0.19) & \nodata (0.57) & \nodata (0.57) &   5.66 (0.19) & \nodata (0.57) & \nodata (0.57) \\ 
Mrk\,450         &   7.00 (1.33) &   4.27 (0.25) & \nodata (0.64) & \nodata (0.64) & \nodata (0.64) &   5.68 (0.21) &   3.03 (0.21) &   1.96 (0.21) \\ 
Mrk\,724         &   3.98 (0.44) &   2.89 (0.44) & \nodata (1.73) & \nodata (1.73) & \nodata (1.73) & \nodata (1.73) & \nodata (1.73) & \nodata (1.73) \\ 
Mrk\,996         &   8.95 (0.32) &   7.31 (0.32) &   1.46 (0.21) &   0.96 (0.32) &   0.90 (0.21) &  16.43 (0.21) &   5.75 (0.21) & \nodata (0.63) \\ 
Mrk\,1315        &   9.74 (0.48) &   5.09 (0.48) & \nodata (0.80) & \nodata (0.80) & \nodata (0.80) &   7.26 (0.27) &   1.81 (0.27) & \nodata (0.80) \\ 
Mrk\,1329        &  15.51 (0.27) &   8.07 (0.32) & \nodata (2.10) & \nodata (2.10) & \nodata (2.10) &   8.59 (0.70) &   2.83 (0.70) & \nodata (2.10) \\ 
SBS\,0917$+$527  &   1.00 (0.12) &   0.67 (0.12) & \nodata (0.76) & \nodata (0.76) & \nodata (0.76) & \nodata (0.76) & \nodata (0.76) & \nodata (0.76) \\ 
SBS\,0946$+$558  & \nodata (1.63) & \nodata (1.63) & \nodata (0.51) & \nodata (0.51) & \nodata (0.51) &   1.56 (0.17) &   0.57 (0.17) & \nodata (0.51) \\ 
SBS\,1030$+$583  & \nodata (1.77) & \nodata (1.77) & \nodata (0.83) & \nodata (0.83) & \nodata (0.83) & \nodata (0.83) & \nodata (0.83) & \nodata (0.83) \\ 
SBS\,1152$+$579  &   2.45 (0.12) &   1.38 (0.14) & \nodata (0.76) & \nodata (0.76) & \nodata (0.76) &   2.68 (0.25) & \nodata (0.76) & \nodata (0.76) \\ 
SBS\,1415$+$437  &   1.22 (0.16) &   1.09 (0.16) &   0.61 (0.13) & \nodata (0.40) & \nodata (0.40) &   1.41 (0.13) &   2.15 (0.13) & \nodata (0.40) \\ 
Tol\,1924$-$416  &  23.99 (0.34) &   7.40 (0.34) &   3.33 (0.61) & \nodata (3.51) & \nodata (1.84) &  18.57 (0.61) &  15.23 (0.74) & \nodata (6.54) \\ 
UM\,311          &  13.75 (0.69) &   2.53 (0.69) & \nodata (1.99) & \nodata (1.99) & \nodata (1.99) & \nodata (1.99) &   4.01 (0.66) & \nodata (1.99) \\ 
\enddata 
\tablenotetext{a}{Fluxes are measured by integrating Gaussian profiles, and
are in units of $10^{-17}$\,W\,m$^{-2}$.
$1\sigma$ uncertainties are given in parentheses; when there is no
flux available, the value reported in parentheses is the $3\sigma$ upper limit.} 
\end{deluxetable}

\begin{deluxetable}{lrrrrrrrr}
\rotate
\tablecaption{\htwo\ Lines\tablenotemark{a} \label{tab:h2}}
\tabletypesize{\footnotesize}
\setlength{\tabcolsep}{0.04in}
\tablewidth{0pt}
\tablecolumns{9}
\tablehead{
\colhead{Name} &
\colhead{S(7)} &
\colhead{S(6)} &
\colhead{S(5)} &
\colhead{S(4)} &
\colhead{S(3)} &
\colhead{S(2)} &
\colhead{S(1)\tablenotemark{b}} &
\colhead{S(0)\tablenotemark{b}} \\
  & \multicolumn{1}{c}{  5.511\micron} 
   & \multicolumn{1}{c}{  6.109\micron} 
   & \multicolumn{1}{c}{  6.910\micron} 
   & \multicolumn{1}{c}{  8.025\micron} 
   & \multicolumn{1}{c}{  9.665\micron} 
   & \multicolumn{1}{c}{ 12.279\micron} 
   & \multicolumn{1}{c}{ 17.035\micron} 
   & \multicolumn{1}{c}{ 28.219\micron} 
}
\startdata
CGCG\,007$-$025  &   0.46 (0.13) & \nodata (0.33) & \nodata (0.33) &   0.72 (0.21) & \nodata (0.33) &   1.67 (0.43) &   5.45 (0.43) & \nodata (1.49) \\ 
Haro\,3          &   2.03 (0.46) & \nodata (1.39) & \nodata (1.39) & \nodata (1.39) & \nodata (1.39) & \nodata (0.82) & \nodata (0.82) & \nodata (2.80) \\ 
HS\,0837$+$4717  &   0.44 (0.19) & \nodata (0.13) & \nodata (0.13) & \nodata (0.13) &   0.18 (0.07) & \nodata (1.23) &   1.07 (0.41) & \nodata (0.92) \\ 
HS\,1442$+$4250  &   0.47 (0.17) &   0.41 (0.14) & \nodata (0.21) & \nodata (0.21) &   0.17 (0.08) & \nodata (1.60) & \nodata (1.60) & \nodata (1.00) \\ 
II\,Zw\,70       & \nodata (0.72) &   1.53 (0.64) & \nodata (0.72) &   1.34 (0.46) &   0.69 (0.31) & \nodata (0.66) & \nodata (0.66) & \nodata (0.51) \\ 
J0519$+$0007     &   0.90 (0.17) & \nodata (0.22) & \nodata (0.22) &   0.34 (0.11) &   0.22 (0.07) & \nodata (1.17) & \nodata (1.17) & \nodata (0.67) \\ 
Mrk\,5           &   0.38 (0.08) &   0.60 (0.08) &   0.35 (0.08) &   0.56 (0.08) & \nodata (0.24) & \nodata (0.39) & \nodata (0.39) & \nodata (0.21) \\ 
Mrk\,36          & \nodata (0.27) & \nodata (0.27) & \nodata (0.27) & \nodata (0.27) &   0.94 (0.14) & \nodata (0.43) & \nodata (0.43) & \nodata (0.91) \\ 
Mrk\,209         & \nodata (0.32) & \nodata (0.32) & \nodata (0.32) & \nodata (0.32) &   0.52 (0.23) &   0.49 (0.16) & \nodata (0.45) & \nodata (0.57) \\ 
Mrk\,450         & \nodata (0.21) &   0.78 (0.08) &   0.58 (0.07) &   0.39 (0.18) &   0.20 (0.08) & \nodata (0.76) & \nodata (0.76) & \nodata (0.64) \\ 
Mrk\,996         & \nodata (0.66) & \nodata (0.66) &   1.19 (0.48) & \nodata (0.66) &   1.15 (0.35) & \nodata (0.97) &   0.74 (0.32) &   0.71 (0.21) \\ 
Mrk\,1315        & \nodata (0.15) & \nodata (0.15) &   0.18 (0.05) & \nodata (0.15) & \nodata (0.15) & \nodata (1.43) & \nodata (1.43) & \nodata (0.80) \\ 
Mrk\,1329        & \nodata (0.46) &   1.05 (0.21) & \nodata (0.46) &   0.78 (0.18) & \nodata (0.46) & \nodata (0.80) & \nodata (0.80) & \nodata (2.10) \\ 
SBS\,0917$+$527  & \nodata (0.12) & \nodata (0.12) & \nodata (0.12) &   0.38 (0.16) & \nodata (0.12) & \nodata (0.36) & \nodata (0.36) & \nodata (0.76) \\ 
SBS\,0946$+$558  & \nodata (0.15) & \nodata (0.15) &   0.21 (0.05) & \nodata (0.15) & \nodata (0.15) & \nodata (1.63) & \nodata (1.63) & \nodata (0.51) \\ 
SBS\,1030$+$583  &   0.53 (0.13) & \nodata (0.10) & \nodata (0.10) & \nodata (0.10) & \nodata (0.10) & \nodata (1.77) & \nodata (1.77) & \nodata (0.83) \\ 
SBS\,1152$+$579  & \nodata (0.40) & \nodata (0.40) & \nodata (0.40) &   0.41 (0.15) &   0.39 (0.13) &   0.30 (0.12) & \nodata (0.36) & \nodata (0.76) \\ 
SBS\,1415$+$437  & \nodata (0.18) & \nodata (0.18) &   2.04 (0.42) & \nodata (0.18) & \nodata (0.18) &   0.43 (0.20) & \nodata (0.49) & \nodata (0.40) \\ 
Tol\,1924$-$416  & \nodata (0.80) & \nodata (0.80) & \nodata (0.80) & \nodata (0.80) &   2.35 (0.27) & \nodata (1.03) & \nodata (1.03) & \nodata (1.84) \\ 
UM\,311          & \nodata (0.29) & \nodata (0.29) &   0.66 (0.27) &   0.73 (0.35) & \nodata (0.29) & \nodata (2.08) &   1.49 (0.69) & \nodata (1.99) \\ 
\enddata
\tablenotetext{a}{Fluxes are measured by integrating Gaussian profiles, and
are in units of $10^{-17}$\,W\,m$^{-2}$.
Unlike other tables, lines shown here as detections are 2$\sigma$.
$1\sigma$ uncertainties are given in parentheses; when there is no
flux available, the value reported in parentheses is the $3\sigma$ upper limit.} 
\tablenotetext{b}{Unlike the shorter-wavelength lines which are fitted with 
Gaussian profiles by PAHFIT, these are
measured by independent Gaussian profiles as described in the text.
}
\end{deluxetable}

\begin{deluxetable}{lrrrrrrrrrr}
\tablecaption{Water and Hydroxyl Lines\tablenotemark{a} \label{tab:molecules}}
\tabletypesize{\footnotesize}
\tablewidth{0pt}
\tablecolumns{6}
\tablehead{
\colhead{Name} 
  & \colhead{OH1/2-3/2}  
  & \colhead{o-H2O}  
  & \colhead{p-H2O}  
  & \colhead{OH19/2-17/2}  
  & \colhead{OH19/2-17/2}  
\\ 
  & \multicolumn{1}{c}{ 28.939\micron} 
  & \multicolumn{1}{c}{ 29.836\micron} 
  & \multicolumn{1}{c}{ 29.885\micron} 
  & \multicolumn{1}{c}{ 30.346\micron} 
  & \multicolumn{1}{c}{ 30.657\micron} 
\\ 
}
\startdata 
Haro\,3          & \nodata (2.80) & \nodata (2.80) &   4.57 (0.93) & \nodata (2.80) & \nodata (2.80) \\ 
II\,Zw\,70       & \nodata (0.51) & \nodata (0.51) & \nodata (0.51) &   0.59 (0.17) & \nodata (0.51) \\ 
Mrk\,5           & \nodata (0.21) & \nodata (0.21) & \nodata (0.21) & \nodata (0.21) &   0.32 (0.07) \\ 
Mrk\,996         & \nodata (0.63) &   2.48 (0.40) & \nodata (0.63) & \nodata (0.63) & \nodata (0.63) \\ 
Mrk\,1315        &   1.65 (0.27) & \nodata (0.80) & \nodata (0.81) & \nodata (0.80) & \nodata (0.81) \\ 
Tol\,1924$-$416  &   2.67 (0.61) & \nodata (1.84) &   2.30 (0.61) & \nodata (1.84) & \nodata (1.84) \\ 
\enddata
\tablenotetext{a}{Fluxes are measured by integrating Gaussian profiles, and
are in units of $10^{-17}$\,W\,m$^{-2}$.
$1\sigma$ uncertainties are given in parentheses; when there is no
flux available, the value reported in parentheses is the $3\sigma$ upper limit.} 
\end{deluxetable}


\clearpage

\setcounter{figure}{0}
\begin{figure}
\centerline{
\hbox{ 
\includegraphics[angle=0,width=0.33\linewidth,bb=40 251 587 713]{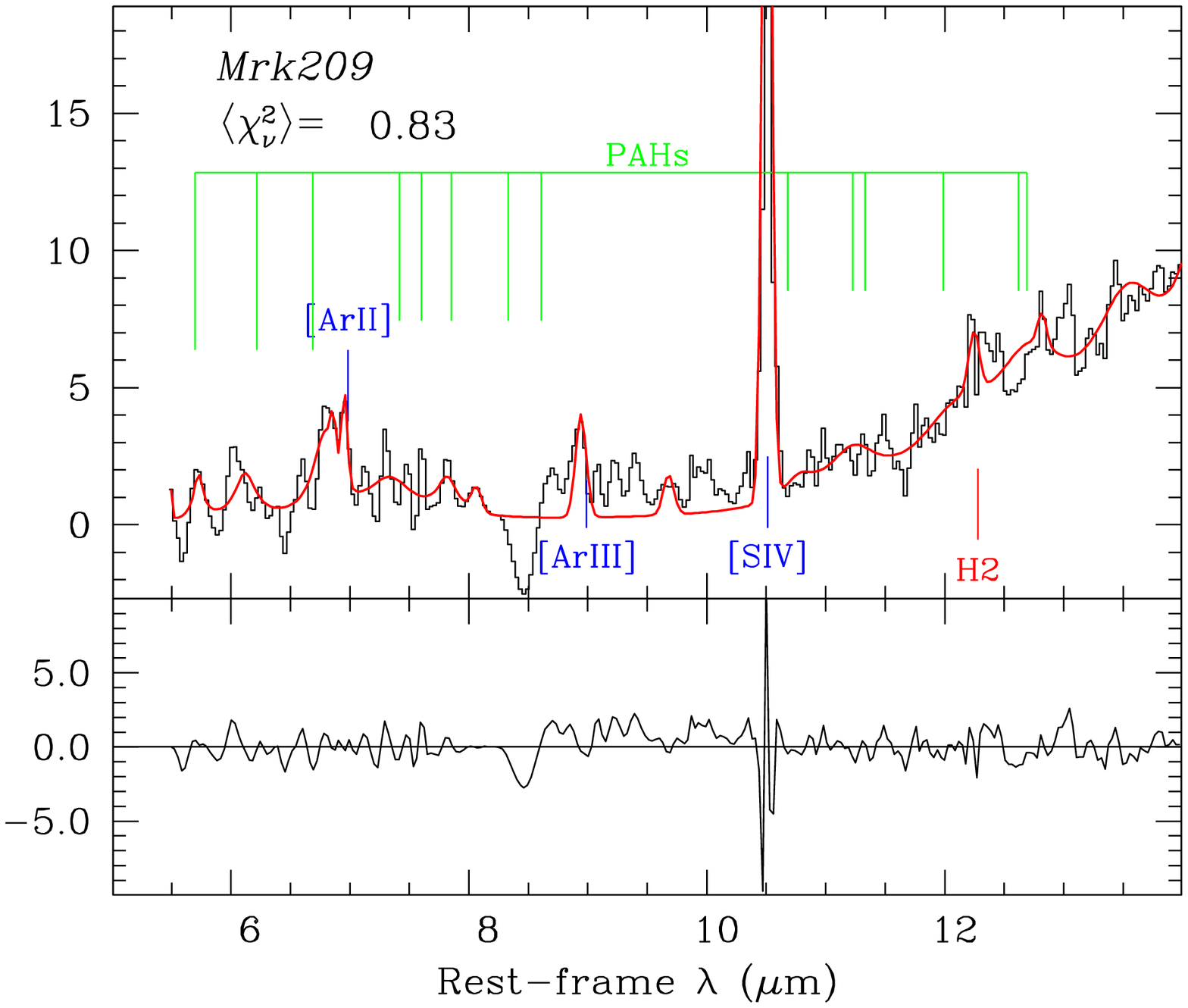} 
\hspace{-0.1cm}
\includegraphics[angle=0,width=0.33\linewidth,bb=40 251 587 713]{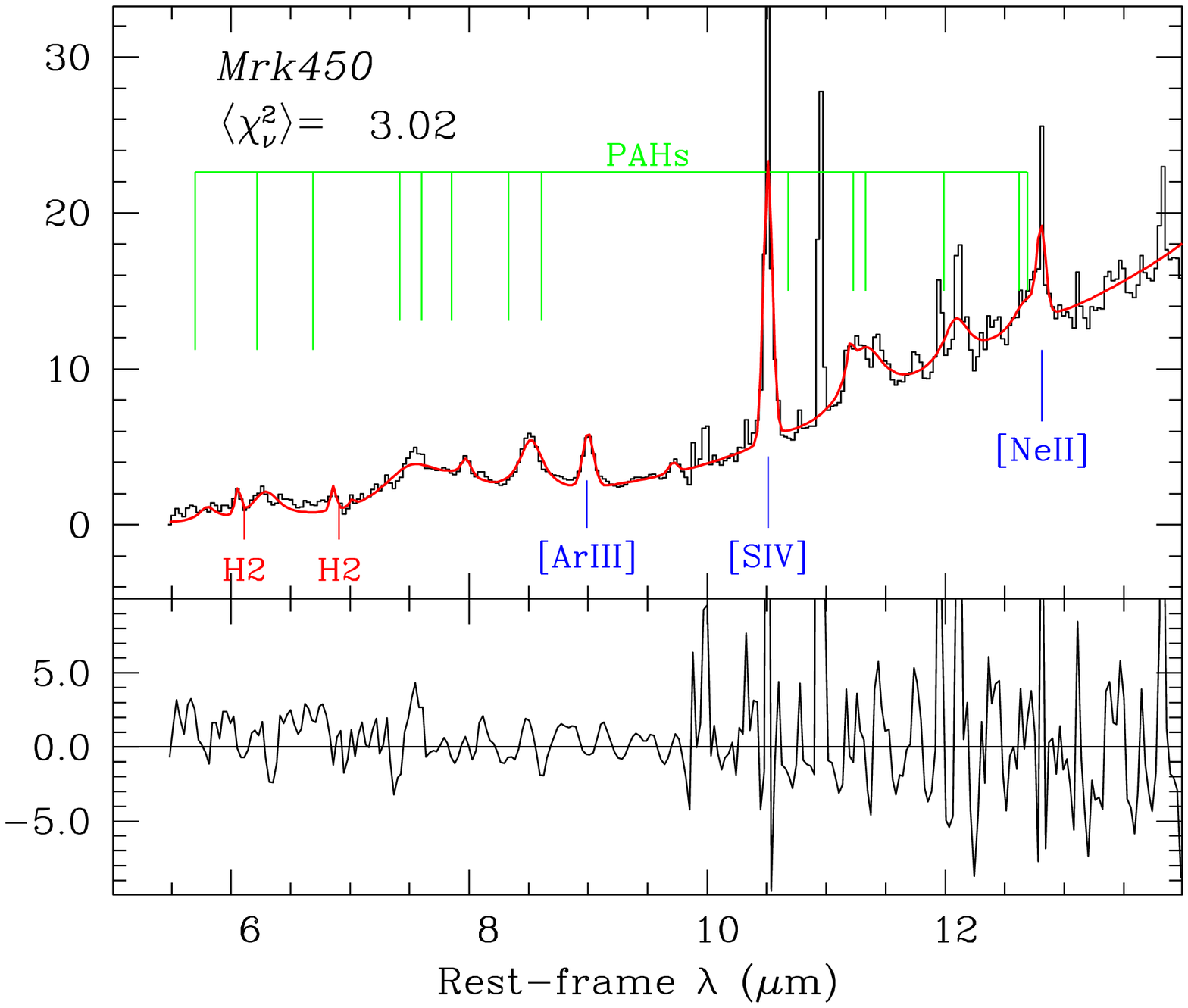}  
\hspace{-0.1cm}
\includegraphics[angle=0,width=0.33\linewidth,bb=40 251 587 713]{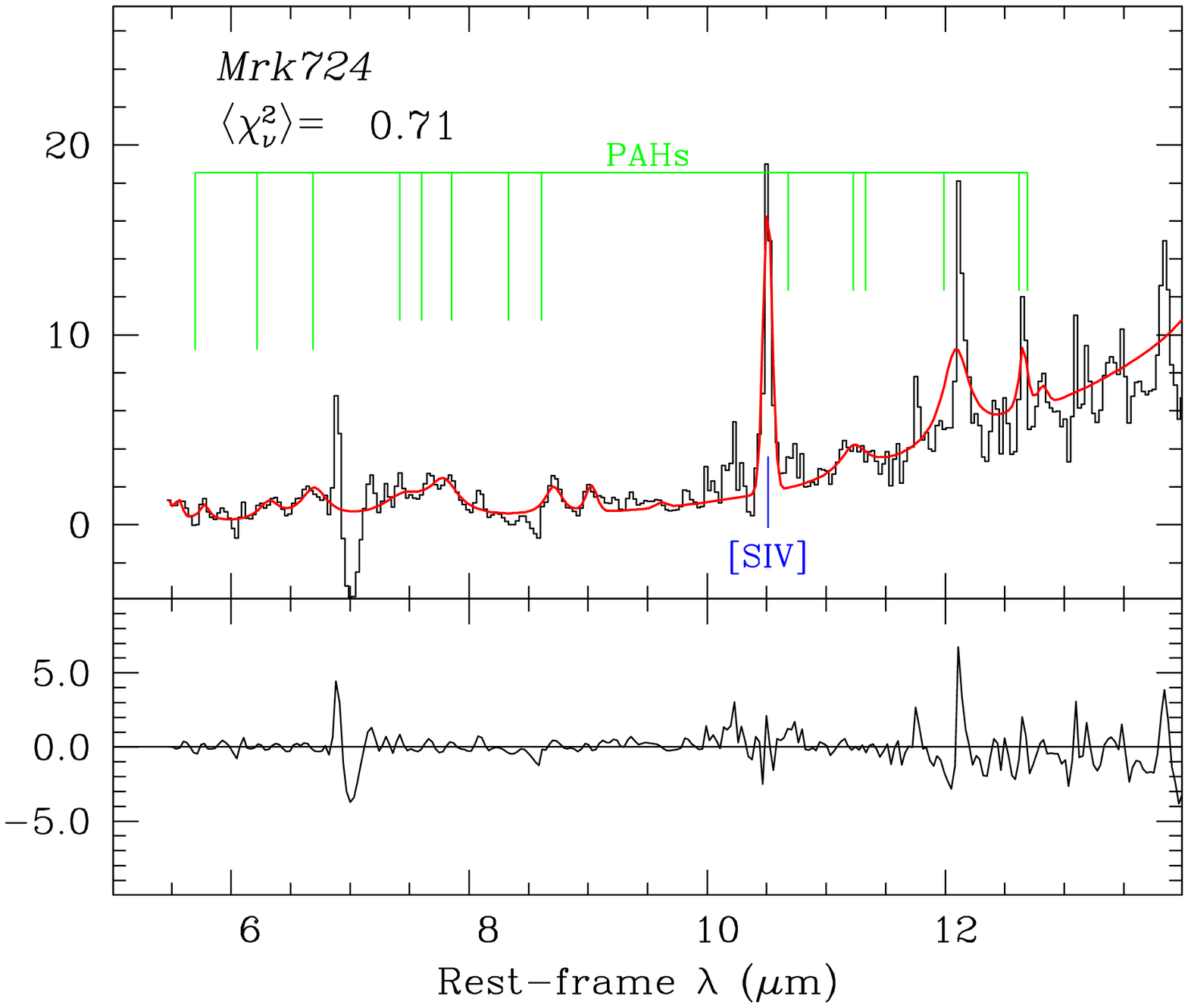}  
}}
\centerline{
\hbox{ 
\includegraphics[angle=0,width=0.33\linewidth,bb=40 251 587 713]{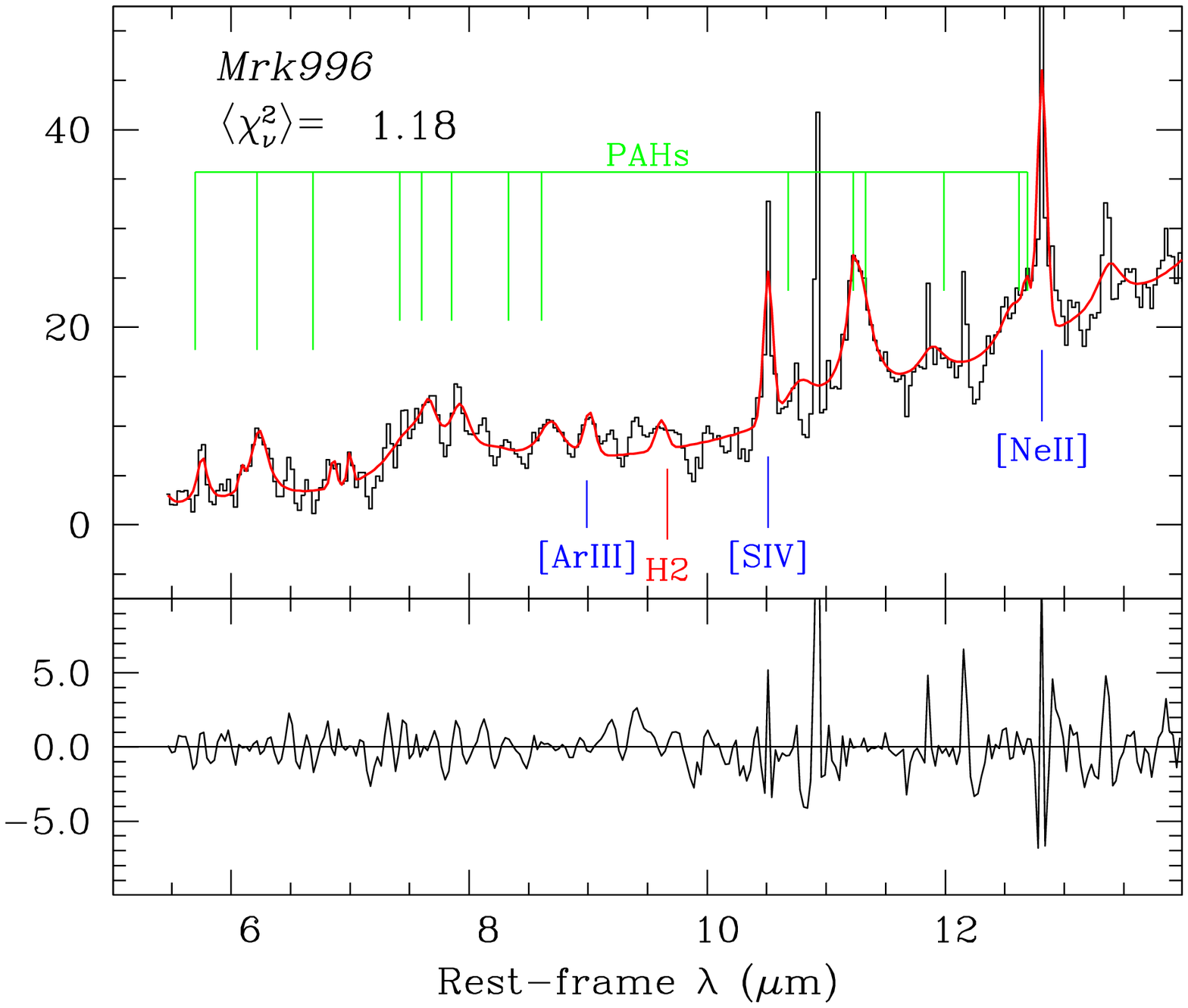} 
\hspace{-0.1cm}
\includegraphics[angle=0,width=0.33\linewidth,bb=40 251 587 713]{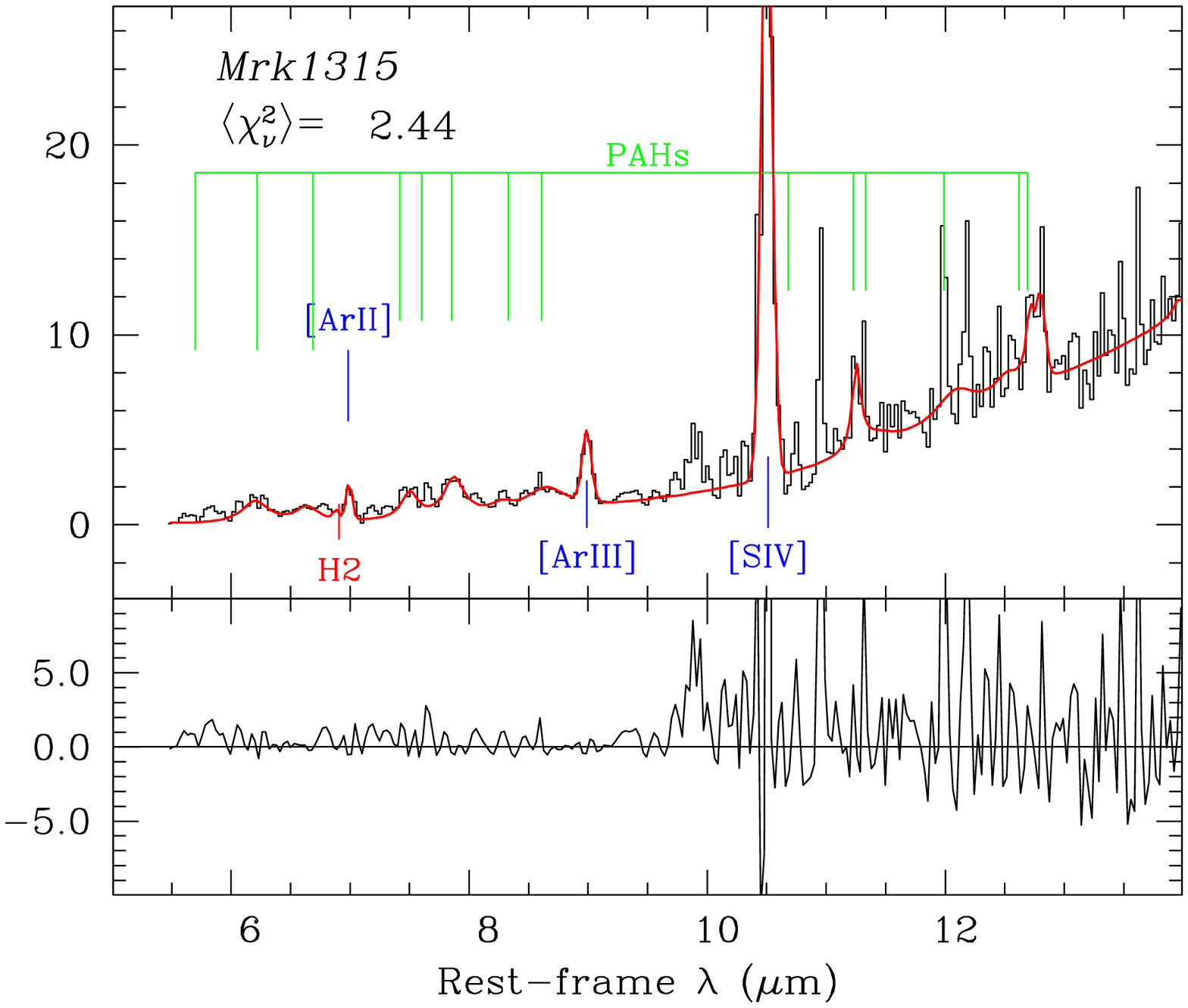} 
\hspace{-0.1cm}
\includegraphics[angle=0,width=0.33\linewidth,bb=40 251 587 713]{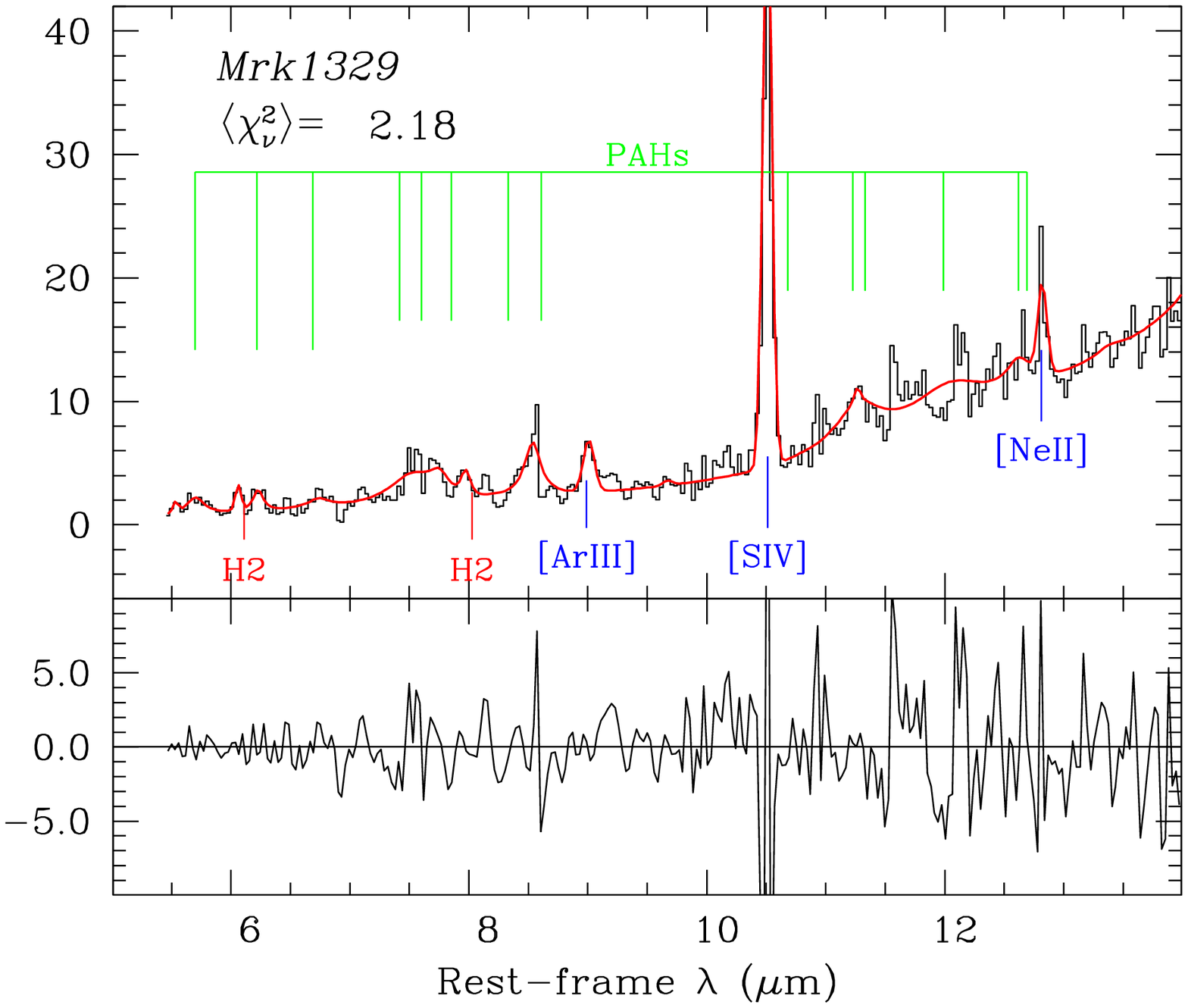} 
}}
\vspace{0.3cm}
\centerline{
\hbox{ 
\includegraphics[angle=0,width=0.33\linewidth,bb=40 251 587 713]{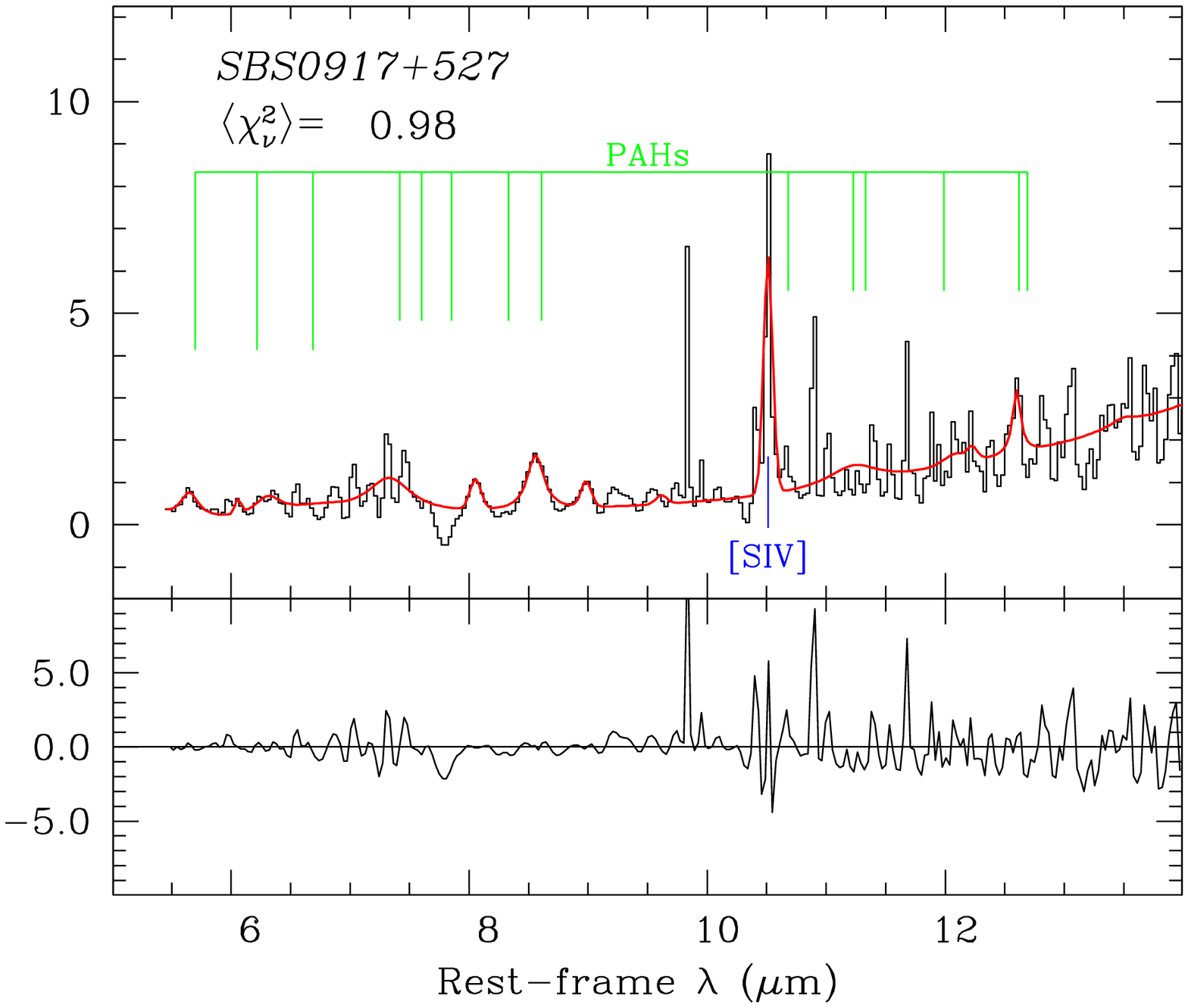} 
\hspace{-0.1cm}
\includegraphics[angle=0,width=0.33\linewidth,bb=40 251 587 713]{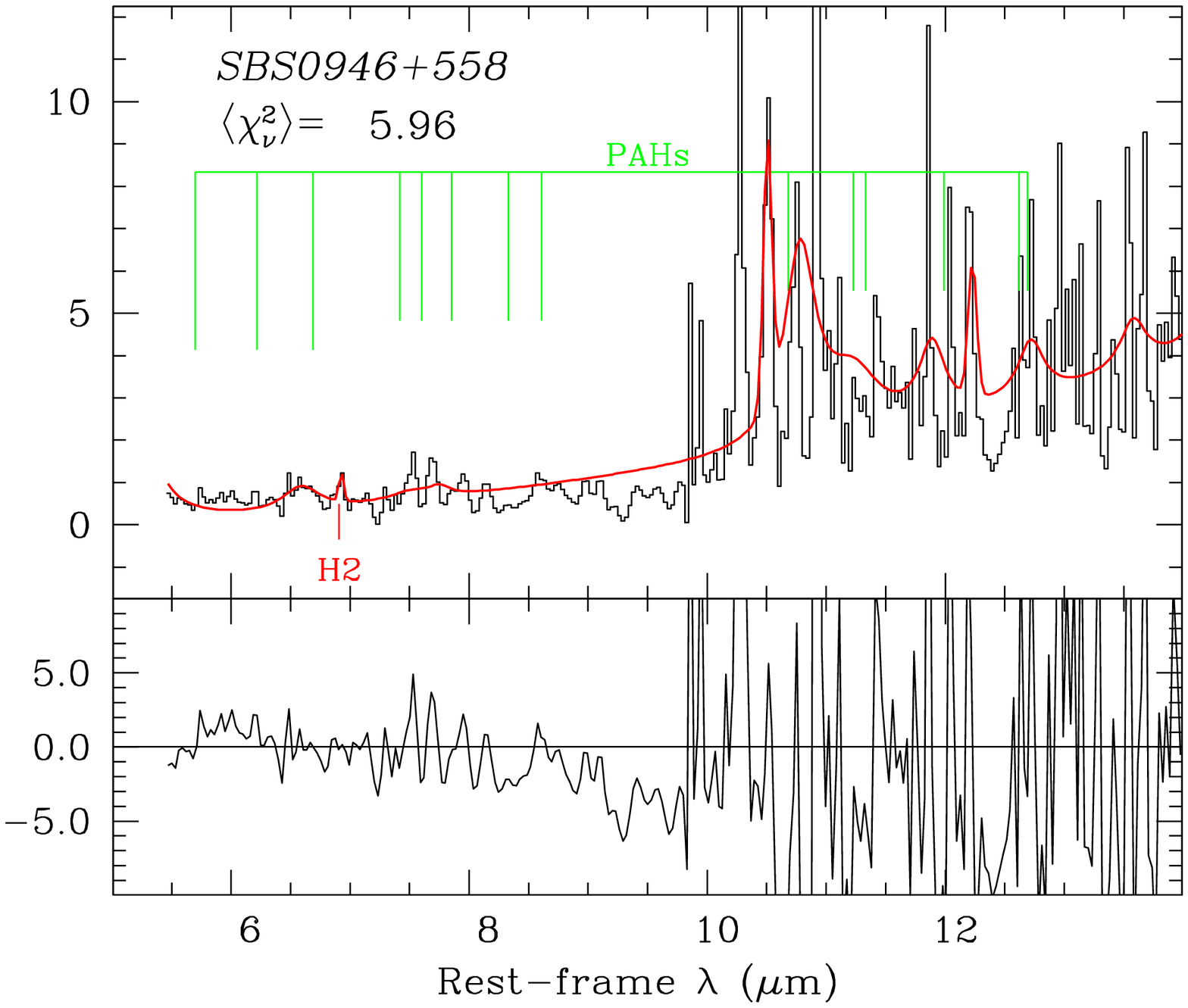} 
\hspace{-0.1cm}
\includegraphics[angle=0,width=0.33\linewidth,bb=40 251 587 713]{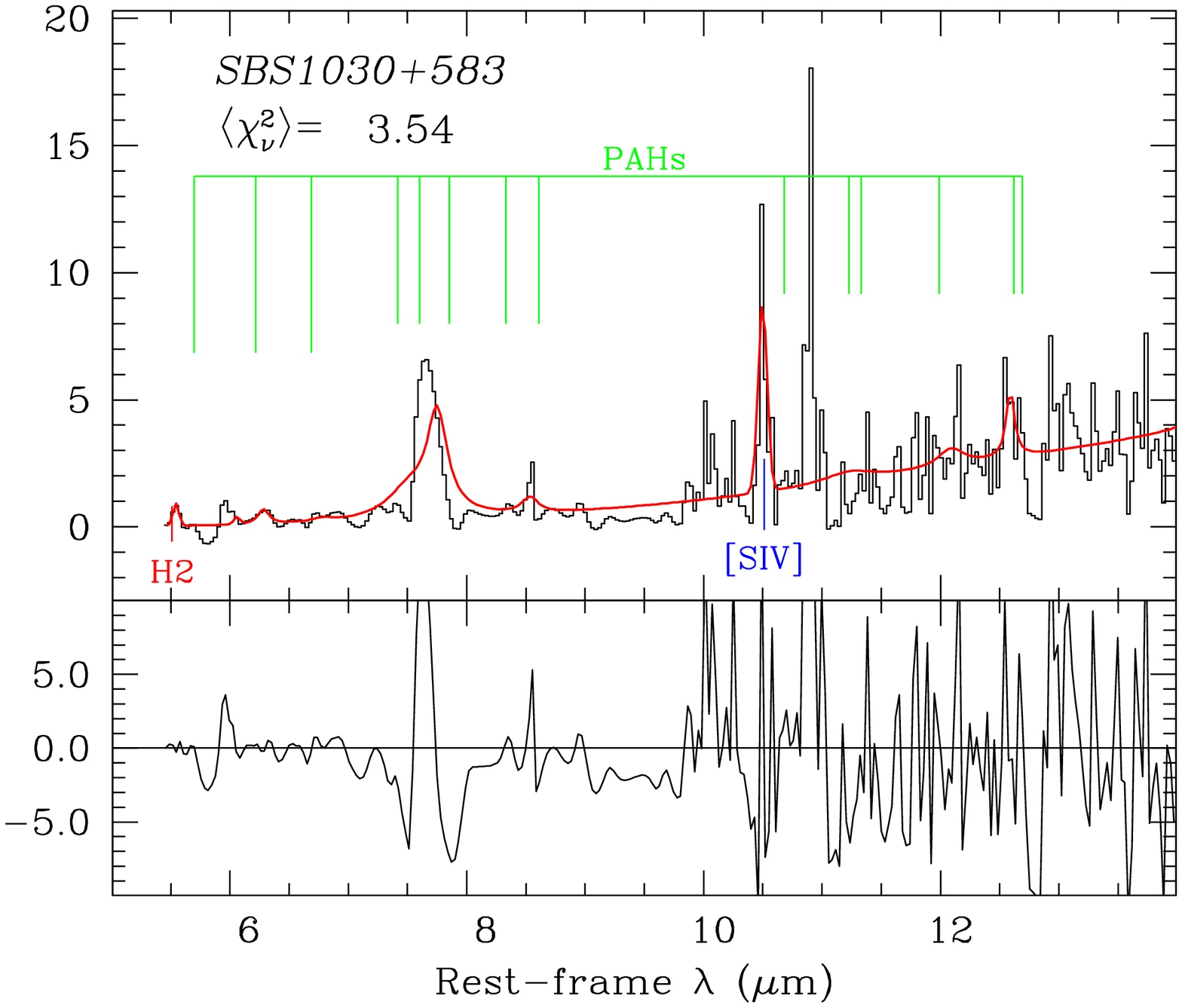}  
}} 
\caption{\ continued.  }
\end{figure}

\clearpage

\setcounter{figure}{0}
\begin{figure}
\centerline{
\hbox{ 
\includegraphics[angle=0,width=0.33\linewidth,bb=40 251 587 713]{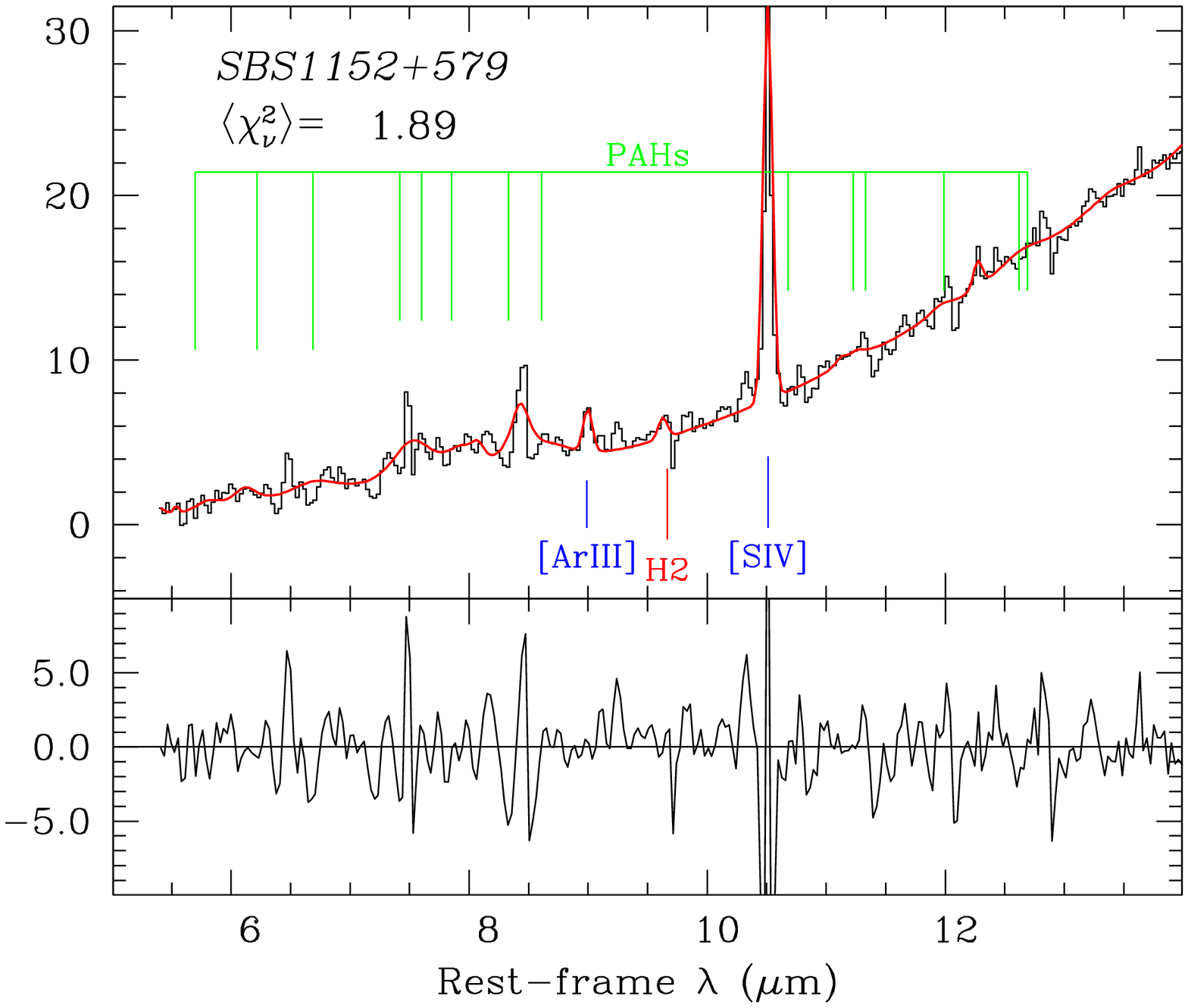} 
\hspace{-0.1cm}
\includegraphics[angle=0,width=0.33\linewidth,bb=40 251 587 713]{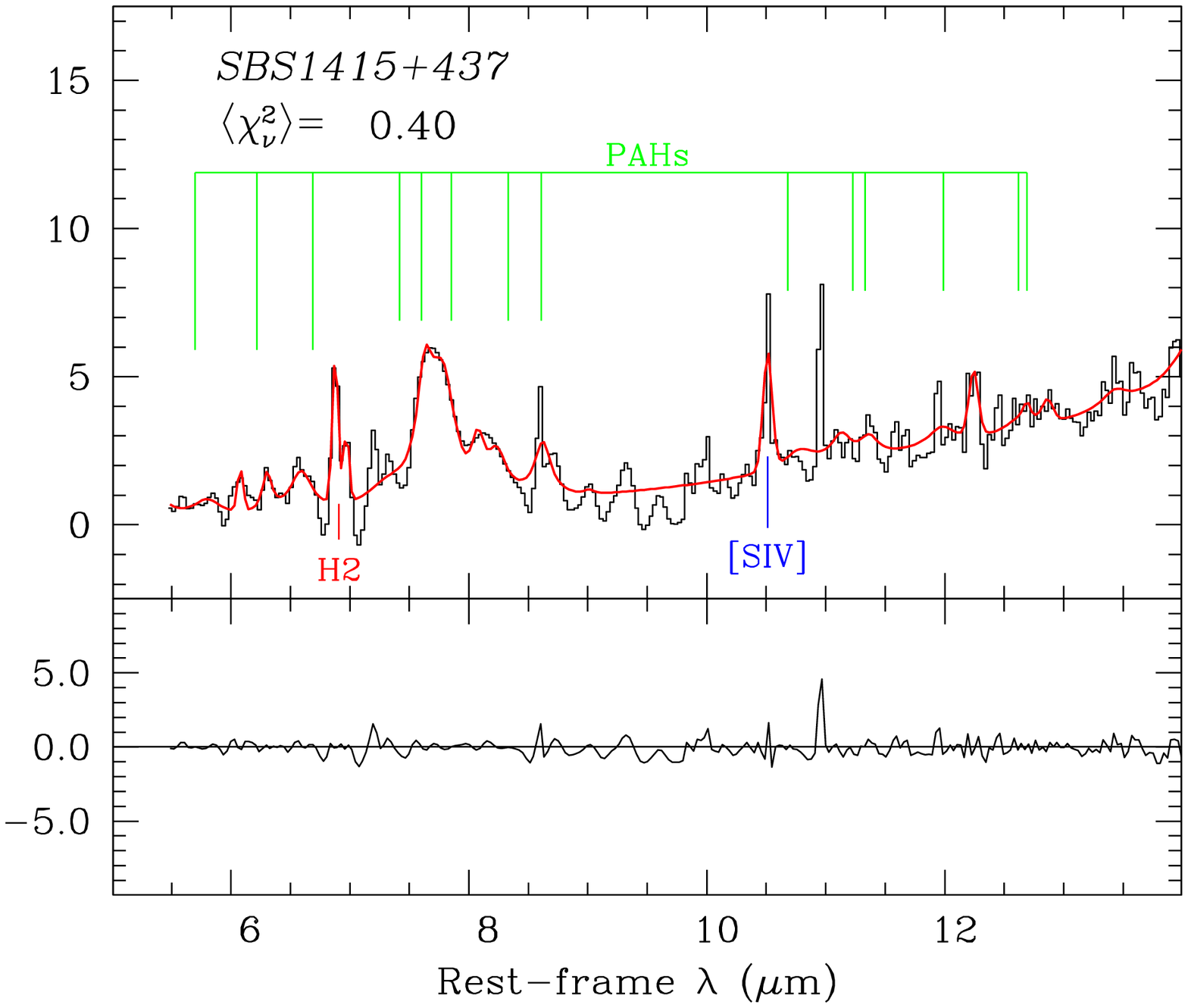} 
\hspace{-0.1cm}
\includegraphics[angle=0,width=0.33\linewidth,bb=40 251 587 713]{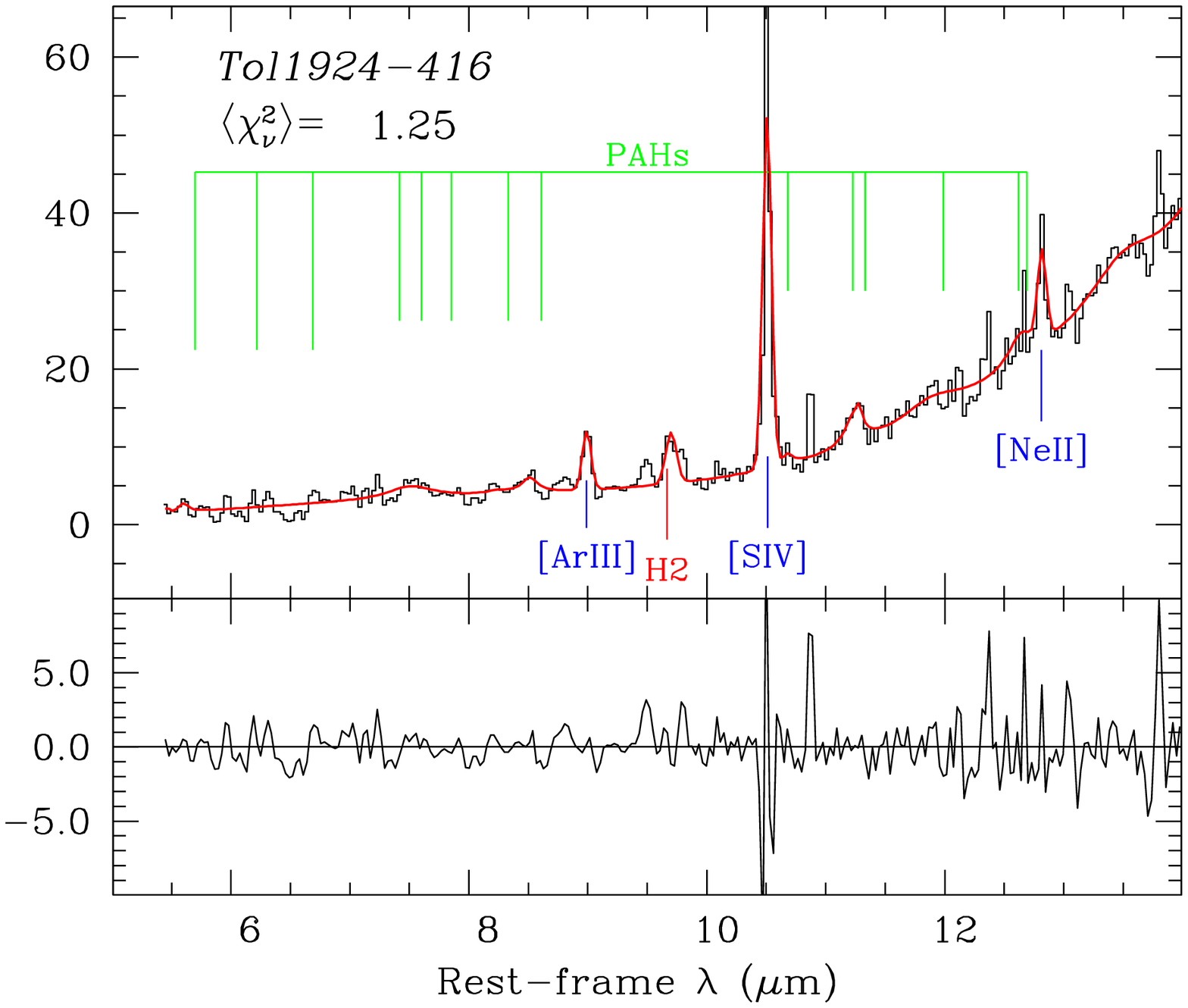}  
}} 
\vspace{0.3cm}
\centerline{
\hbox{ 
\includegraphics[angle=0,width=0.33\linewidth,bb=40 251 587 713]{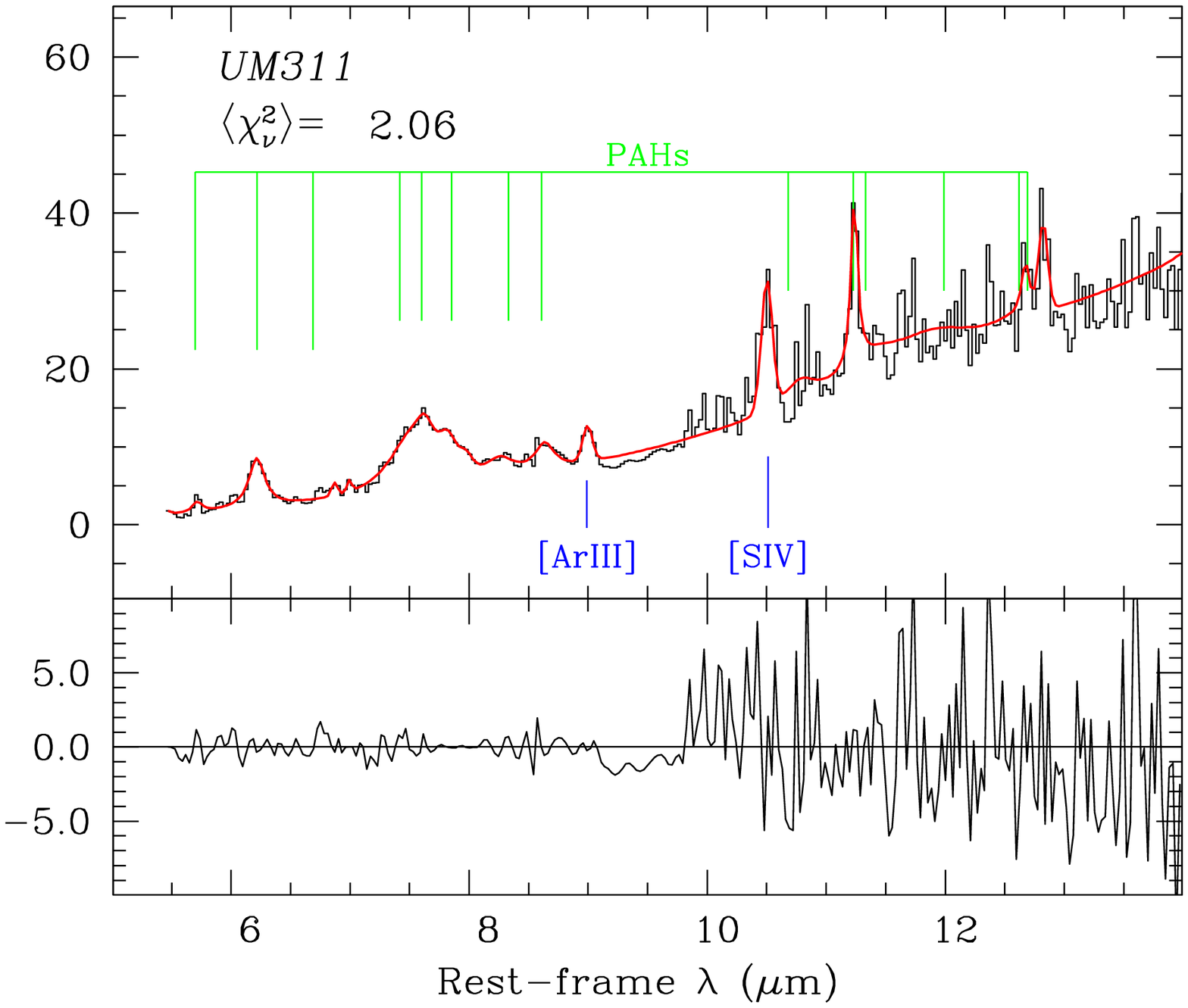}
}} 
\caption{\ continued.  }
\end{figure}


\begin{figure}
\centerline{
\hbox{ 
\includegraphics[angle=0,width=0.33\linewidth,bb=17 162 590 570]{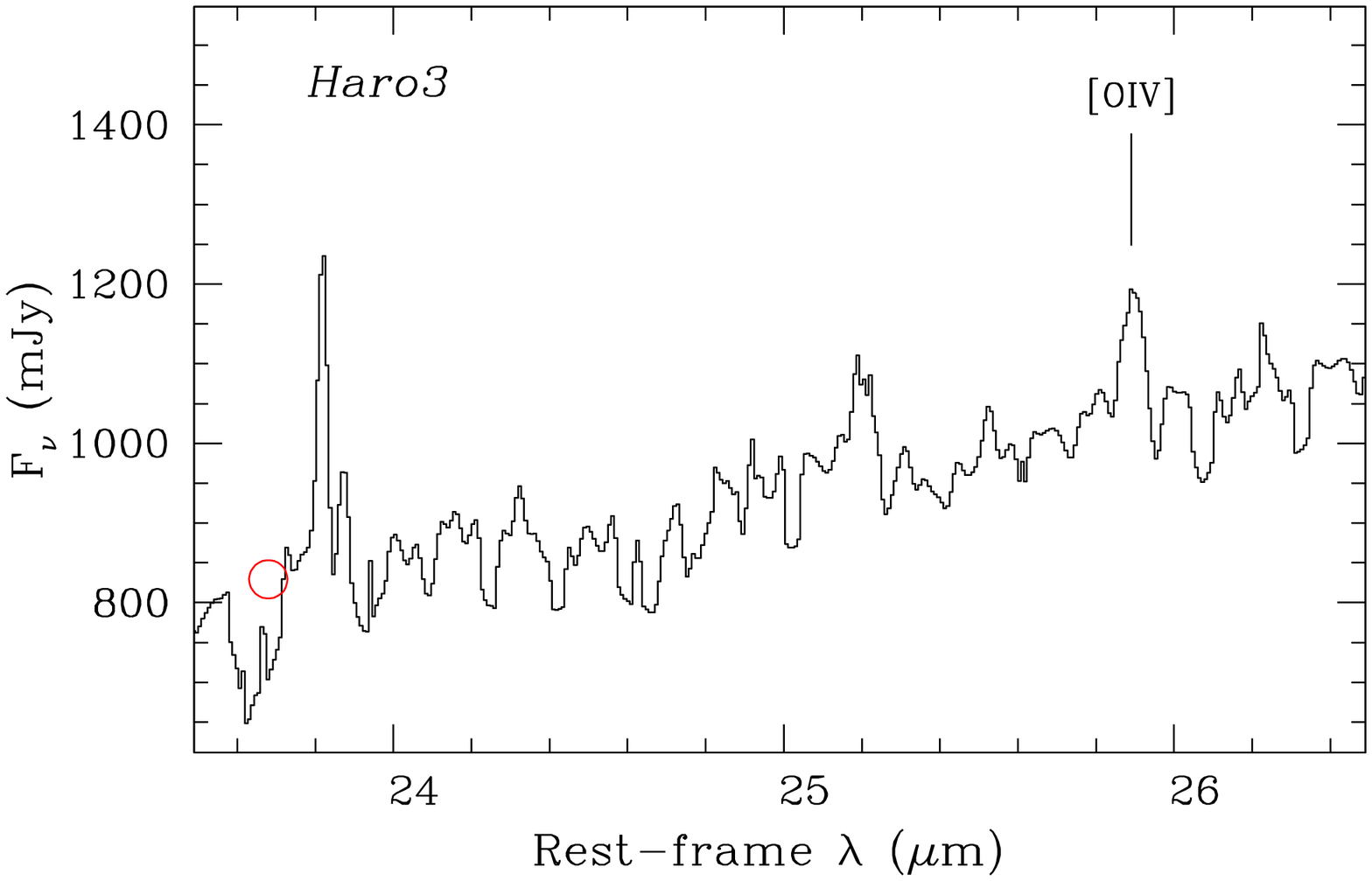} 
\includegraphics[angle=0,width=0.33\linewidth,bb=17 162 590 570]{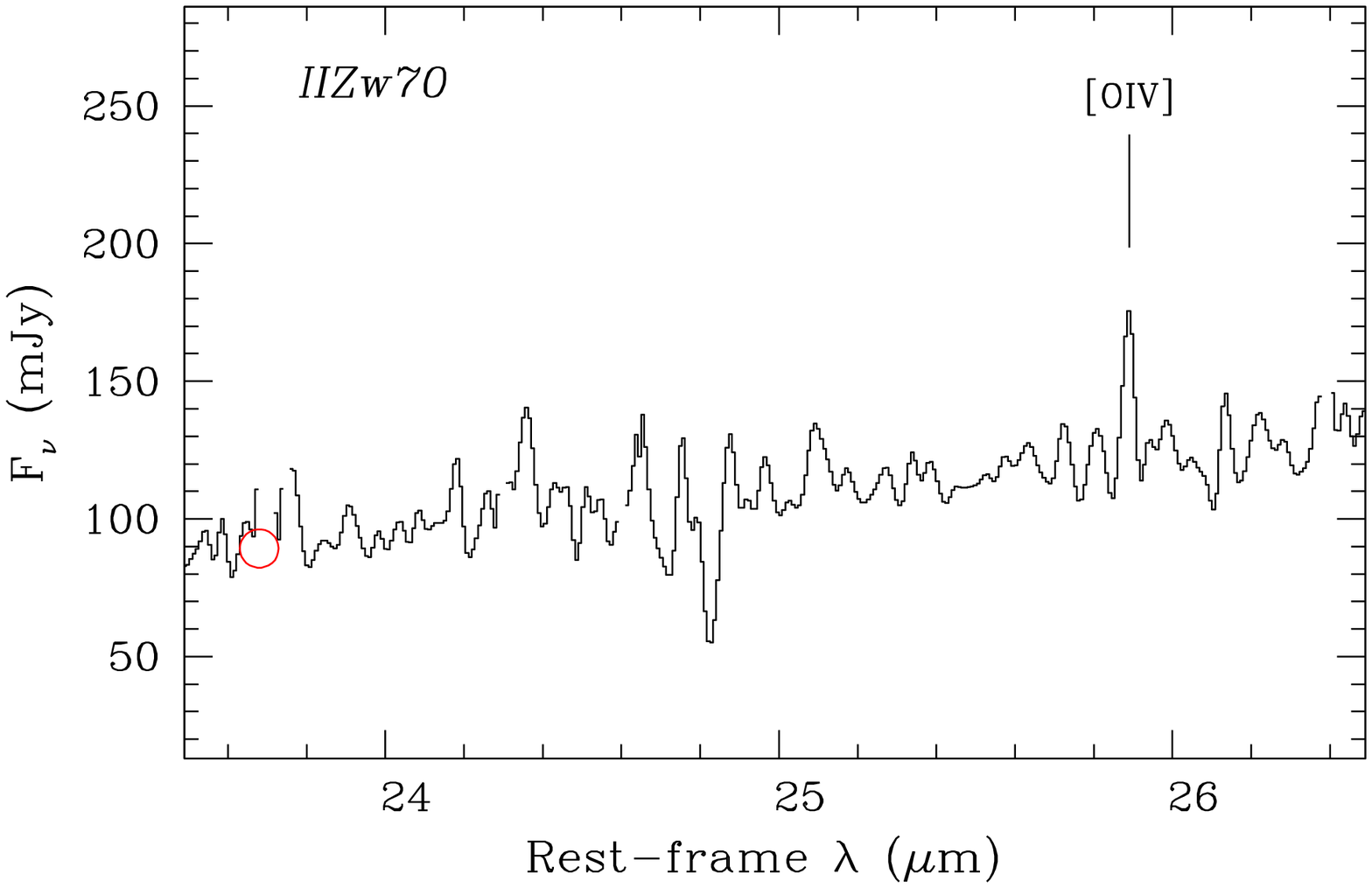} 
\includegraphics[angle=0,width=0.33\linewidth,bb=17 162 590 570]{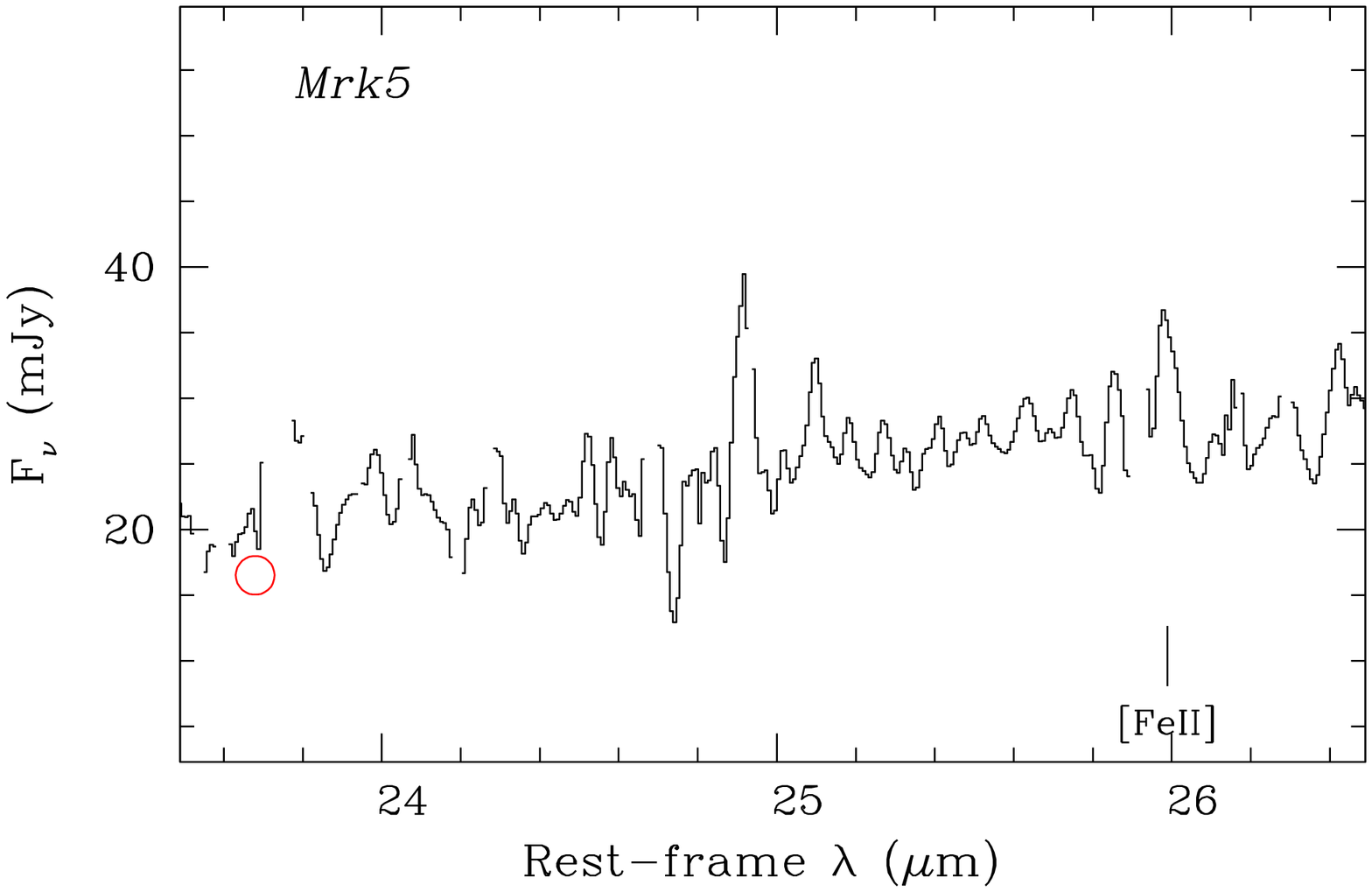}  
}}
\centerline{
\hbox{ 
\includegraphics[angle=0,width=0.33\linewidth,bb=17 162 590 570]{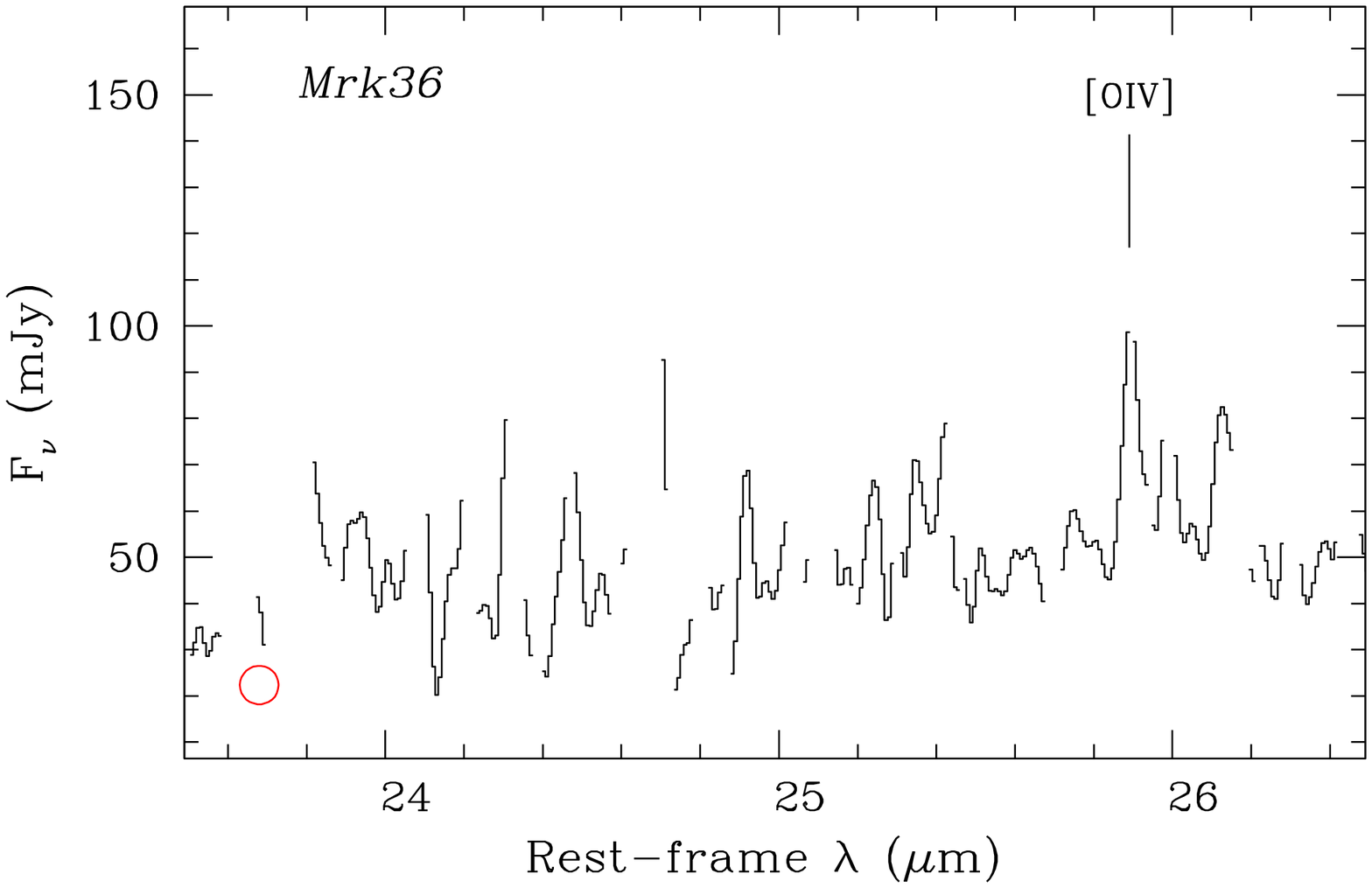} 
\includegraphics[angle=0,width=0.33\linewidth,bb=17 162 590 570]{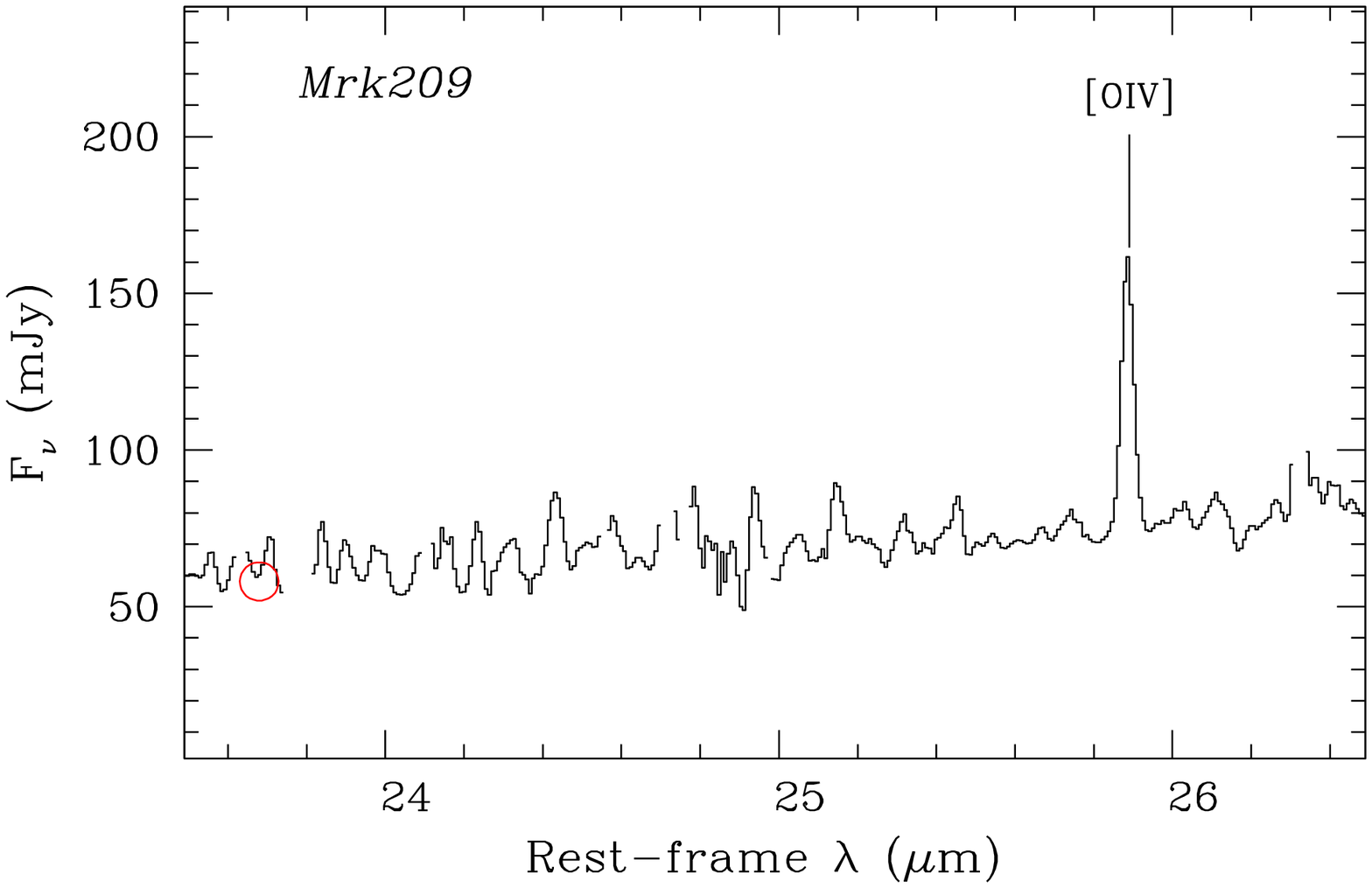} 
\includegraphics[angle=0,width=0.33\linewidth,bb=17 162 590 570]{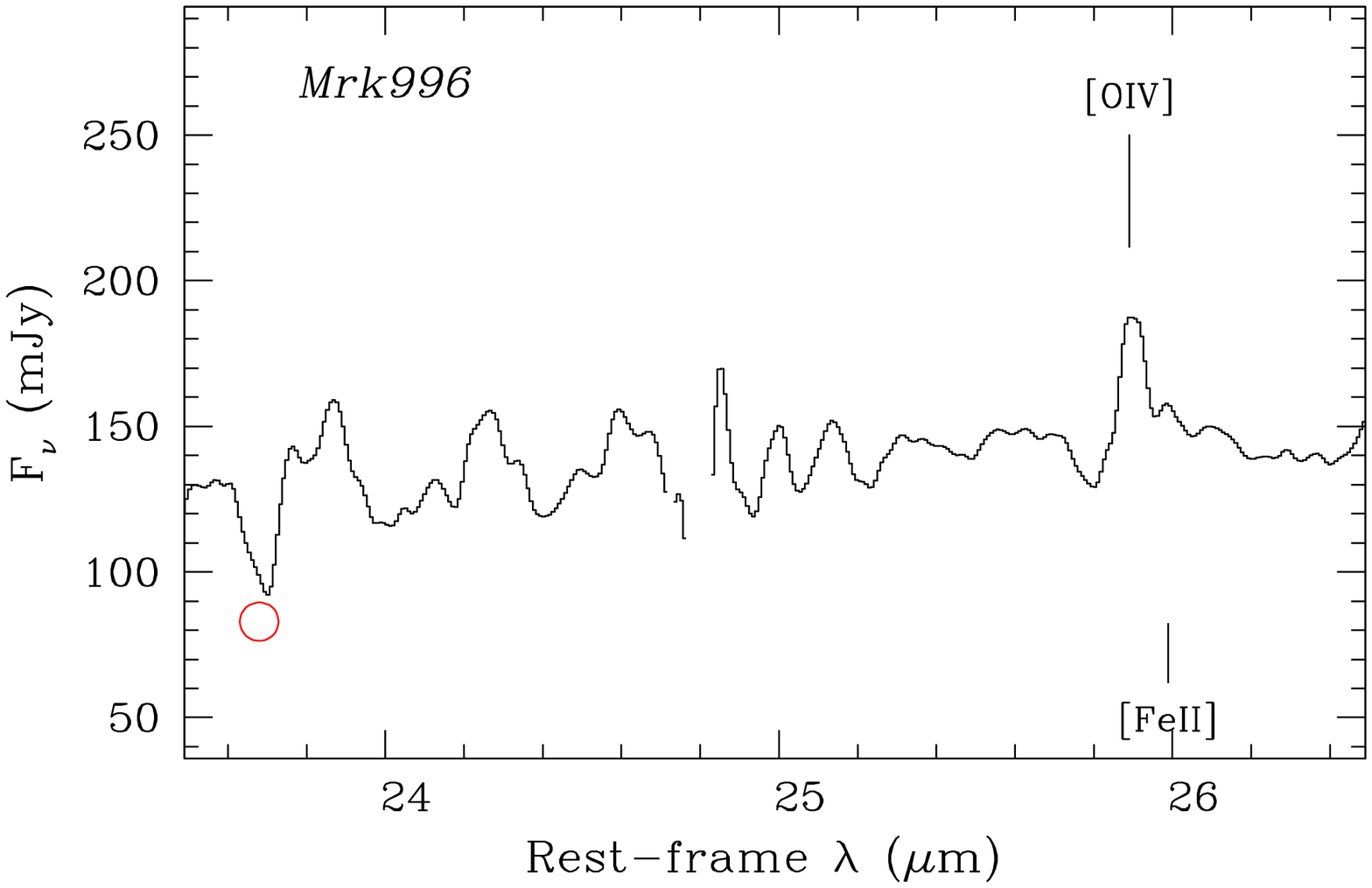} 
}}
\centerline{
\hbox{ 
\includegraphics[angle=0,width=0.33\linewidth,bb=17 162 590 570]{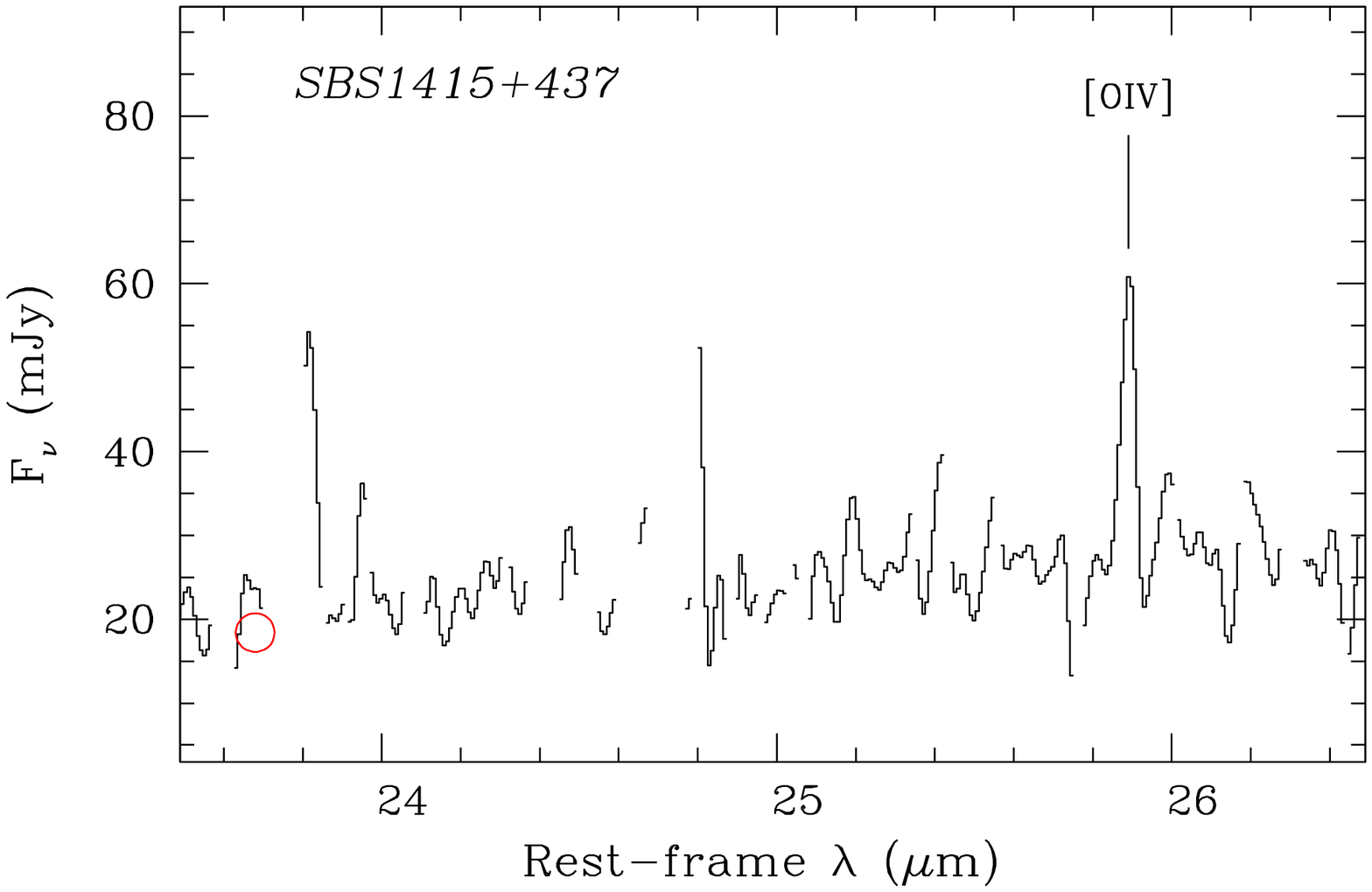} 
\includegraphics[angle=0,width=0.33\linewidth,bb=17 162 590 570]{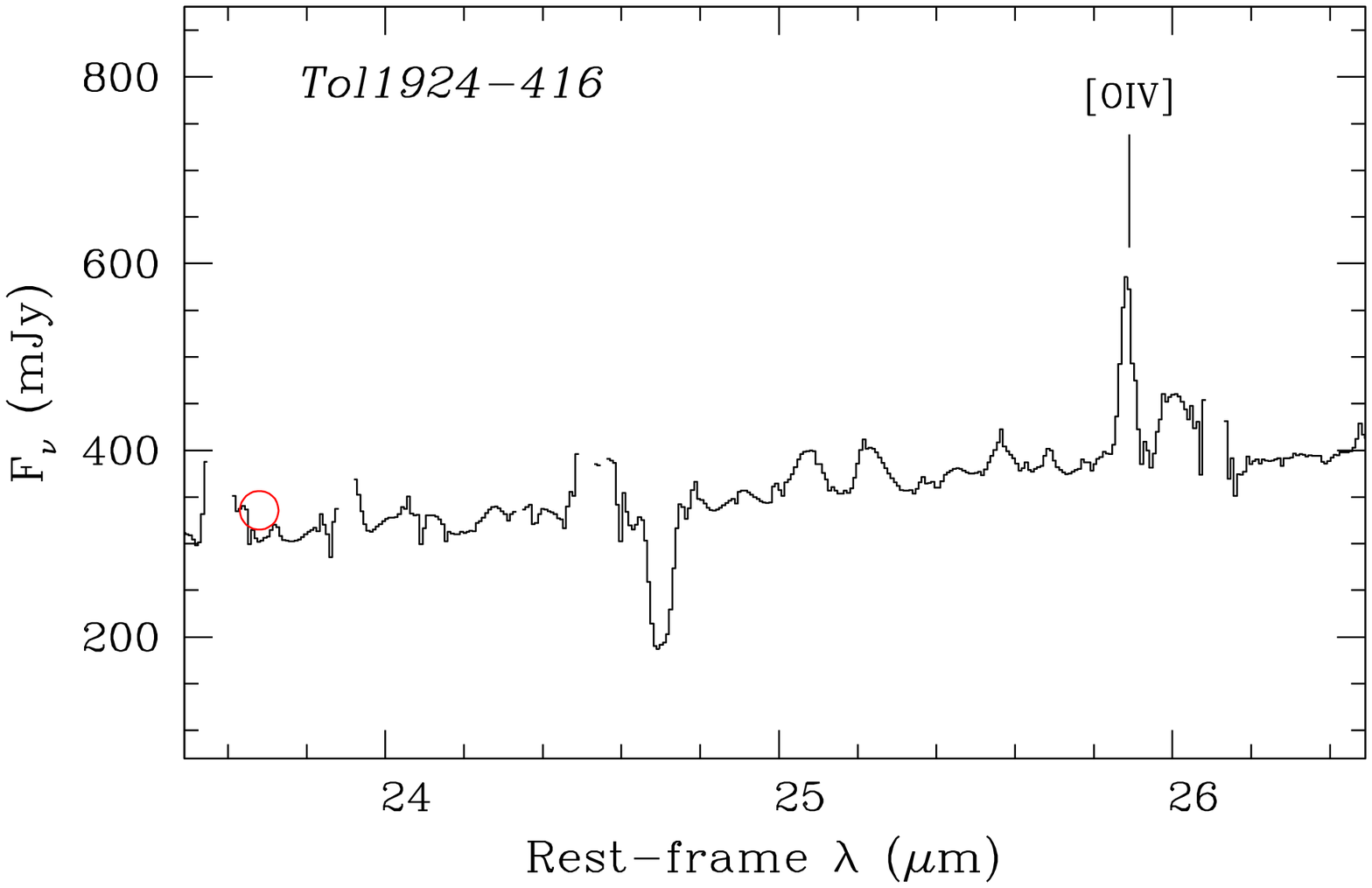}  
}}
\caption{IRS LH spectra of the 26\micron\ region around \oiv\ and \feii\
for those objects with either \oiv\ or \feii\ detections.
As in Fig. \ref{fig:pahfit1}, all marked features are considered significant
(see Table \ref{tab:fsgauss}).
The MIPS24 total flux is marked with a red open circle.
Only those spectral points with S/N$>$3 are shown.
\label{fig:oiv}}
\end{figure}

\clearpage



\begin{figure}
\centerline{
\hbox{ 
\includegraphics[angle=0,width=0.33\linewidth,bb=17 162 590 570]{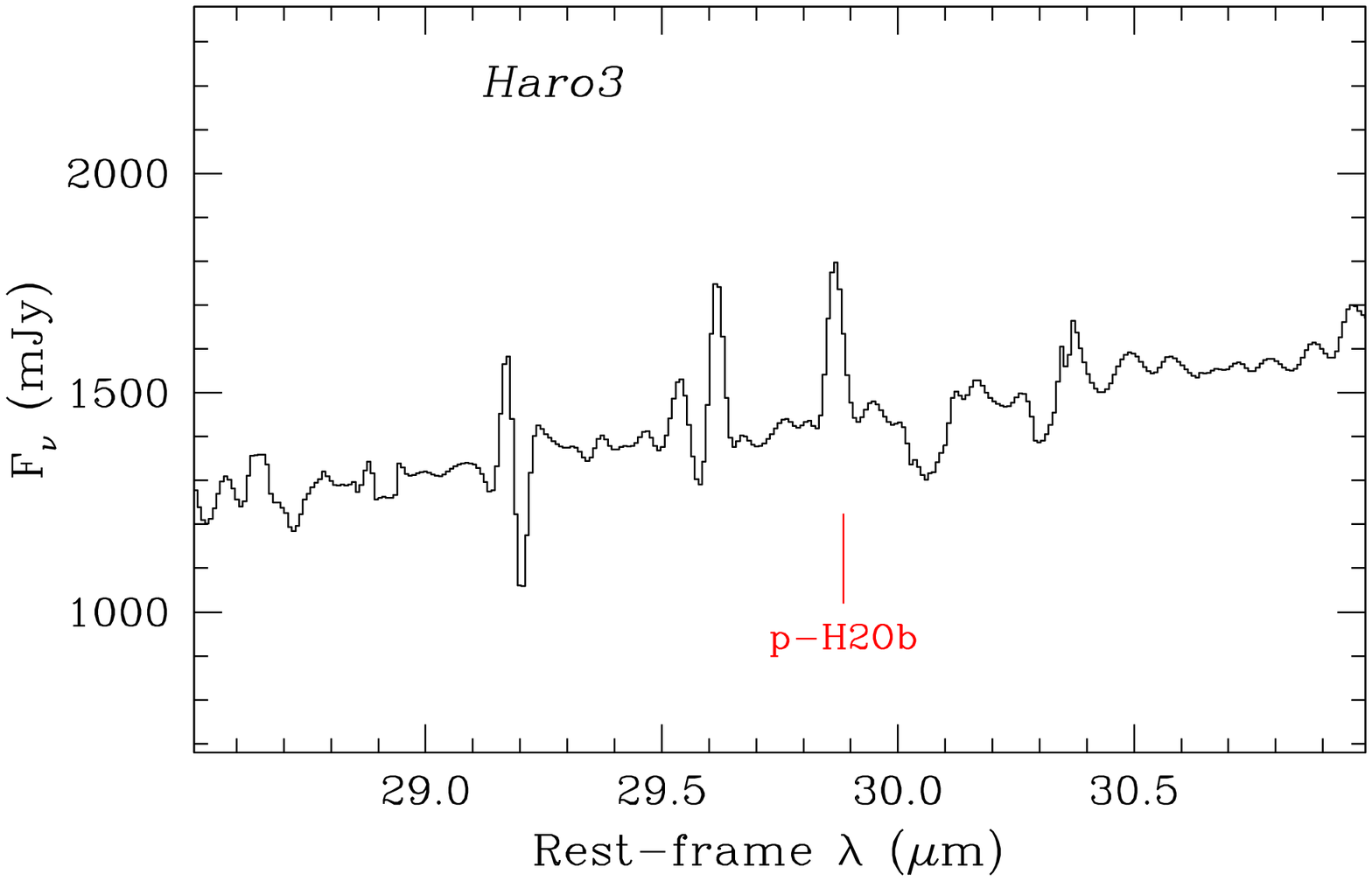} 
\includegraphics[angle=0,width=0.33\linewidth,bb=17 162 590 570]{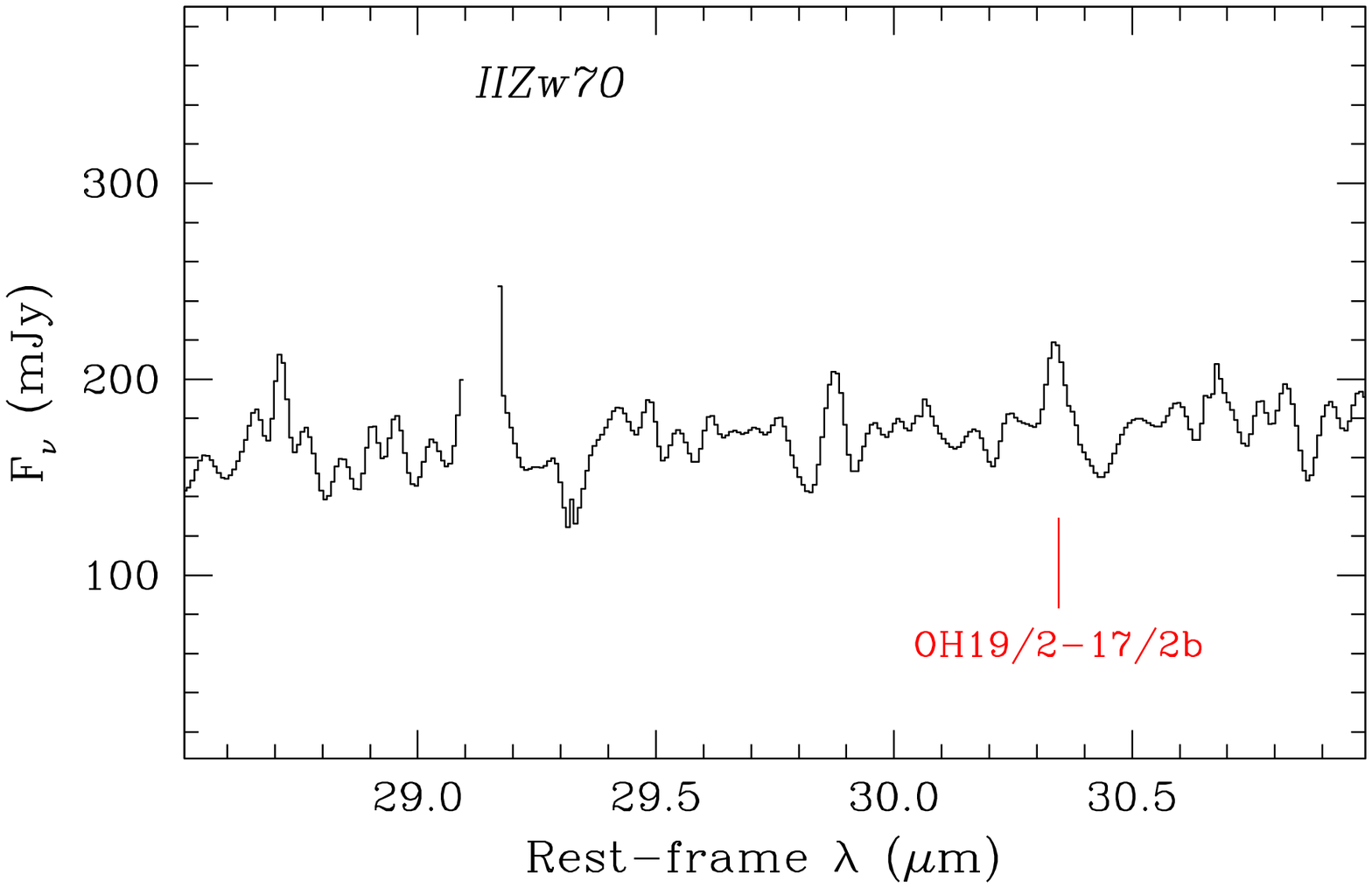}
\includegraphics[angle=0,width=0.33\linewidth,bb=17 162 590 570]{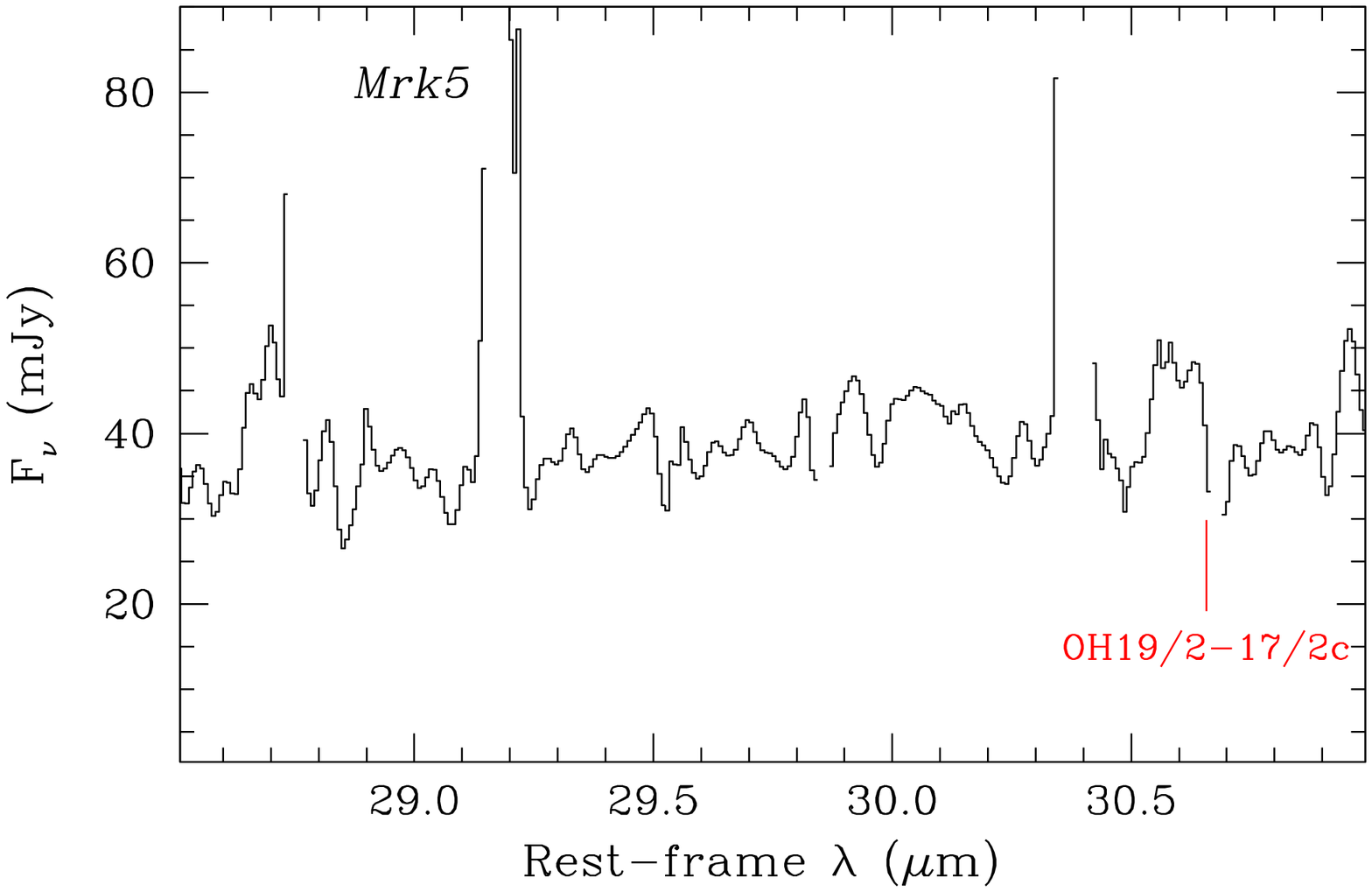}
}}
\centerline{
\includegraphics[angle=0,width=0.33\linewidth,bb=17 162 590 570]{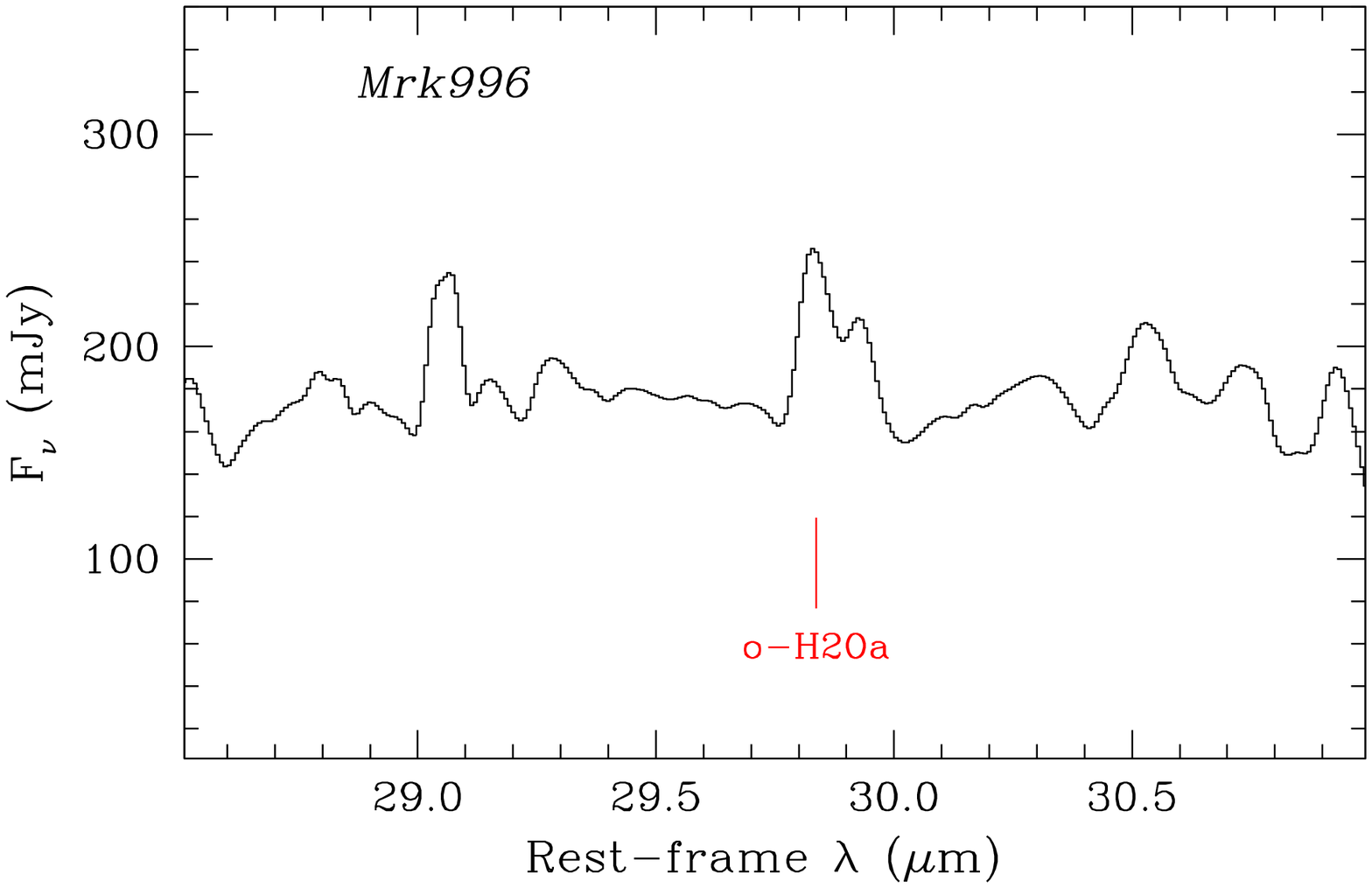}
\includegraphics[angle=0,width=0.33\linewidth,bb=17 162 590 570]{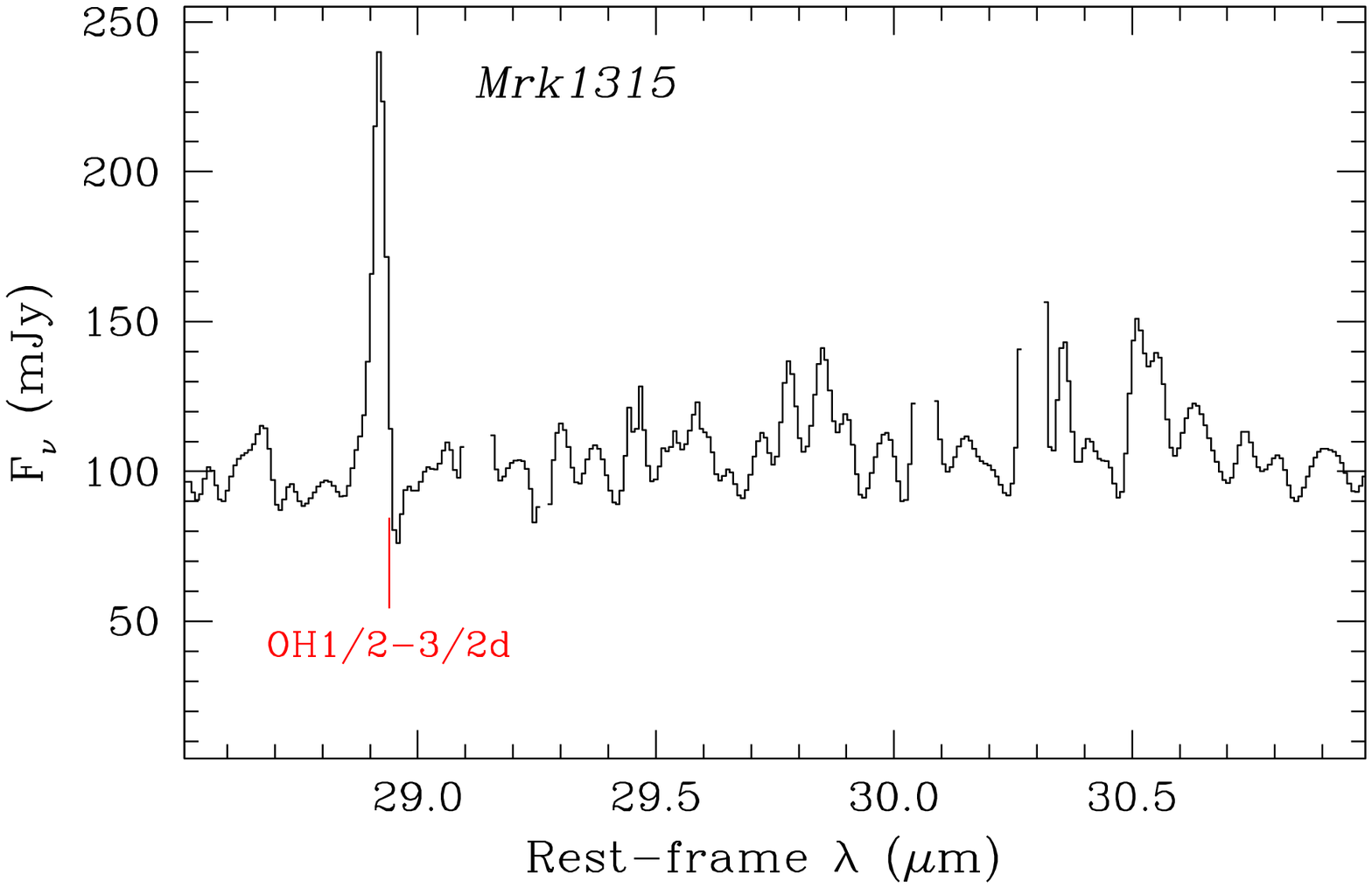}
\includegraphics[angle=0,width=0.33\linewidth,bb=17 162 590 570]{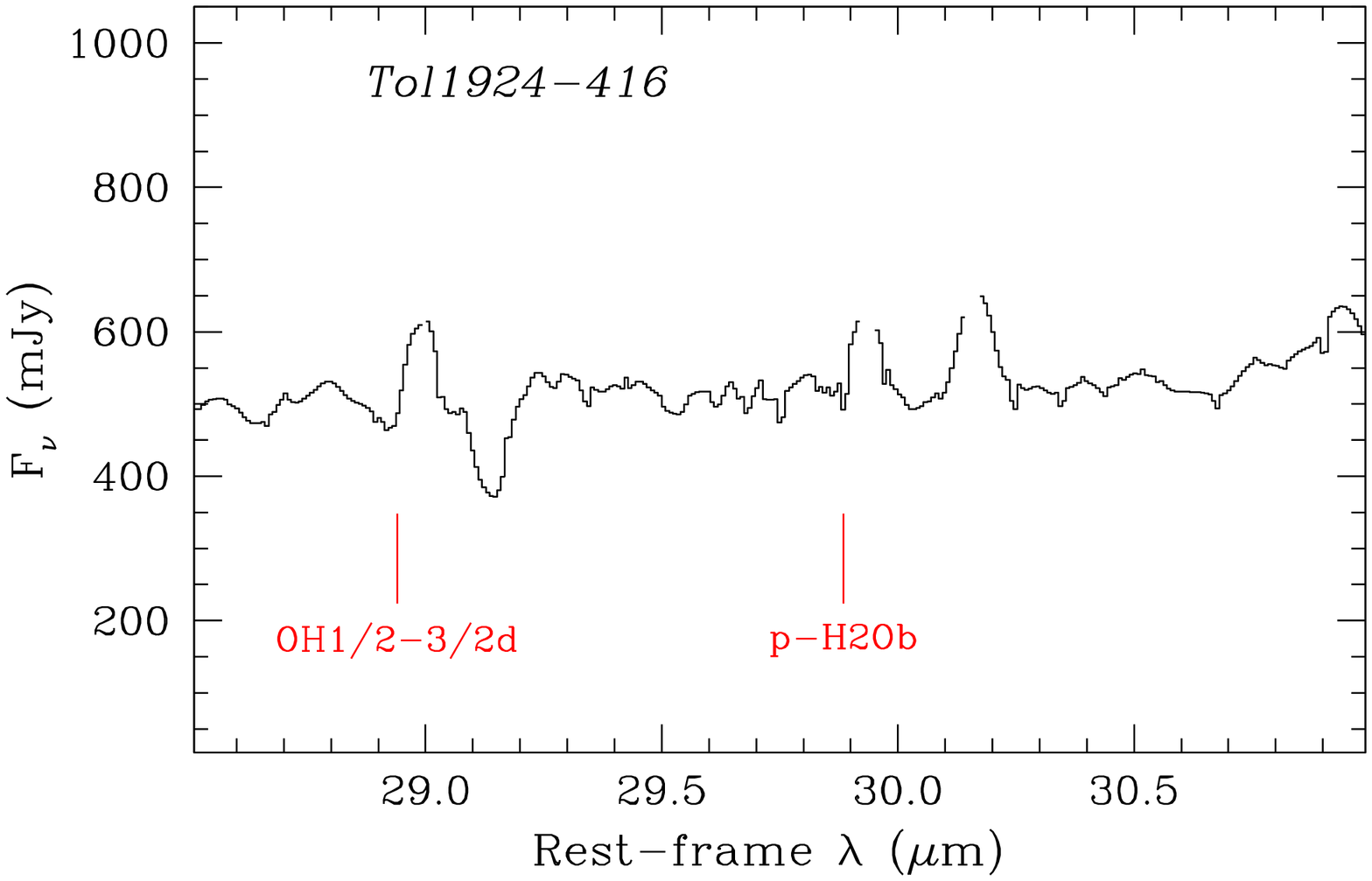}
\hbox{ 
}}
\caption{Close-up of tentative \water\ (29.8\micron)
and OH (28.9\,\micron) detections.
All marked lines are considered to be significant detections (see Table \ref{tab:molecules}).
Only those spectral points with S/N$>$3 are shown.
\label{fig:h2o_oh}}
\end{figure}

\clearpage


\begin{figure}
\bigskip
\centerline{
\includegraphics[angle=0,width=0.6\linewidth]{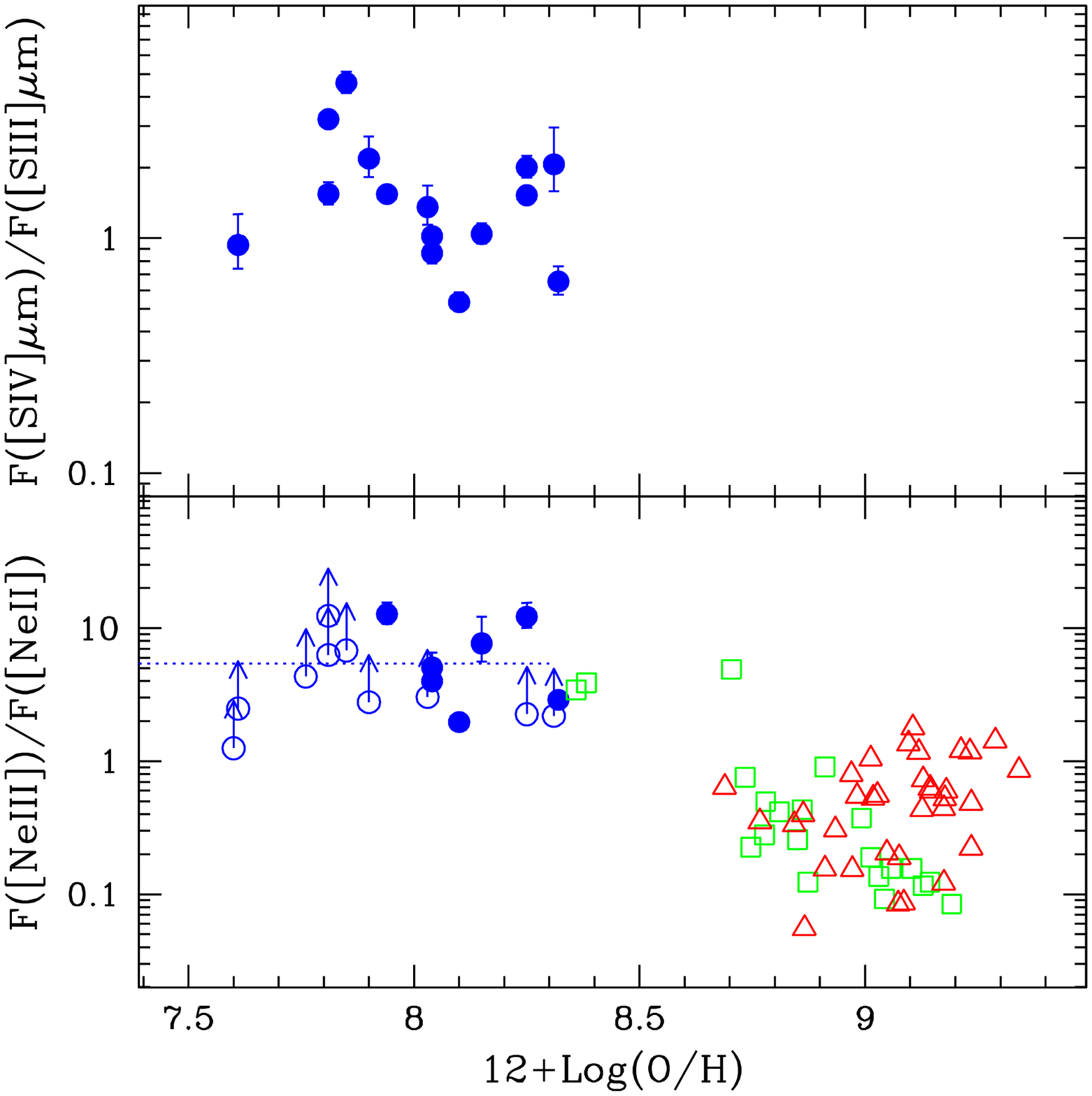} }
\caption{\siv/\siii\ (18.7\,\micron) flux ratios (top panel) and \neiii/\neii\ ratios
(bottom) plotted against the nebular oxygen abundance, \logoh.
The BCDs in our sample are shown as filled circles,
while the SINGS galaxies \citep{dale09}
as open squares (\hii\ region-nuclei) and open triangles (AGN).
The horizontal dotted line in the lower panel shows the mean
neon ratio (5.42) for \logoh$\leq$8.3 (not considering lower limits).
\label{fig:nes_oh}}
\end{figure}

\begin{figure}
\centerline{
\includegraphics[angle=0,width=0.6\linewidth]{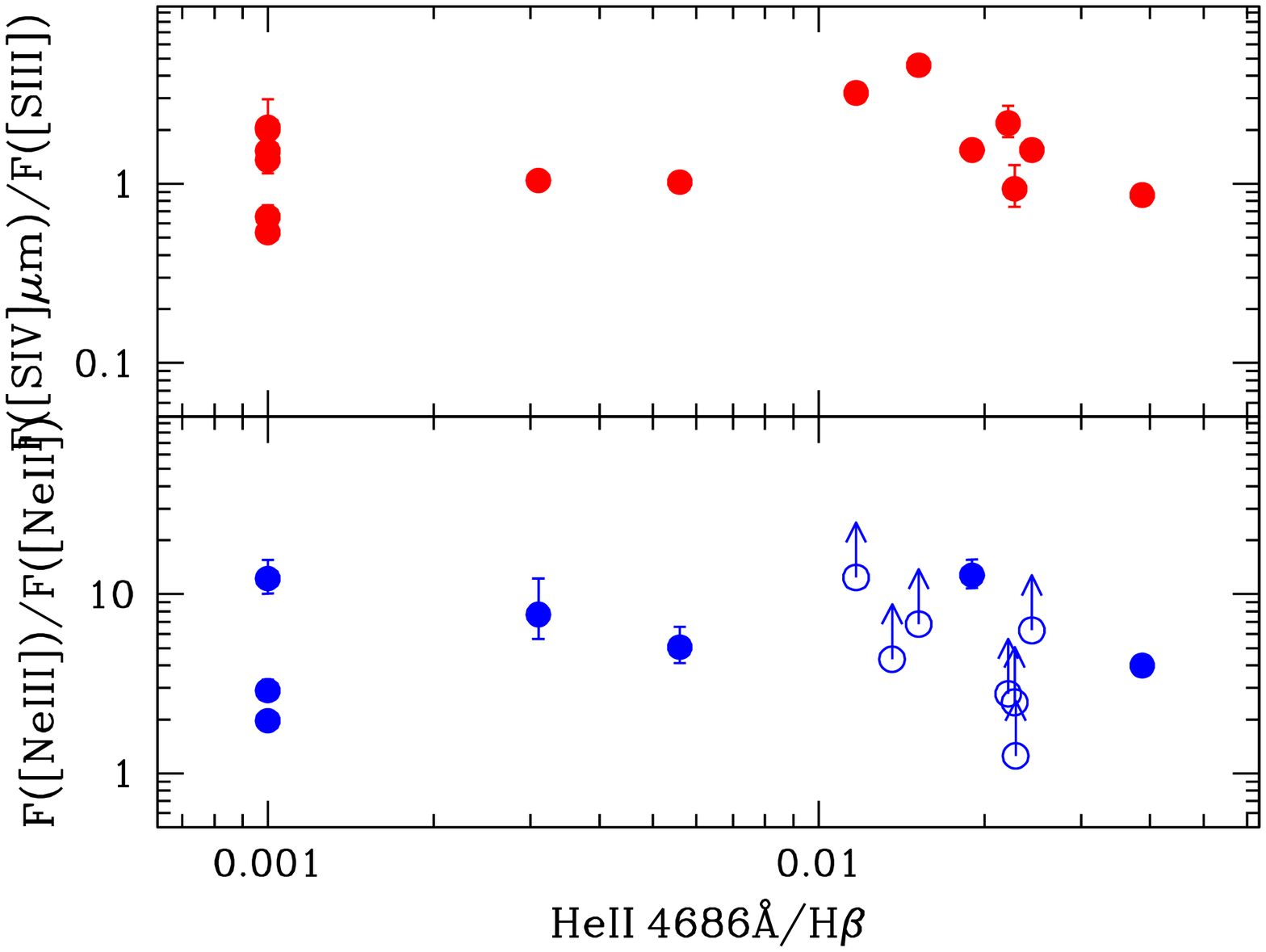} }
\caption{\siv/\siii\ (18.7\,\micron) flux ratios (top panel) and \neiii/\neii\ ratios
(bottom) plotted against the optical \heii/\hb\ ratio.
\label{fig:nes_heii}}
\end{figure}

\bigskip

\begin{figure}
\centerline{
\includegraphics[angle=0,width=0.5\linewidth,bb=19 159 588 547]{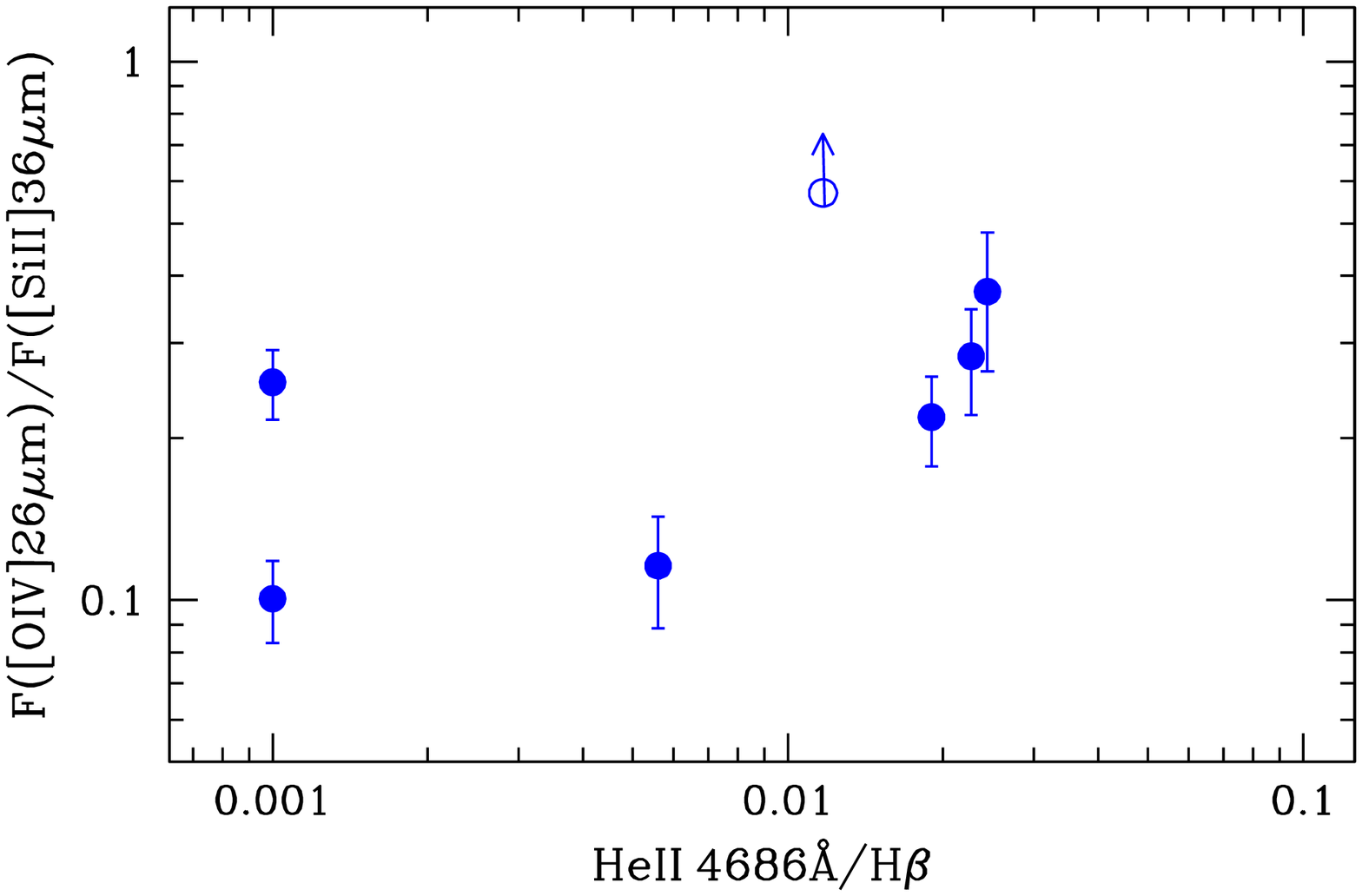} 
\hspace{-0.094\linewidth}
\includegraphics[angle=0,width=0.5\linewidth,bb=19 159 588 547]{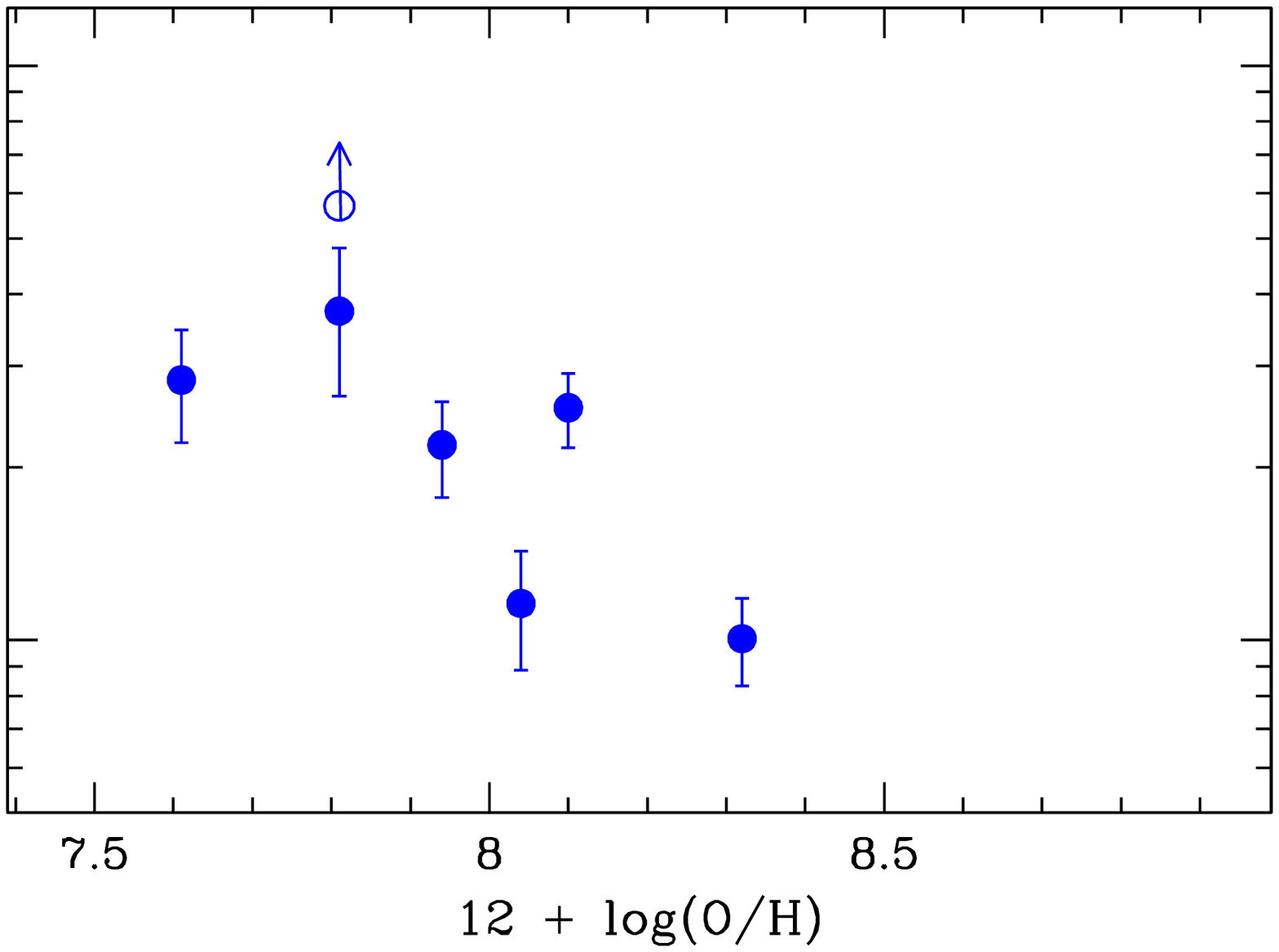} }
\caption{\oiv/\SiII\ flux ratios plotted against the 
optical \heii/\hb\ ratio (left panel) and \logoh\ (right).
The figure shows only those objects with 3$\sigma$ detections or better
in the \oiv\ line.
\label{fig:oiv_heii_oh}}
\end{figure}

\begin{figure}
\centerline{
\includegraphics[angle=0,width=0.6\linewidth]{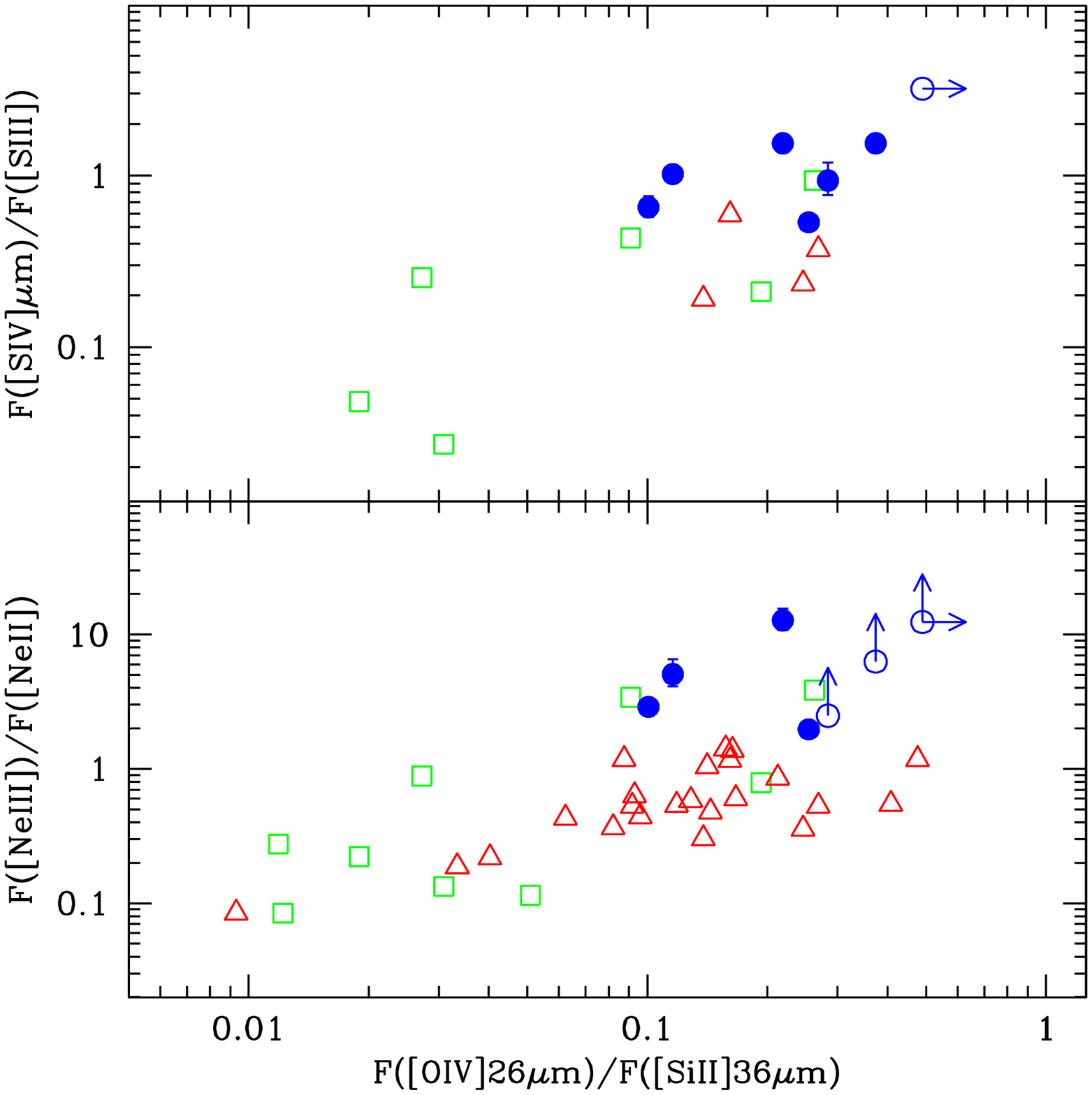} }
\caption{\neiii/\neii/ and \siv/\siii\ (18.7\,\micron) plotted against the \oiv/\SiII\ flux ratios. 
As in Fig. \ref{fig:nes_oh}, BCDs are shown as filled circles,
and the SINGS galaxies \citep{dale09}
as open squares (\hii\ region-nuclei) and open triangles (AGN).
The figure shows only those objects with 3$\sigma$ detections or better
in the \oiv\ line.
\label{fig:nes_oiv}}
\end{figure}

\clearpage

\begin{figure}
\centerline{
\includegraphics[angle=0,width=0.6\linewidth]{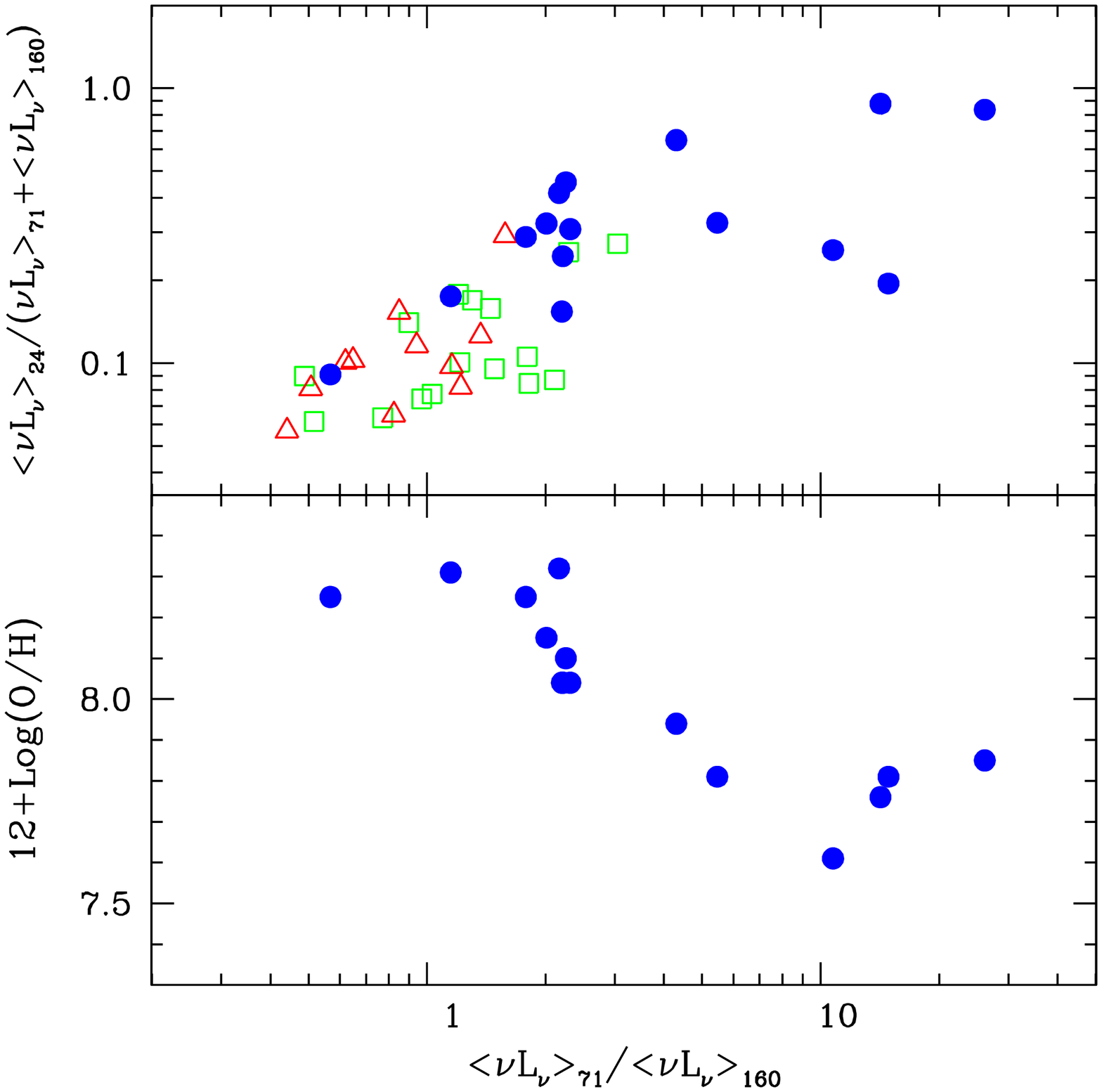} }
\caption{\nfn $_{24}$/(\nfn $_{71}+$ \nfn $_{160}$ (upper panel) 
and \logoh\ (lower) plotted against the MIPS ratio \nfn $_{71}$/\nfn $_{160}$.
As in Fig. \ref{fig:nes_oh}, BCDs are shown as filled circles,
and the SINGS galaxies \citep{dale09}
as open squares (\hii\ region-nuclei) and open triangles (AGN).
Both panels show correlations significant within the BCD sample
at $\simgt$99\% confidence level, as described in the text.
\label{fig:p24oh_r71}}
\end{figure}

\begin{figure}
\centerline{
\includegraphics[angle=0,width=0.6\linewidth]{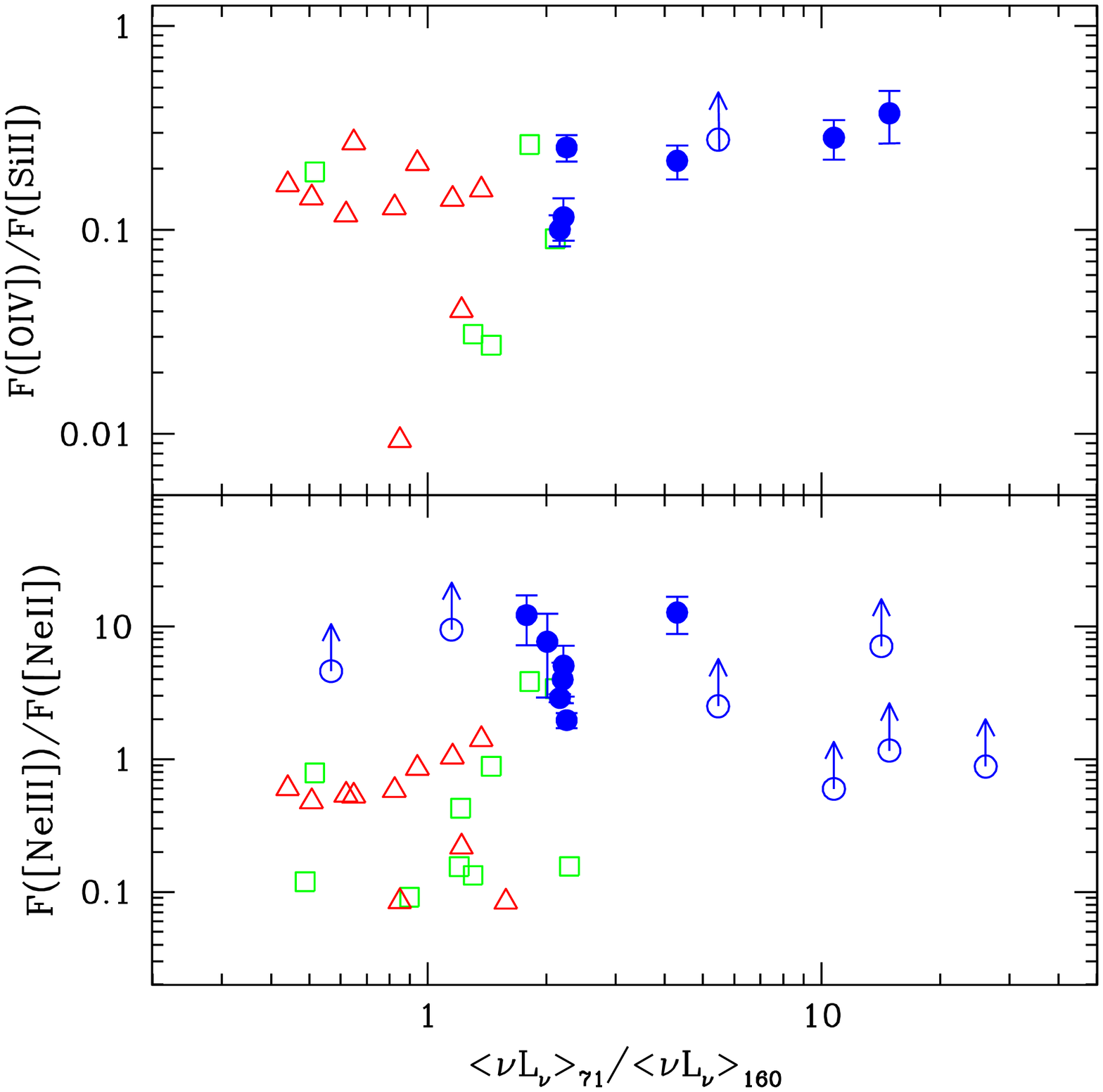} }
\caption{\neiii/\neii/ (lower panel) and \oiv/\SiII\  (upper) plotted against 
MIPS ratio \nfn $_{71}$/\nfn $_{160}$.
As in Fig. \ref{fig:nes_oh}, BCDs are shown as filled circles,
and the SINGS galaxies \citep{dale09}
as open squares (\hii\ region-nuclei) and open triangles (AGN).
In the upper panel,
the $3\sigma$ \SiII\ upper limit (with \oiv\ detection) is shown
as an open circle. 
\label{fig:neoiv_r71}}
\end{figure}

\begin{figure}
\centerline{
\includegraphics[angle=0,width=0.6\linewidth]{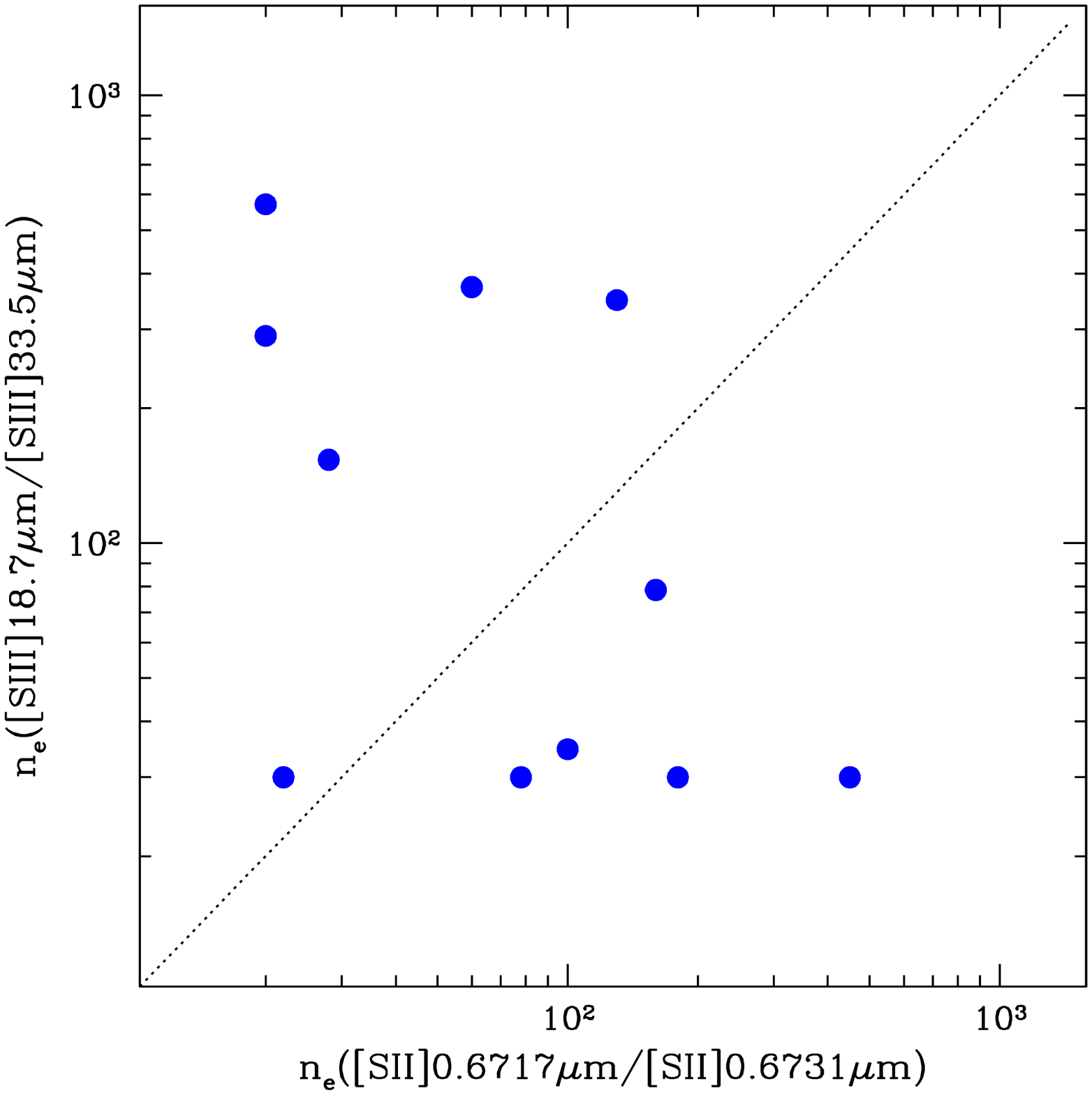} }
\caption{Electron densities \Ne\ derived from
IR \siii\ line ratios vs. those inferred from optical \siii\ line
ratios.
The dotted line denotes equality.
\label{fig:densities}}
\end{figure}

\clearpage

\begin{figure}
\centerline{
\includegraphics[angle=0,width=0.6\linewidth,bb=19 171 590 545]{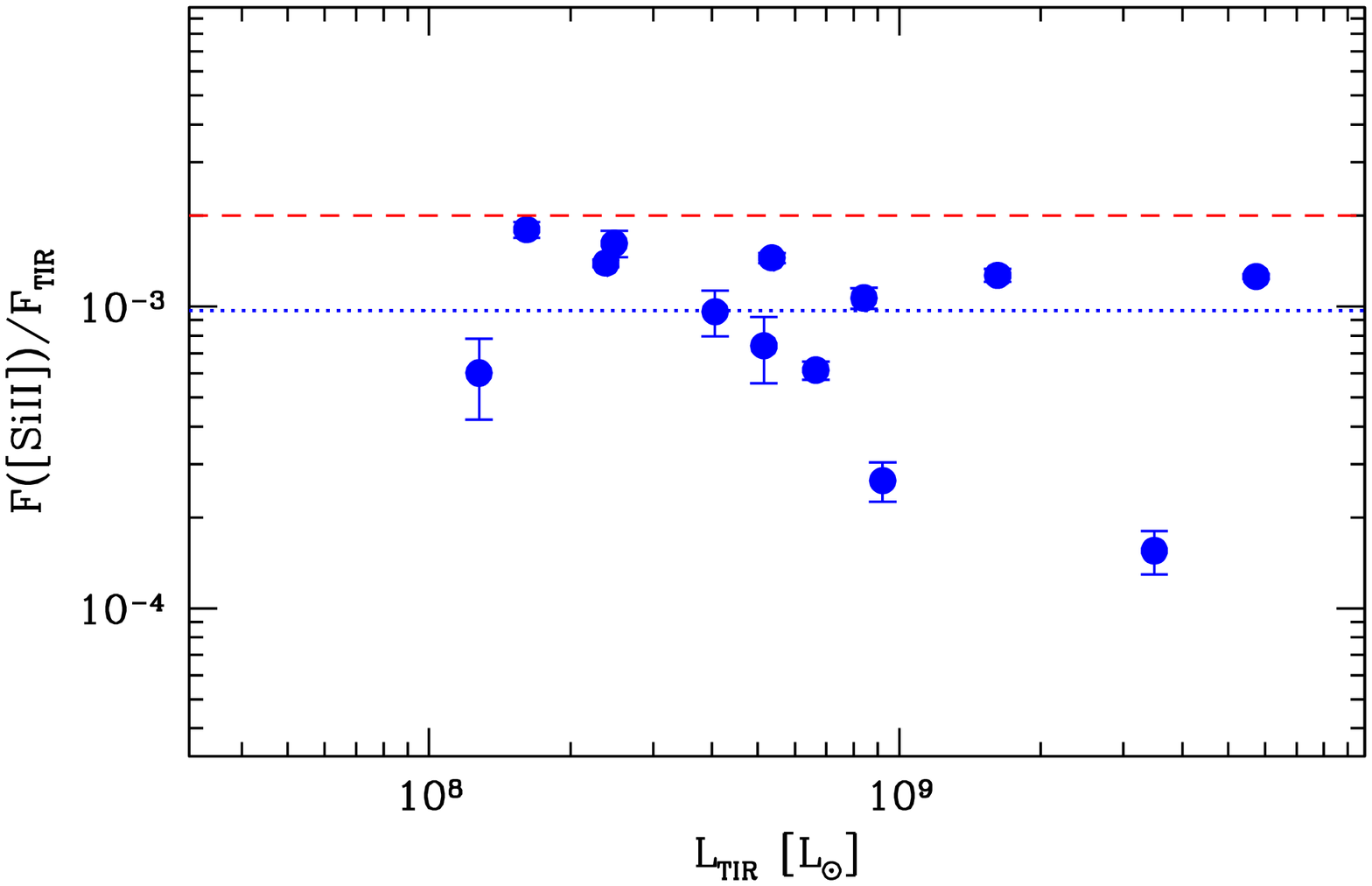} }
\caption{\SiII\ emission normalized by and plotted against TIR.
The blue dotted horizontal line indicates the BCD mean,
without any MIPS ULs and 
without UM\,311 (\ltir$\sim 3\times10^{9}$\lsun), since its TIR is overestimated
because of crowding (see text).
The red dashed line corresponds to the 
mean for the SINGS sample \citep{roussel07}.
\label{fig:tir} }
\end{figure}


\begin{figure}
\centerline{
\includegraphics[angle=0,width=0.8\linewidth,bb=18 156 591 593]{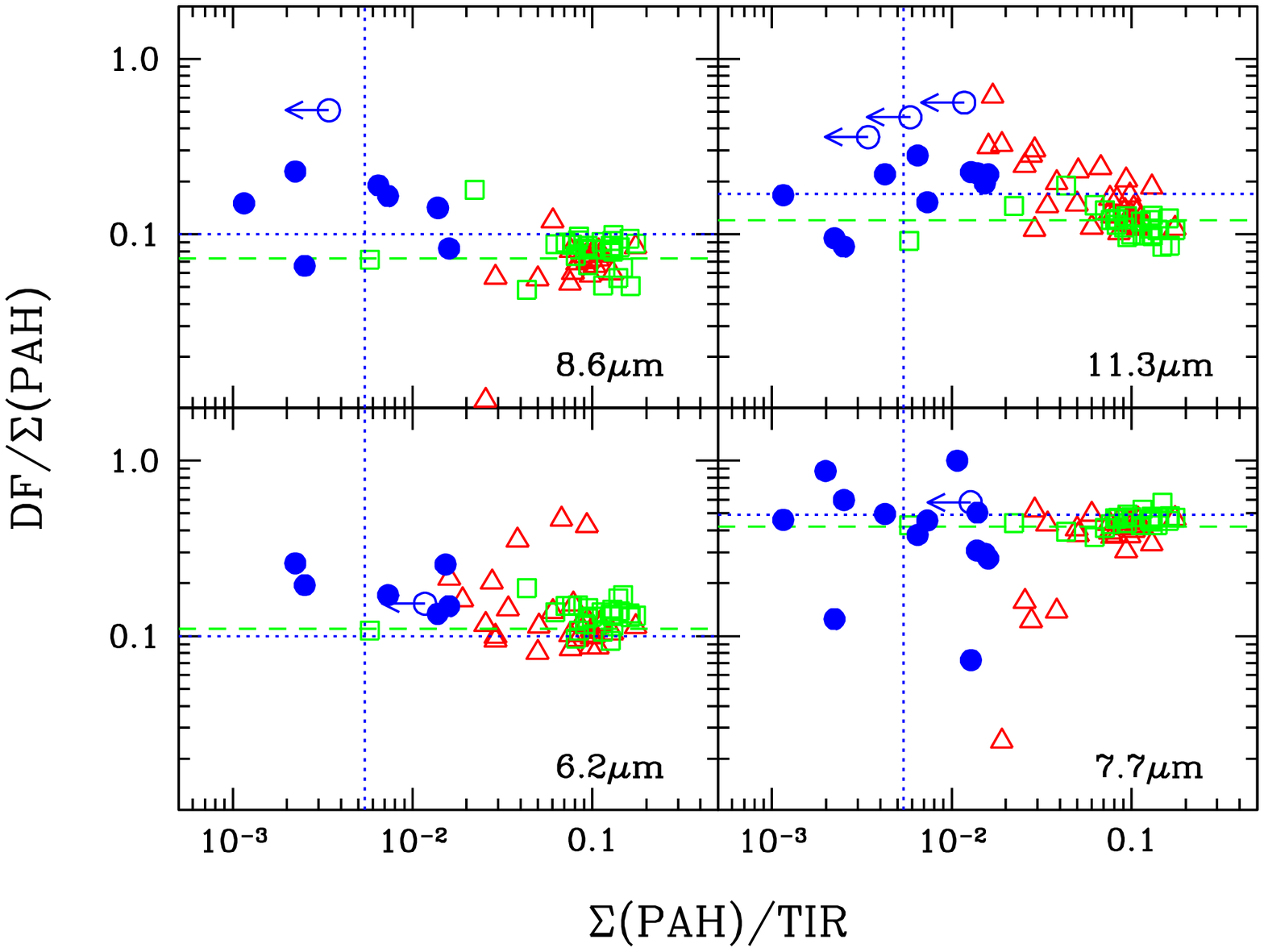} }
\caption{PAH strengths relative to total PAH luminosity vs. the total
PAH power normalized to TIR.
The four panels show the the following dust features (DF):
6.2 \micron\ (bottom left), 7.7 \micron\ (bottom right),
8.6 \micron\ (top left), 11.3 (top right) \micron\ features.
BCDs are plotted as filled (blue) circles, with TIR lower limits shown as open circles;
the TIR lower limits do not affect the vertical axis, only the placement along the horizontal one.
Open (green) squares correspond to SINGS HII nuclei, and open (red) triangles
to SINGS AGN \citep{smith07}.
The horizontal dashed lines give the SINGS sample medians \citep{smith07}, 
and the dotted ones the means for the BCD sample,
taking into account all objects with 7.7\,\micron\ detections.
The vertical dotted line corresponds to the BCD mean PAH power normalized to TIR.
\label{fig:pahfrac}}
\end{figure}

\begin{figure}
\centerline{
\includegraphics[angle=0,width=0.6\linewidth,bb=19 171 590 545]{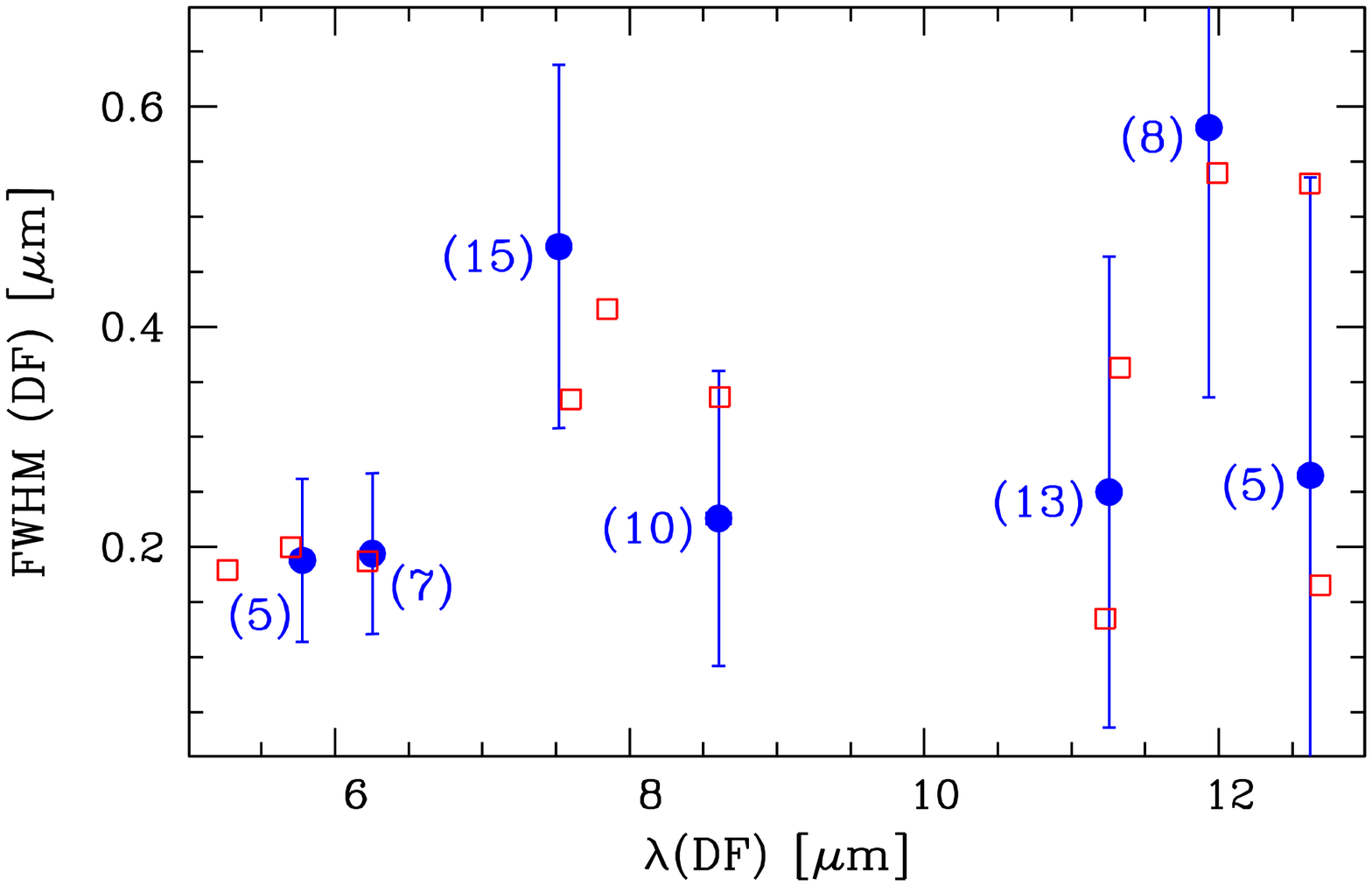} }
\caption{Median Drude profile widths (FWHM) and central wavelengths
for seven aromatic features detected ($\simgt 3\sigma$) in our sample.
The error bars are the standard deviations of the measurements,
and the numbers in parentheses are the numbers of BCDs with
significant detections.
Open squares correspond to SINGS sample means \citep{smith07}.
\label{fig:pahfwhm}}
\end{figure}

\begin{figure}
\centerline{
\includegraphics[angle=0,width=0.6\linewidth]{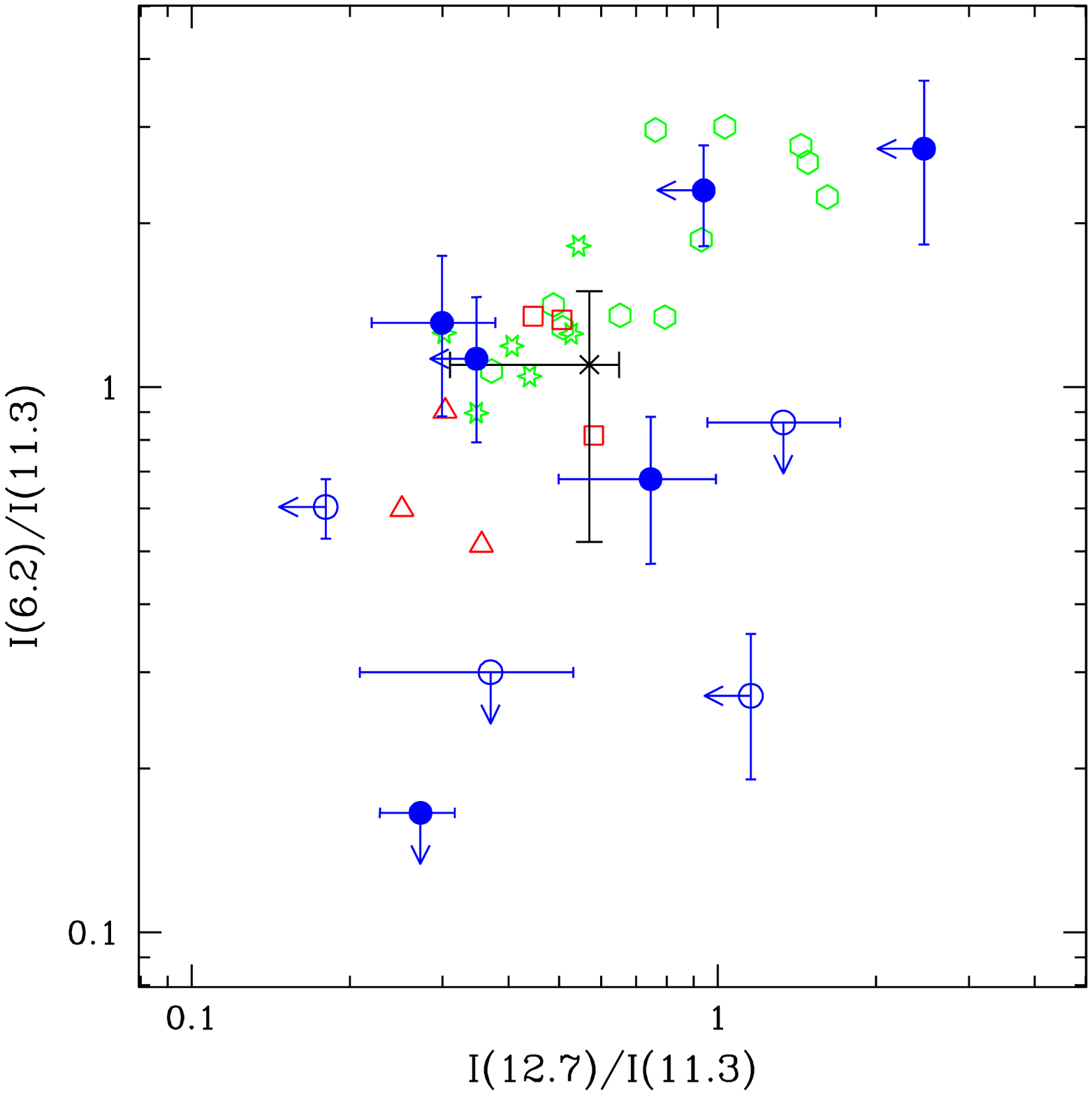} }
\caption{The integrated flux in the PAH features at 6.2\,\micron,
11.2\,\micron, and 12.7\,\micron, plotted in two sets of ratios:
6.2/11.2 vs. 12.7/11.2.
Only those BCDs with at least one ratio with 11.2\,\micron\ are shown:
filled (blue) circles correspond to those BCDs with \logoh$\geq$8.1; 
open ones to those objects with lower metallicities. 
The other data are taken from \citet{hony01} and correspond
to planetary nebulae (PNe, open red triangles),
reflection nebulae (RNe, open red squares),
intermediate-mass star-forming regions (six-sided open green stars),
and \hii\ regions (open green hexagons).
The SINGS average is shown as a (black) $\times$, with error
bars reporting standard deviations over the sample \citep{smith07}.
\label{fig:pah62127}}
\end{figure}

\begin{figure}
\centerline{
\includegraphics[angle=0,width=0.6\linewidth]{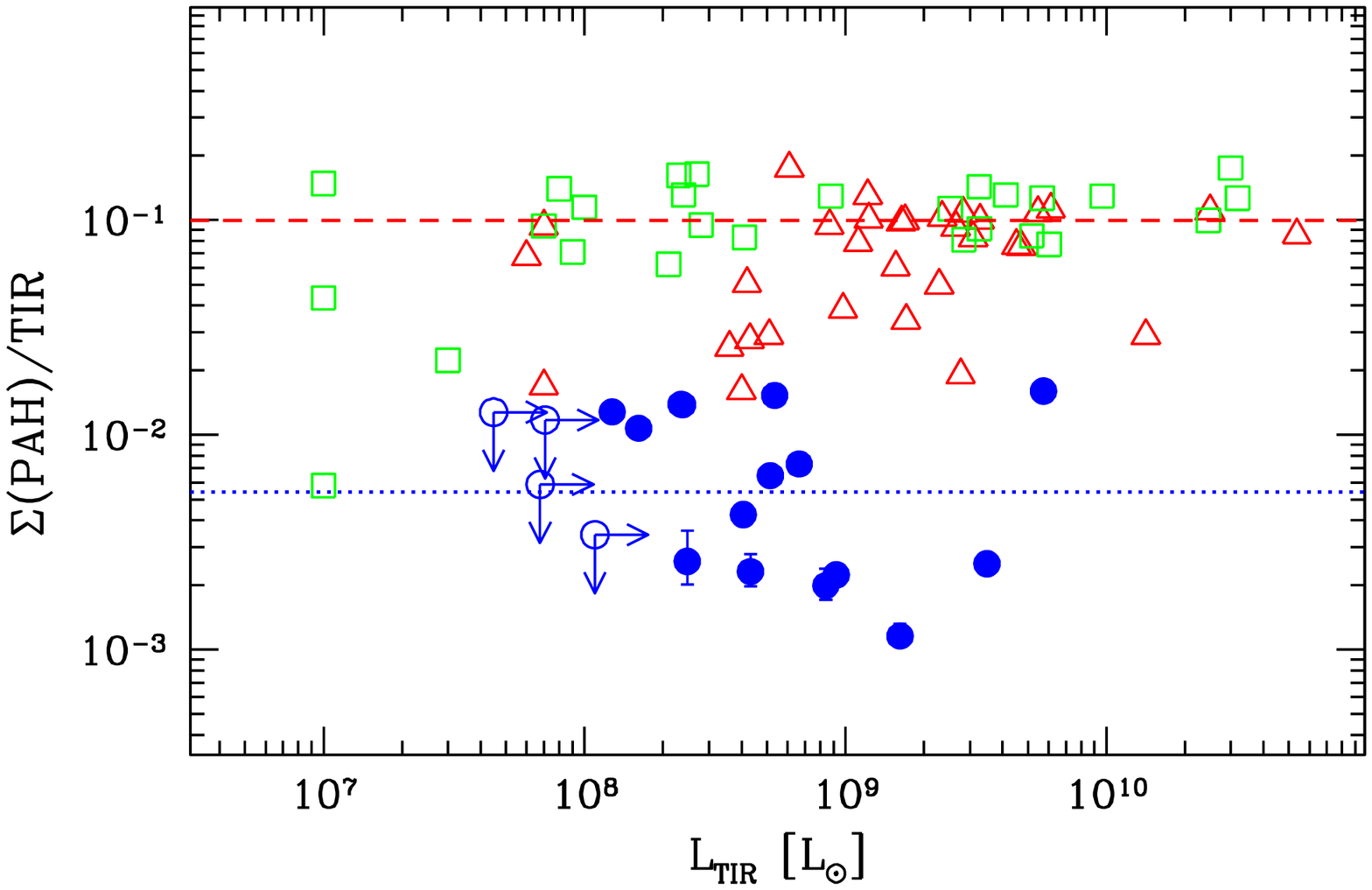} }
\caption{The integrated flux in all the PAH features detected at $\geq 3\sigma$,
relative to the total IR flux, plotted against the 
the total IR flux as described in the text.
BCDs are plotted as filled blue circles.
The non-AGN SINGS galaxies are plotted as open green squares, and the
AGN SINGS galaxies (Seyferts and LINERs) as open red triangles.
The blue dotted line shows the median $\Sigma$(PAH)/TIR
for our sample of 0.54\%; the red dashed line shows the
mean  $\Sigma$(PAH)/TIR for the SINGS sample of $\sim$10\% \citep{smith07}.
\label{fig:pahtot_tir}}
\end{figure}

\begin{figure}
\centerline{
\includegraphics[angle=0,width=0.5\linewidth,bb=19 159 588 547]{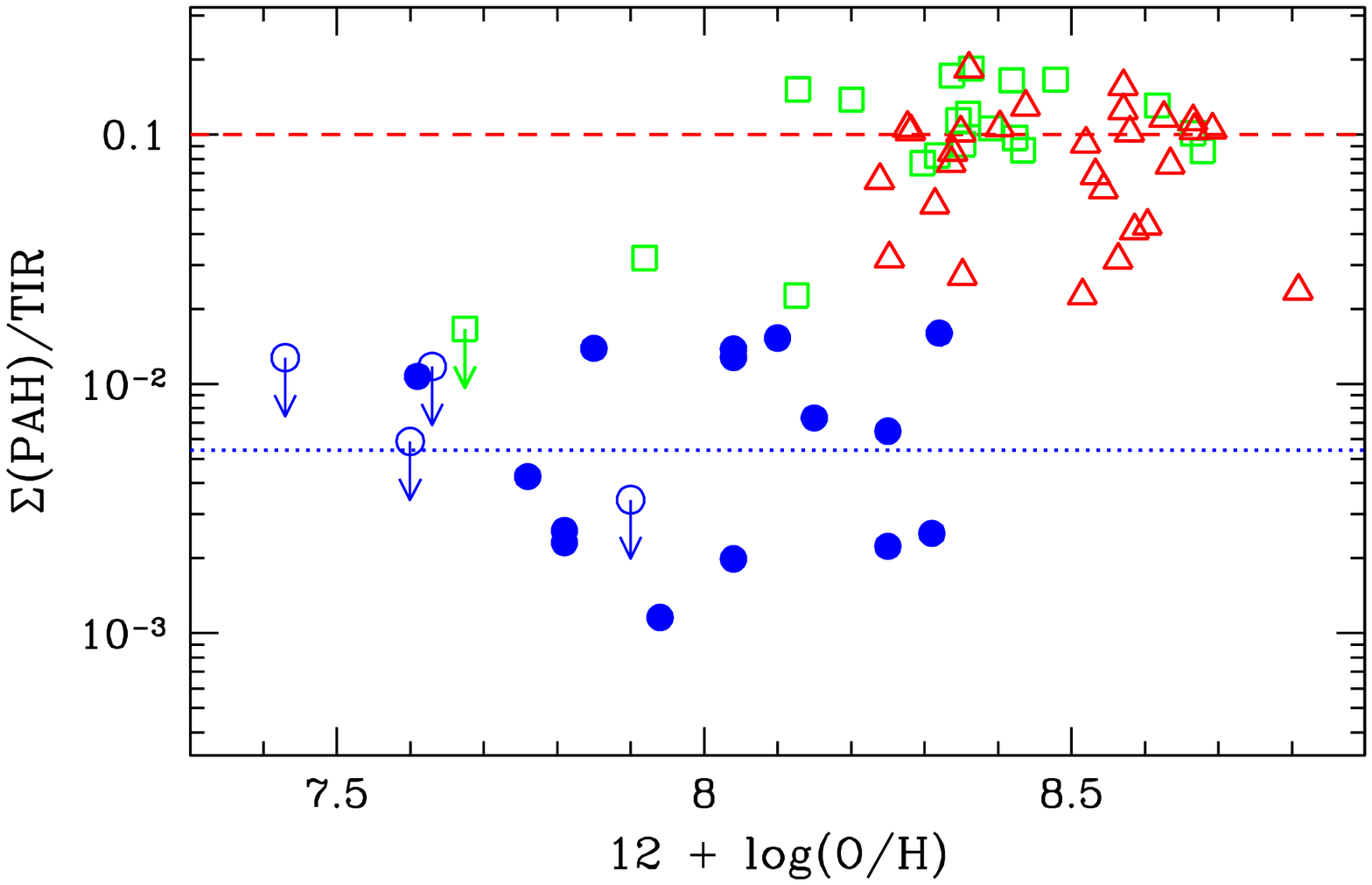} 
\hspace{-0.094\linewidth}
\includegraphics[angle=0,width=0.5\linewidth,bb=19 159 588 547]{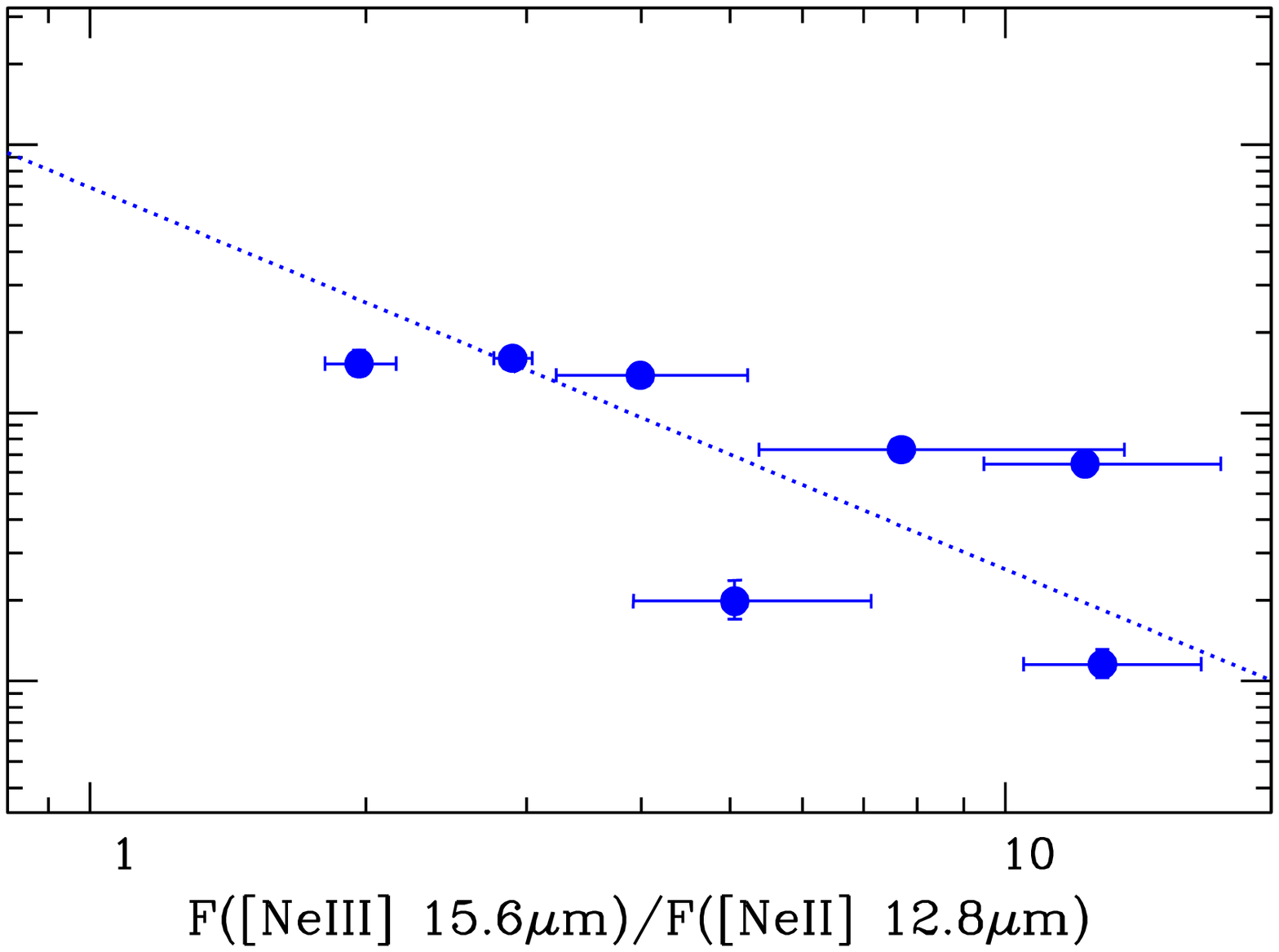} }
\caption{The integrated flux in all the PAH features detected at $\geq 3\sigma$,
relative to the total IR flux, plotted against the 
the oxygen abundance (left panel) and the \neiii/\neii\ flux ratio (right).
BCDs are shown as filled blue circles,
the non-AGN SINGS galaxies as open green squares, and the
AGN SINGS galaxies (Seyferts and LINERs) as open red triangles.
The horizontal dotted line shows the BCD sample mean $\Sigma$(PAH)/TIR of 0.54\%,
and the dashed line the SINGS mean of 10\%.
The right panel shows only those objects with 3$\sigma$ detections or 
better in each line; the best-fit linear regression is shown as a dotted line.
\label{fig:pahtot_ohne}}
\end{figure}

\begin{figure}
\centerline{
\includegraphics[angle=0,width=0.6\linewidth,bb=19 159 588 547]{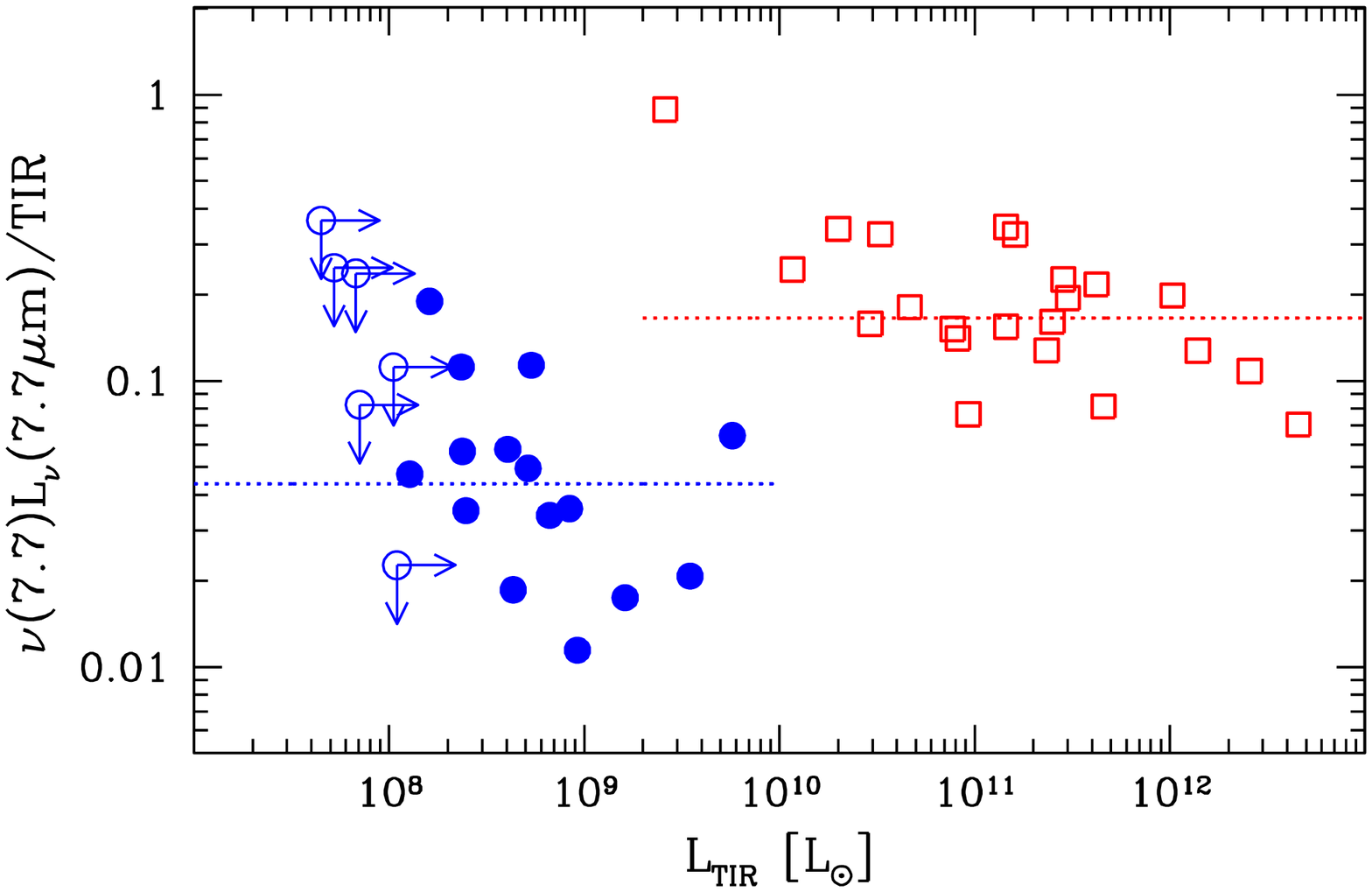} }
\caption{$\nu L_\nu$(7.7\micron)/TIR plotted again \ltir\ for the local
starbursts in \citet{houck07} (shown as open red squares)
and the BCDs (filled blue circles).
The dotted horizontal line segments correspond to the sample means:
0.044 for the BCDs, and 0.167 for the starbursts.
\label{fig:pah77}}
\end{figure}

\clearpage

\begin{figure}
\centerline{
\includegraphics[angle=0,width=0.6\linewidth]{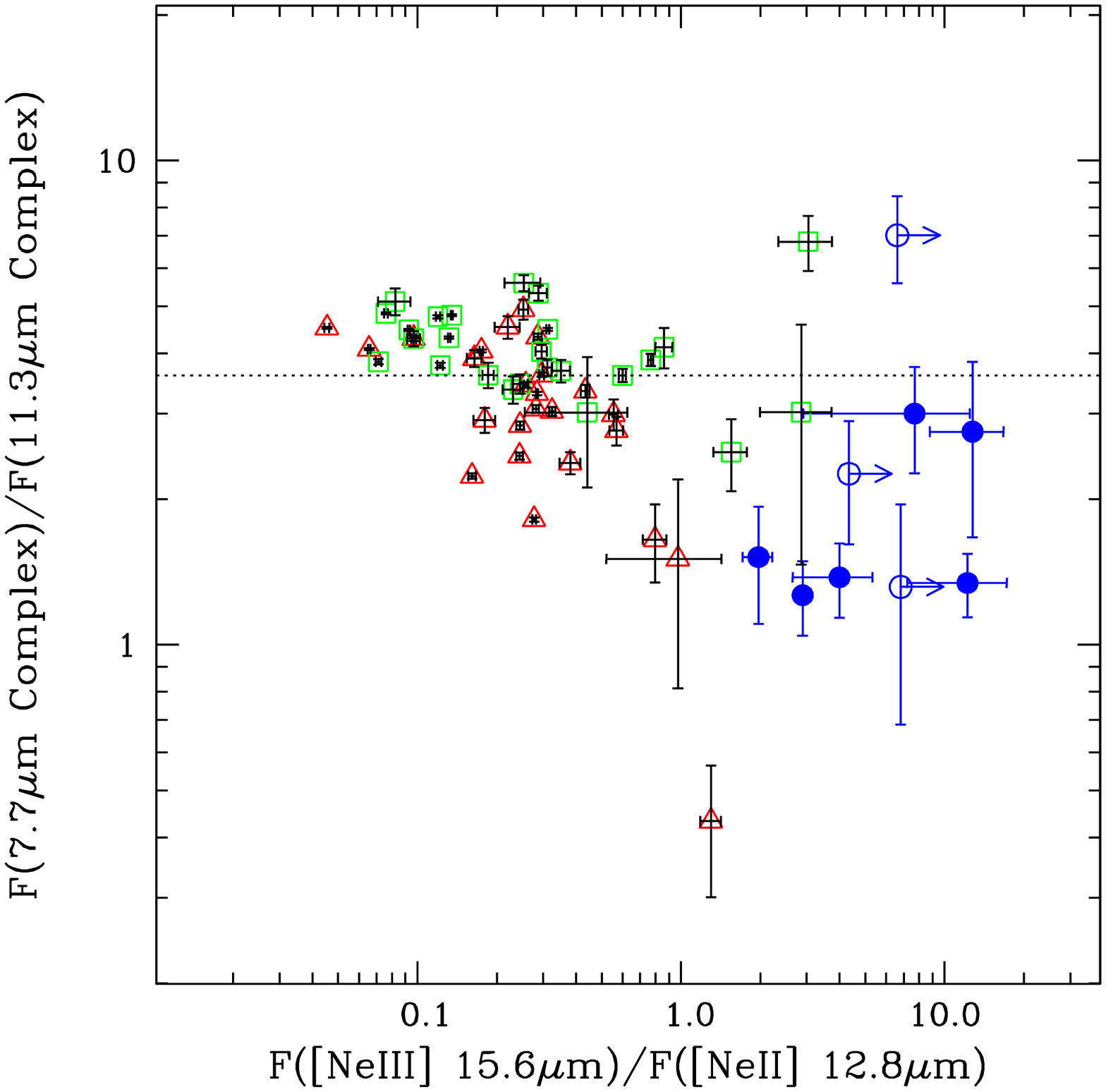} }
\caption{The ratios of the 7.7/11.3\,\micron\ bands plotted against \neiii/\neii.
BCDs with $\simgt 3\sigma$ detections are shown as filled blue circles,
SINGS AGN as open red triangles, and SINGS HII nuclei as open green squares.
The mean 7.7/11.3\,\micron\ ratio for the SINGS sample is shown as a horizontal
dotted line \citep{smith07}.
\label{fig:pah_ne}
}
\end{figure}

\clearpage

\begin{figure}
\centerline{
\includegraphics[angle=0,width=0.6\linewidth]{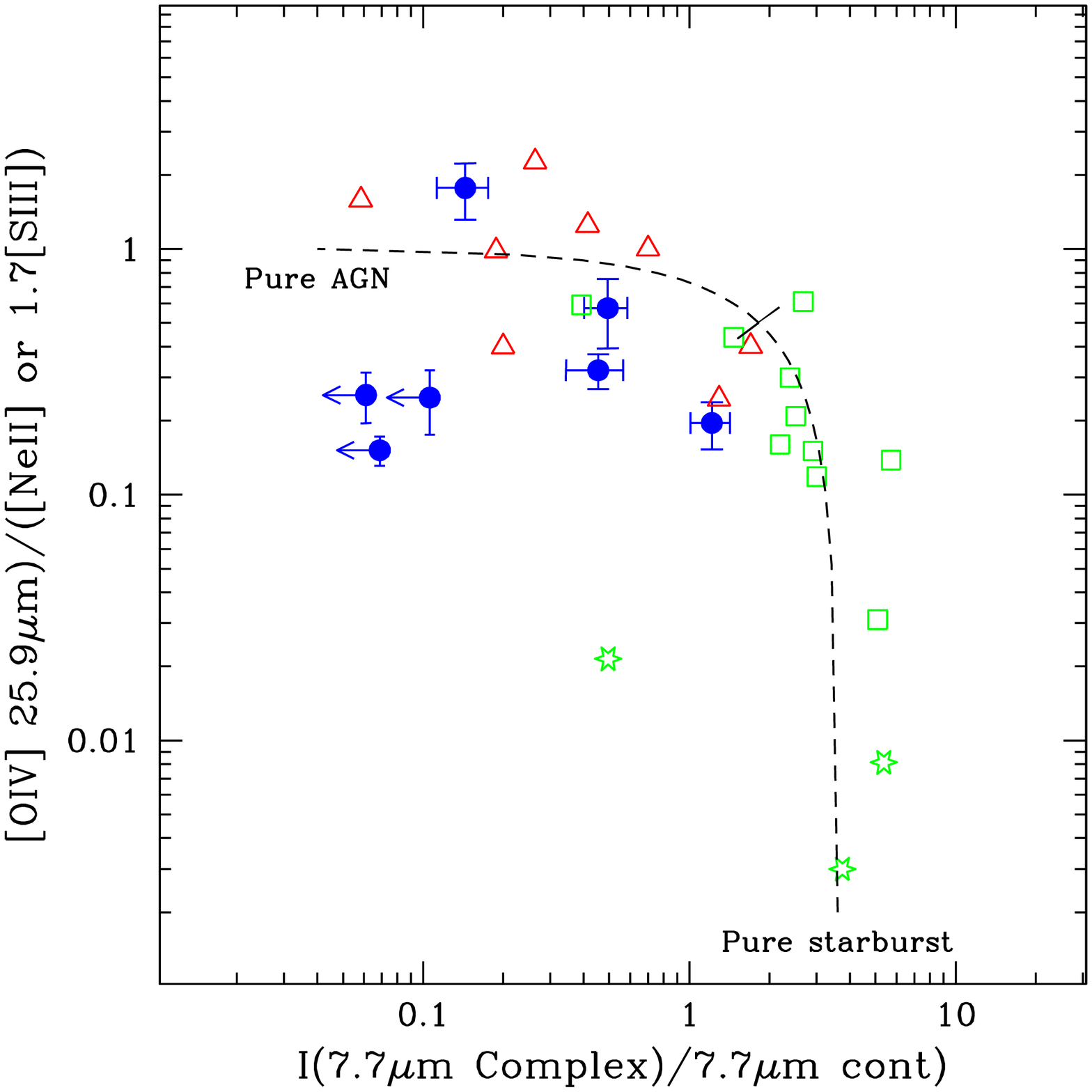} }
\caption{\oiv\/\neii\ (or 1.7$\times$33.5\micron\ \siii) vs. the strength of
the 7.7\,\micron\ PAH feature.
The data for starburst galaxies, ULIRGs, and AGN are taken from \citet{genzel98},
and are marked by green open stars, green open squares, and red open triangles, respectively.
Our BCDs as filled blue circles, with upper limits for the PAH features shown
by left-pointing arrows; only the (8) BCDs with $\simgt 3\sigma$ \oiv\ detections
are shown.
The ``mixing curve'' shown by a dashed line is made following \citet{genzel98};
various fractions of total luminosity in an AGN and a starburst are combined,
assuming a pure AGN has \oiv/\neii/ $\sim$1, PAH strength $\sim$0.04,
and a pure starburst \oiv\neii/ $\sim$0.002, PAH strength $\sim$3.6.
The 50\% mark is shown as a line segment along the mixing curve.
\label{fig:genzel_diag} }
\end{figure}

\begin{figure}
\centerline{
\includegraphics[angle=0,width=0.6\linewidth,bb=19 159 588 547]{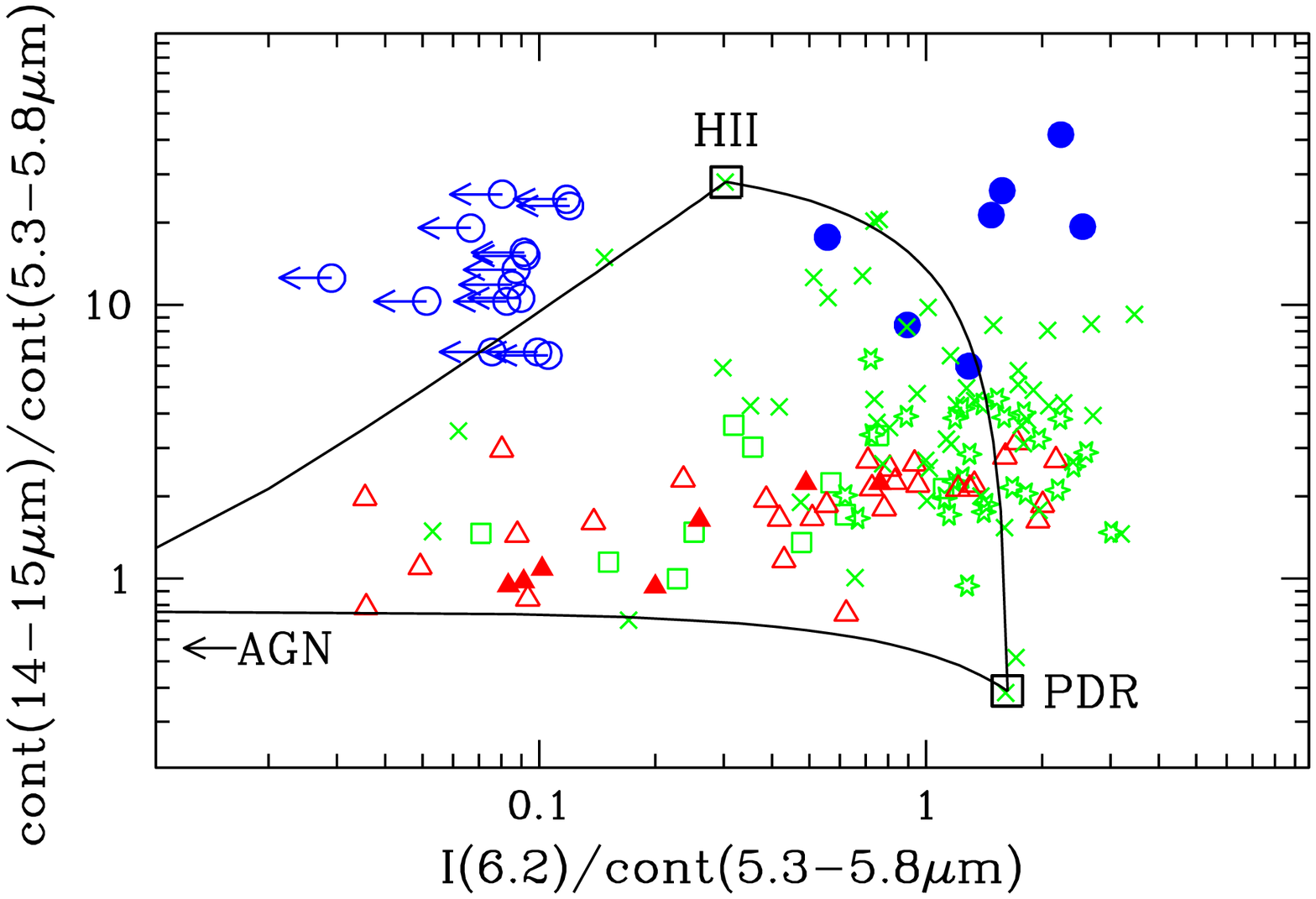} }
\caption{Diagnostic diagram as in \citet{peeters04a}, adapted from \citet{laurent00}.
The templates for the ``mixing'' curves are empirical,
defined to be the positions of M\,17 (\hii\ region), NGC\,7023
(an exposed PDR), and a pure AGN (assumed to have negligible 6.2\,\micron\ flux
in the PAH band).
Data are taken from \citet{peeters04a}, and include
Seyfert 1s (filled red triangles), Seyfert 2s (open red triangles),
ULIRGs (open green squares),
starburst galaxies (open green stars), and
Galactic star-forming regions (green $\times$).
Again, BCDs are shown as filled blue circles, with upper limits
for the 6.2\,\micron\ PAH feature shown as left-pointing arrows.
\label{fig:laurent_diag} }
\end{figure}

\clearpage


\begin{figure}
\centerline{
\includegraphics[angle=0,width=0.6\linewidth,bb=19 159 588 547]{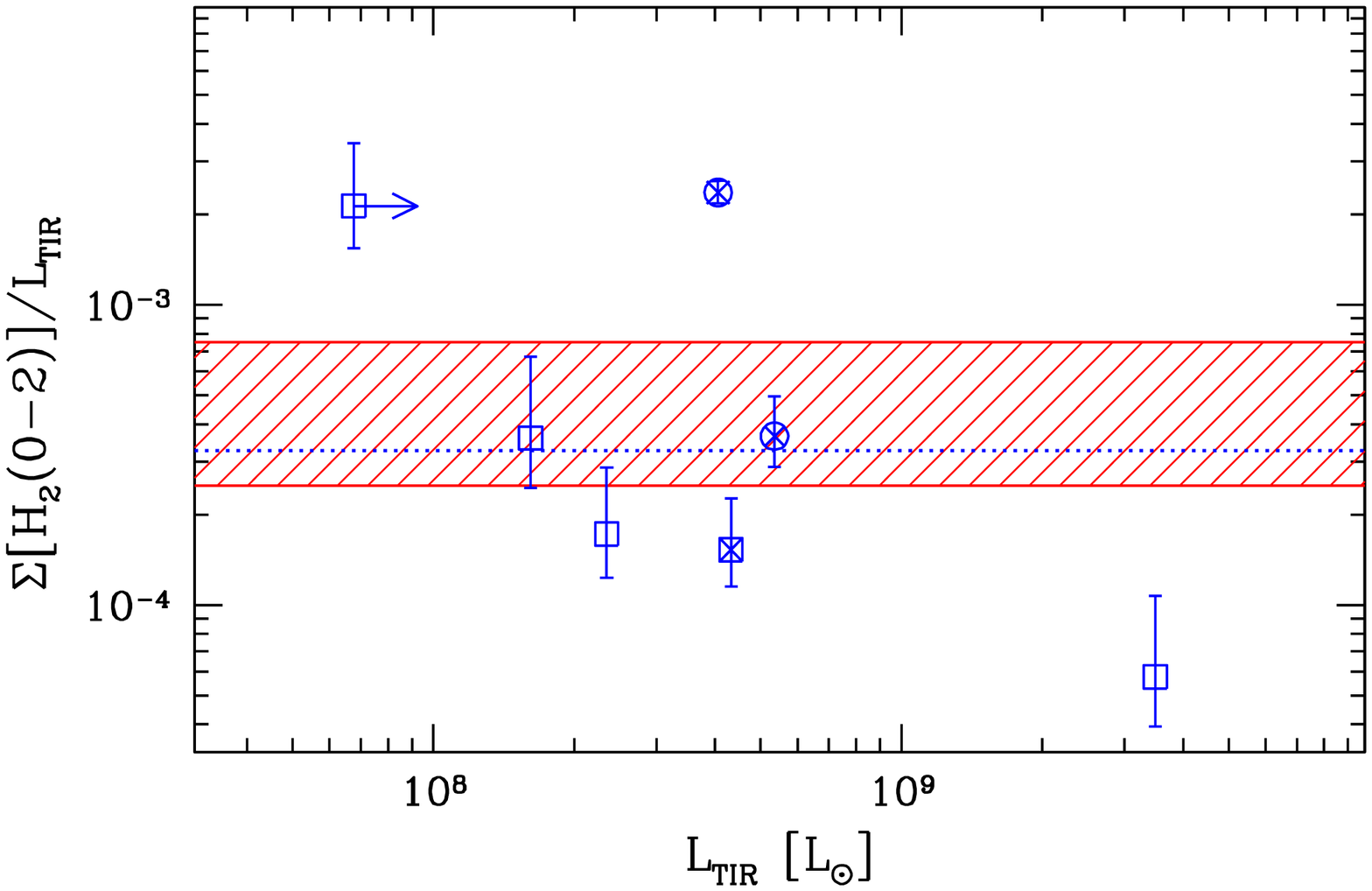} }
\caption{Sum of S(0) to S(2) \htwo\ luminosity 
normalized by and plotted against TIR.
All detections with $\simgt2\sigma$ are plotted; a $\times$
denotes the 3$\sigma$ detections.
Different symbols distinguish which \htwo\ lines are
considered in the sum: 
Mrk\,996 with S(0) and S(1), and \cgcg\ with S(1) and S(2), 
are shown by open circles;
the remaining BCDs with only S(1) or S(2) by open squares.
The red hatched area shows the range for SINGS galaxies
\citep{roussel07}, and
the blue dotted horizontal line indicates the BCD mean,
without UM\,311 (\ltir$\sim 3\times10^{9}$\lsun) because
of the likely overestimate of its TIR because of source crowding (see text).
\label{fig:h2tir} }
\end{figure}

\begin{figure}
\centerline{
\includegraphics[angle=0,width=0.28\linewidth,bb=19 145 588 712]{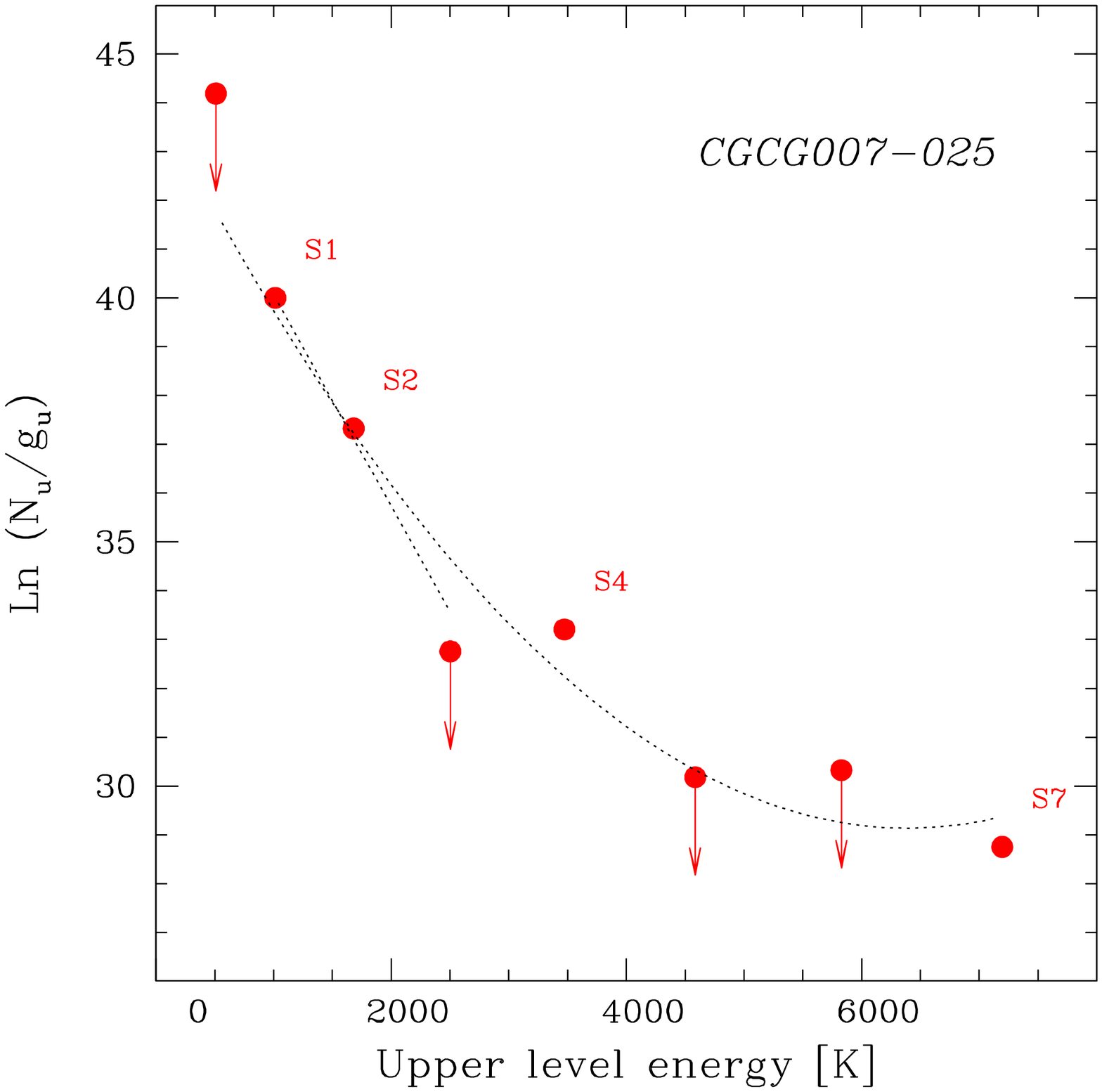} 
\hspace{-0.9cm}
\includegraphics[angle=0,width=0.28\linewidth,bb=19 145 588 712]{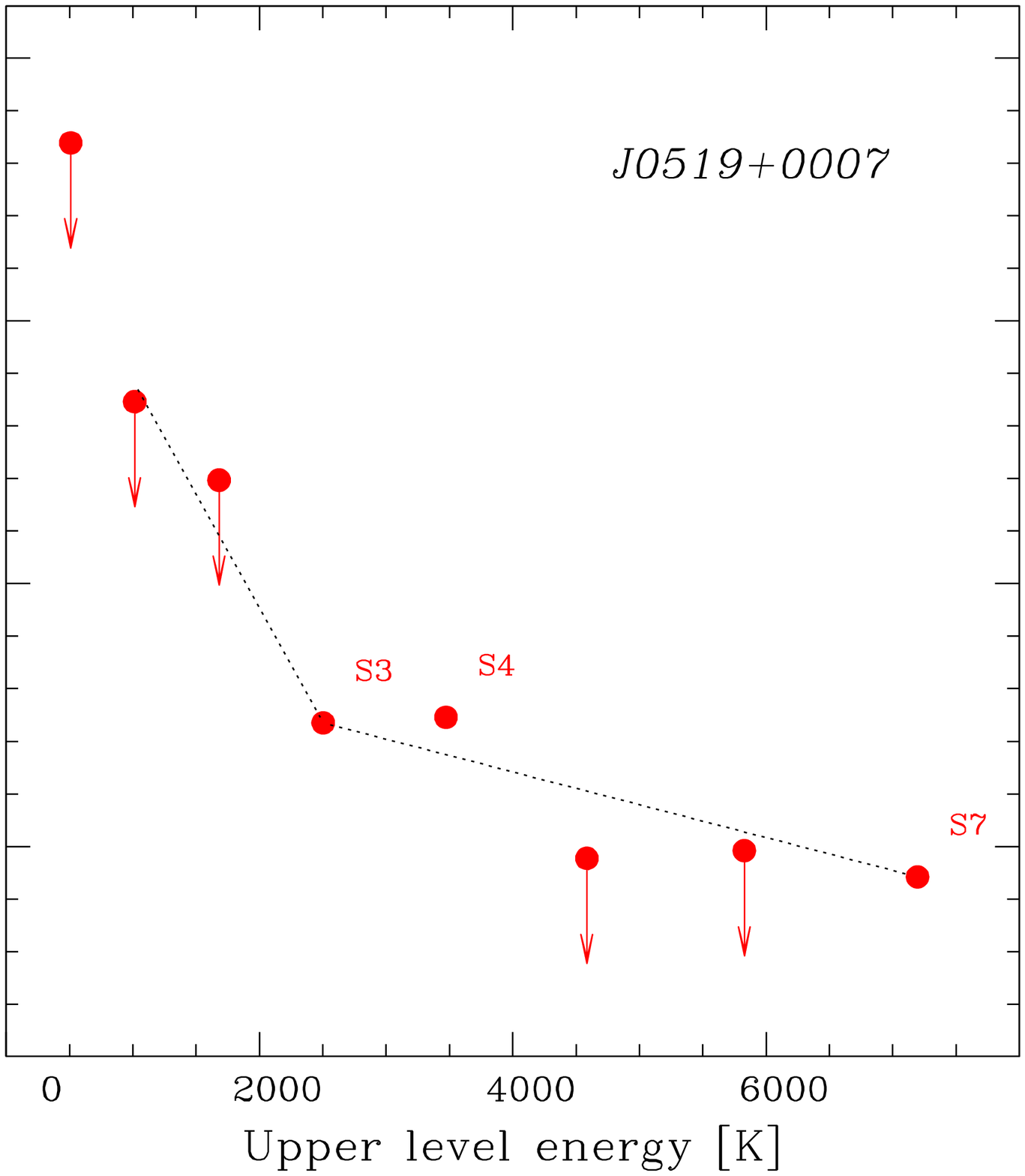} 
\hspace{-0.9cm}
\includegraphics[angle=0,width=0.28\linewidth,bb=19 145 588 712]{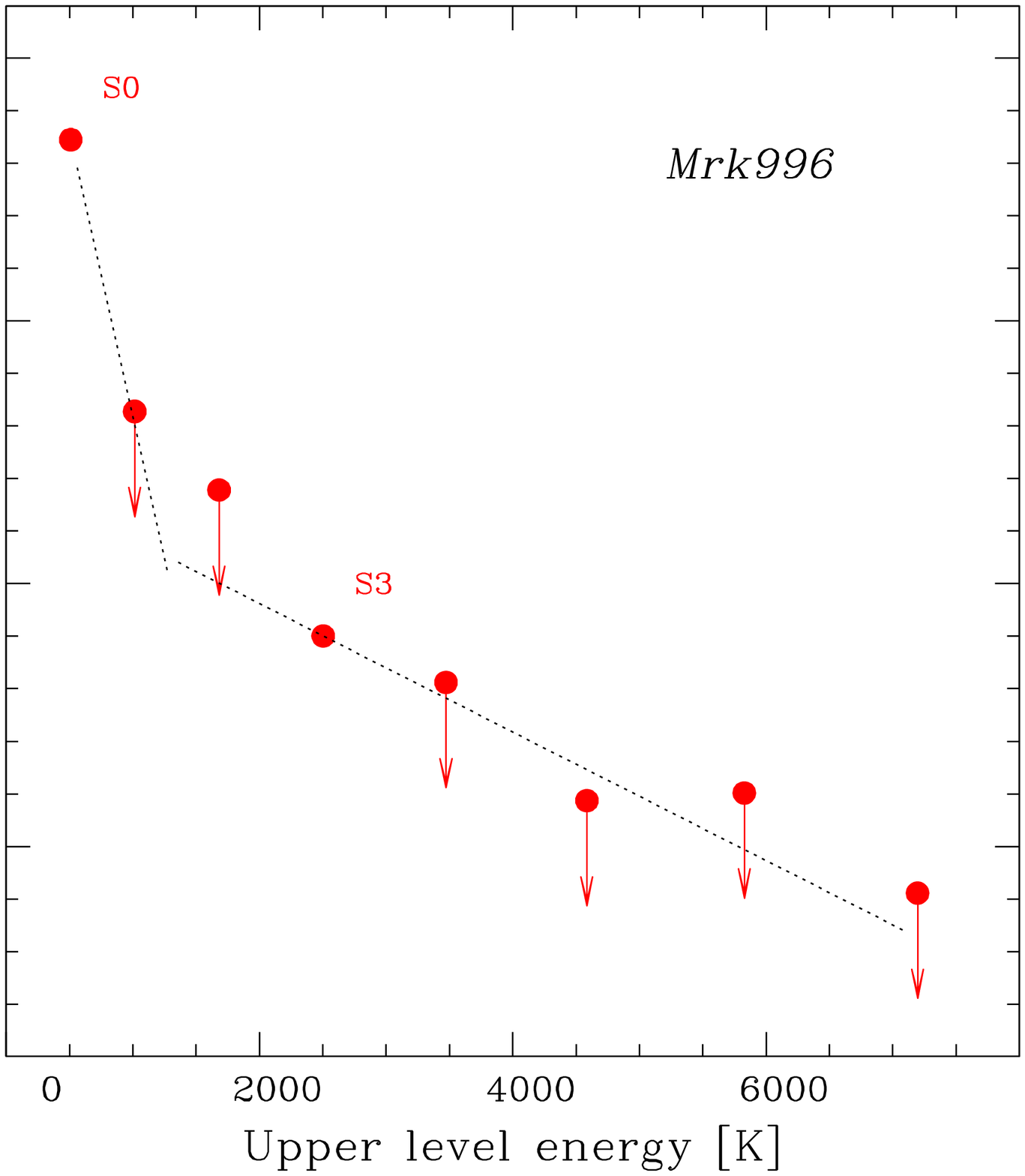} 
\hspace{-0.9cm}
\includegraphics[angle=0,width=0.28\linewidth,bb=19 145 588 712]{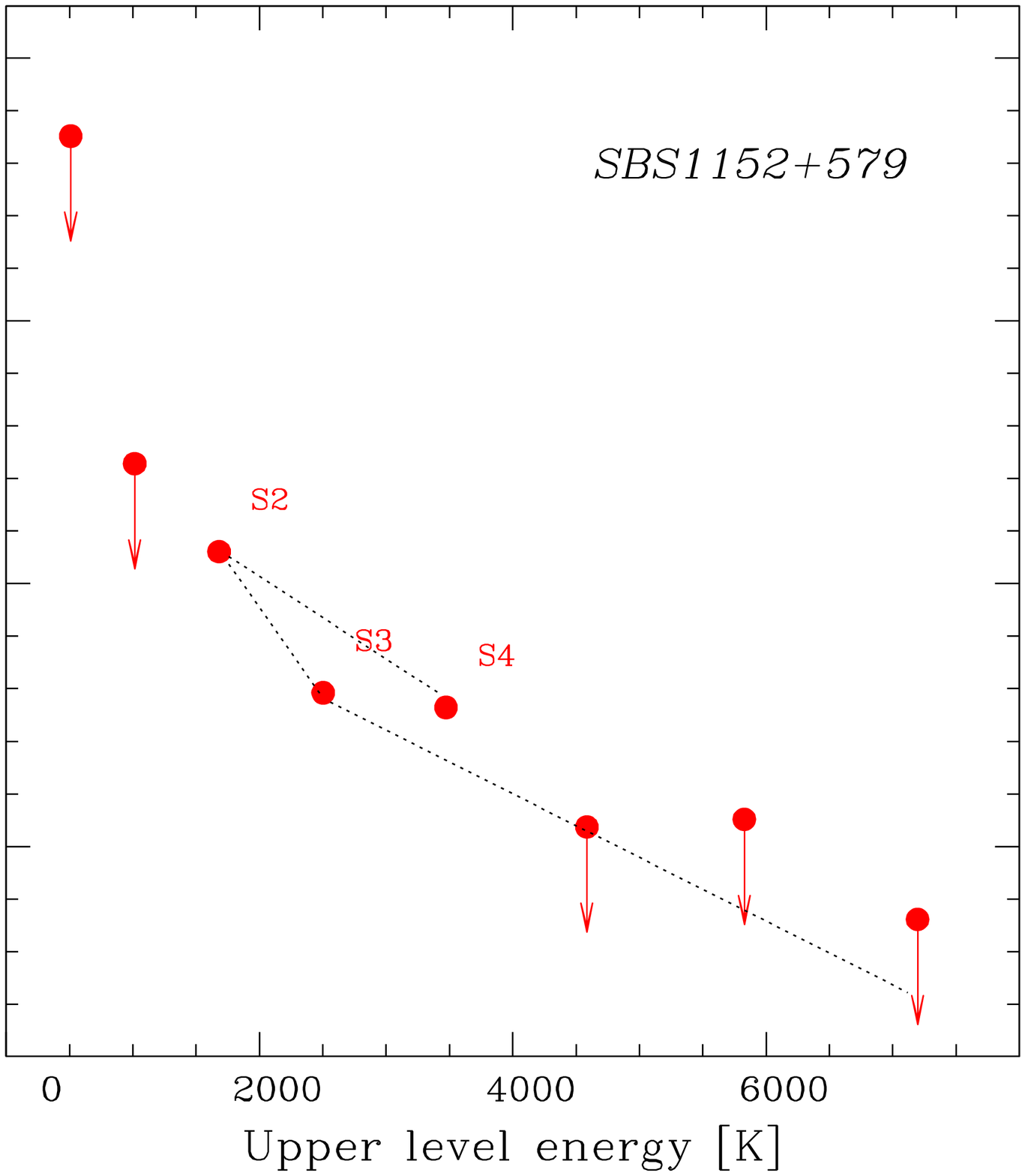}
}
\caption{Excitation diagrams for the four BCDs with multiple significant \htwo\
detections and at least one significant ($\simgt 2.5\sigma$ for SBS\,1152$+$579)
detection in a low-order transition (see text).
3$\sigma$ upper limits are shown with vertical arrows.
The dotted lines correspond to best-fit estimates of the temperature of the 
line-emitting gas (see text).
The parabolic fit for \cgcg\ is an experiment to show the
continuity of the data, and thus the continuous range of gas temperatures. 
\label{fig:h2excitation} }
\end{figure}

\clearpage


\end{document}